\documentclass[12pt,preprint]{aastex}

\usepackage{dutchcal}

\usepackage{natbib}

\usepackage{amsmath}
\numberwithin{equation}{section}

\usepackage[margin=1.00in]{geometry}

\begin{document}

\title{A Mathematical Model For the Spread of a Virus}

\author{N. R. Sheeley, Jr}
\affil{Alexandria VA November 13, 2020; revised May 28, 2021}

\begin{abstract}
This paper describes a mathematical model for the spread of a virus through an isolated population
of a given size.  The model uses three, color-coded components, called molecules (red for infected
and still contagious, green for infected, but no longer contagious, and blue for uninfected).  In
retrospect, the model turns out to be a digital analog of the well-known SIR-model of
Kermac \& McKendrick (1927).  In our RGB-model, the accumulated infections go through three phases, beginning at a very low level, then changing to a transition ramp of rapid growth, and ending in a
plateau of final values.  Consequently, the differential change or growth rate begins at 0, rises to
a peak corresponding to the maximum slope of the transition ramp, and then falls back to 0 in a time comparable to the time to reach the peak.  The properties of these time variations, including
the slope, duration, and height of the transition ramp, and the width and height of the infection rate,
depend on a single
parameter.  This parameter is the time that a red molecule is contagious divided by the average time
between collisions or encounters of the molecules.  Various milestones, including the starting
time of the transition ramp, the time that the accumulating number of infections obtains its maximum
slope, and the location of the peak of the infection rate all depend on the size of the population in
addition to the contagious lifetime ratio.  Explicit formulas for these quantities are derived and
summarized.  Finally, Appendix E has been added to describe the effect of vaccinations.
\pagebreak
\end{abstract}

\tableofcontents
\pagebreak

\section{Introduction}
The spread of the corona virus and its infection as covid-19 has become a central issue in all of our lives.
 Curves showing the accumulated infections as well as the
infection rate occur in the news every day.  Most of us are `social distancing' in
order to `flatten the curve' and we are receiving a variety of advice about how long it will take before it is
safe to return to our normal activities.  Consequently, I thought it would be both interesting and useful to
calculate these curves using a simple model for the infection.  It would be
interesting just to see if I could actually  do the calculation, and it would be useful if the solution made
sense and gave me an appreciation of how the spreading takes place.  The idea is not to perform a
fully realistic calculation, but to learn what the relevant parameters are and to gain some insight about
how they affect the spread of the disease.  This paper is a summary of what I have learned.

\section{The Model}
Let us regard the affected population as a gas of colliding `molecules' of a given number, $N_{0}$, not
nearly as big as Avogadro's number, $N_{0}= 6 \times 10^{23}$ atoms/mole, but $\sim 10^{3}-10^{8}$,
comparable to populations of various regions of the country and the world.  So we imagine that a population
is like a gas of $N_{0}$ molecules, colliding with each other in a region of constant volume.  Presumably,
this gas will have characteristics of a real gas like temperature, pressure, and mean
free path that can be determined.  However, for the moment, we will bypass this
interesting question and move forward to calculate the spread of the infection.  
\medskip

We assume the problem starts with $N_{0}-1$ molecules that are uninfected and one molecule that is
both infected and contagious, in the sense that it will infect another molecule when it collides with that
molecule.  For simplicity, we suppose that there
is an average time, ${\tau}_{s}$, between collisions, which we will use as the step
time in our calculation.  Referring to the left panel of Figure~1, we begin at step 0
with the index $j=0$ when a contagious (red)
\begin{figure}[h!]
 \centerline{%
 \fbox{\includegraphics[bb=105 18 525 785,clip,angle=90,width=0.95\textwidth]{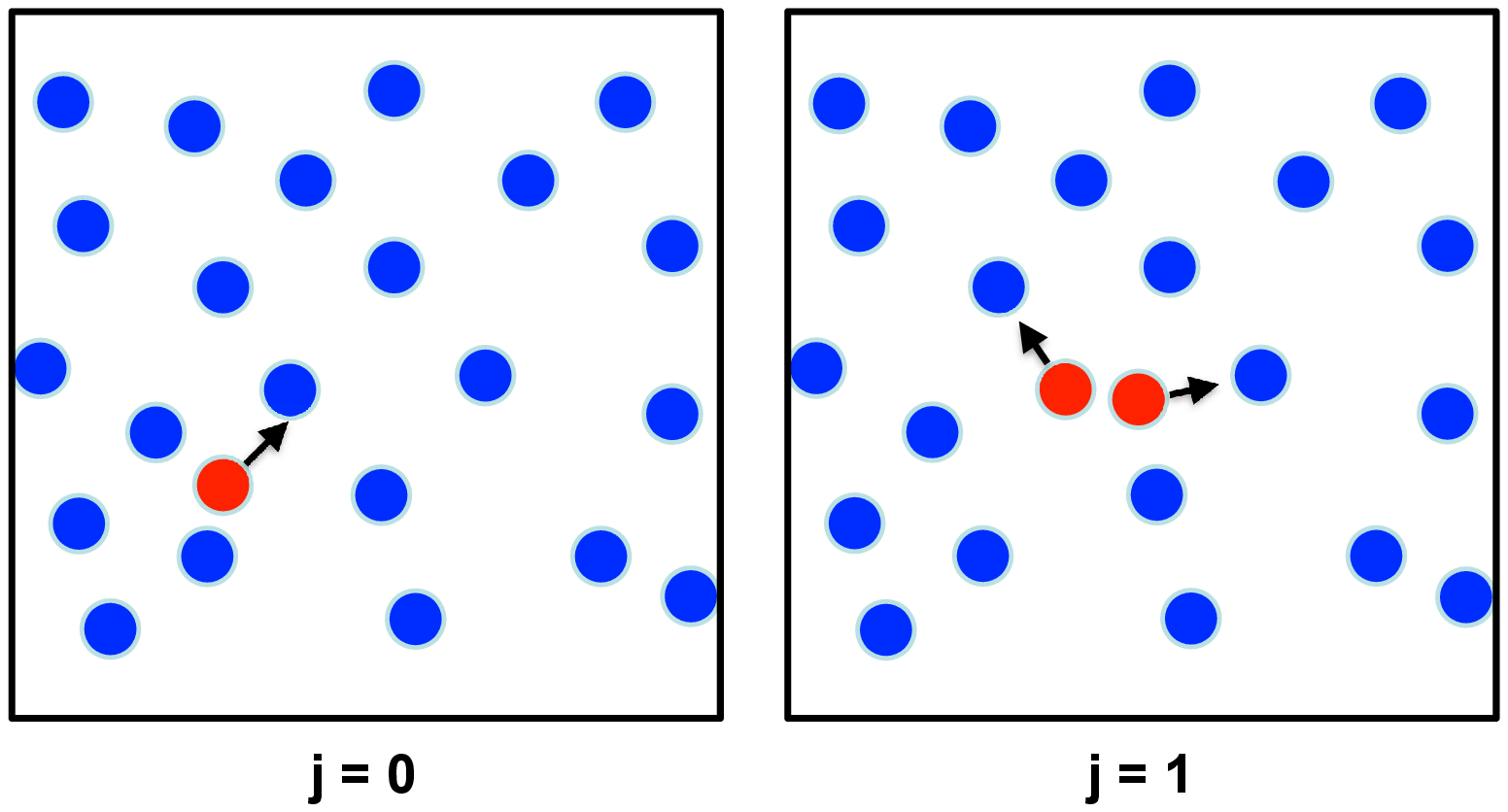}}}
 \caption{(left) A gas of uninfected (blue) molecules with a contagious (red) molecule inserted at the
initial time $j=0$.  (right) The same gas after one collision time when the red molecule has encountered
a blue molecule and infected it (\textit{i.e.} changed it to red).}
\end{figure}
\noindent
molecule is inserted into the gas of $N_{0}-1$
uninfected (blue) molecules.  By the next step in the right
panel of Figure~1, the red molecule has collided with a blue one, making it red, so
that two red molecules can go on to infect two other blue molecules in the next collision.
\medskip

An important aspect of this problem is the finite time that a red
molecule remains contagious.  For the corona virus, we have heard that the
lifetime is on the order of 2 weeks, the time that some potentially infected
people have been told to quarantine.  However, we will delay our treatment of
this effect until later in this paper when we will have gained some insight and
experience by solving the simpler problem for a permanently contagious virus.
\medskip

\section{Permanently Contagious Red Molecules}
\subsection{Deriving the Infection-Spreading Equation}
Let us assume that the problem has advanced to the $j-1$ step at which
point the number of infected molecules is represented by the symbol,
$n_{inf}^{(j-1)}$, where the exponent, $j-1$, is a tag and not a power to
which $n_{inf}$ is to be raised.  The question is then what is the number
of infected molecules in the next step, as represented by the symbol
$n_{inf}^{(j)}$?  For the first few steps with $j \sim 1$, the probability, $p_{j-1}$,
of colliding with an uninfected molecule will be close to 1 so that $n_{inf}^{(j)}$
would be $2n_{inf}^{(j-1)}$ (\textit{i.e.} just twice the present number of
infected molecules before the collision).  However, for larger values of $j$,
we must consider the case of $p_{j-1}~{\neq}~1$.
\medskip

The probability that a red molecule will collide with another infected molecule is
just $(n_{inf}^{(j-1)}-1)/(N_{0}-1)$, so the probability that it will collide with
an uninfected blue molecule is
\begin{equation}
p_{j-1}~=~1-\frac{n_{inf}^{(j-1)}-1}{N_{0}-1}.
\end{equation}
 Consequently,
\begin{equation}
n_{inf}^{(j)}~=~n_{inf}^{(j-1)}~+~n_{inf}^{(j-1)}p_{j-1},
\end{equation} 
and therefore
\begin{equation}
n_{inf}^{(j)}~=~n_{inf}^{(j-1)}~+~n_{inf}^{(j-1)} \left \{ 1-\frac{n_{inf}^{(j-1)}-1}{N_{0}-1}
\right \}.
\end{equation}
Now, for simplicity, we normalize the number of molecules to the total number,
$N_{0}$, in the gas, and define ${\nu}_{j}~=~n_{inf}^{(j)}/N_{0}$ as the normalized quantity.
Thus, ${\nu}_{0}~=1/N_{0}$, and
\begin{equation}
{\nu}_{j}~=~{\nu}_{j-1}~+~(\frac{1}{1-{\nu}_{0}}){\nu}_{j-1}(1-{\nu}_{j-1})~=~
{\nu}_{j-1} \left \{1+(\frac{1}{1-{\nu}_{0}})(1-{\nu}_{j-1}) \right \}.
\end{equation}

\subsection{Plotting the Spread}
Next, in Figure~2, we plot the growth curve, ${\nu}_{j}$, and the growth rate,
${\Delta}{\nu}_{j}={\nu}_{j}-{\nu}_{j-1}$, versus the number of steps ($j$) or collision
times ($j{\tau}_{s}$) that have elapsed since the start of the infection.  The difference
equation Eq.(3.4) gives the sequence of points that are plotted as red dots.  (The
continuous blue curves are plotted from the analytical solution to Eq. (3.4)
derived in the next subsection.)  As one can see, for this population of
$N_{0}=1 \times 10^5$, the
\begin{figure}[h!]
 \centerline{%
 \fbox{\includegraphics[bb=88 260 530 720,clip,width=0.46\textwidth]{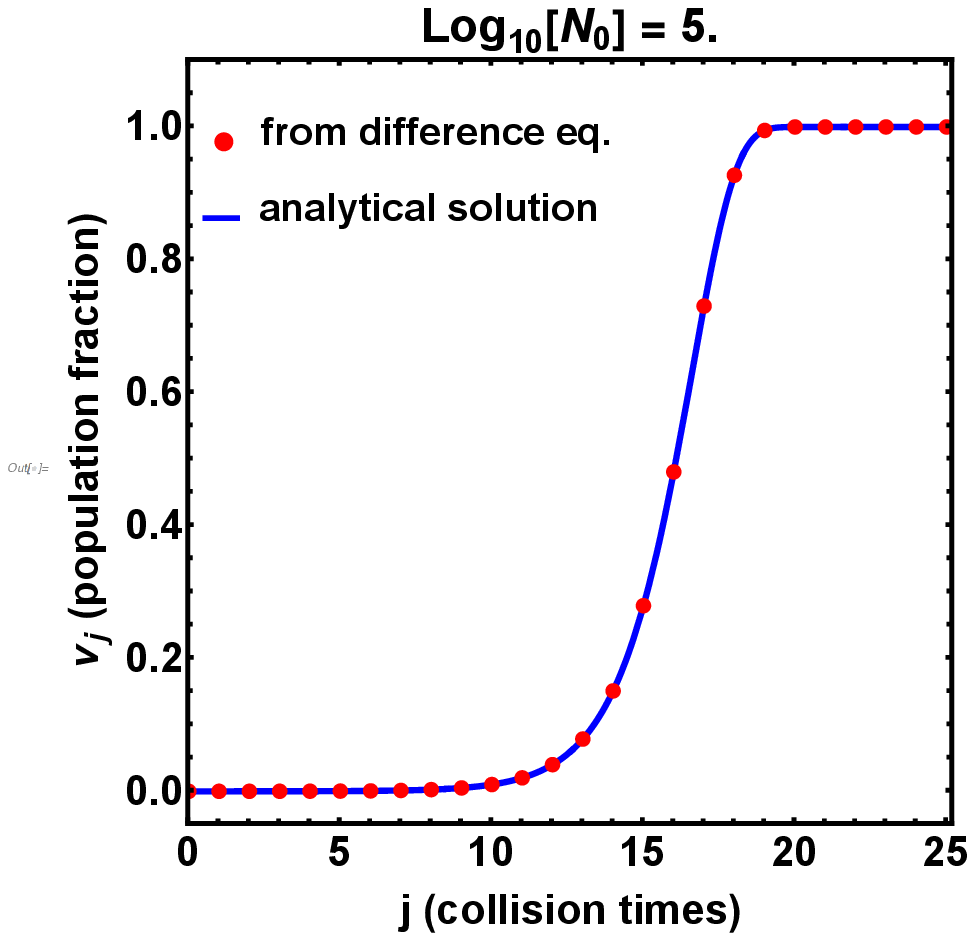}}
\hspace{0.01in}
 \fbox{\includegraphics[bb=88 260 540 720,clip,width=0.47\textwidth]{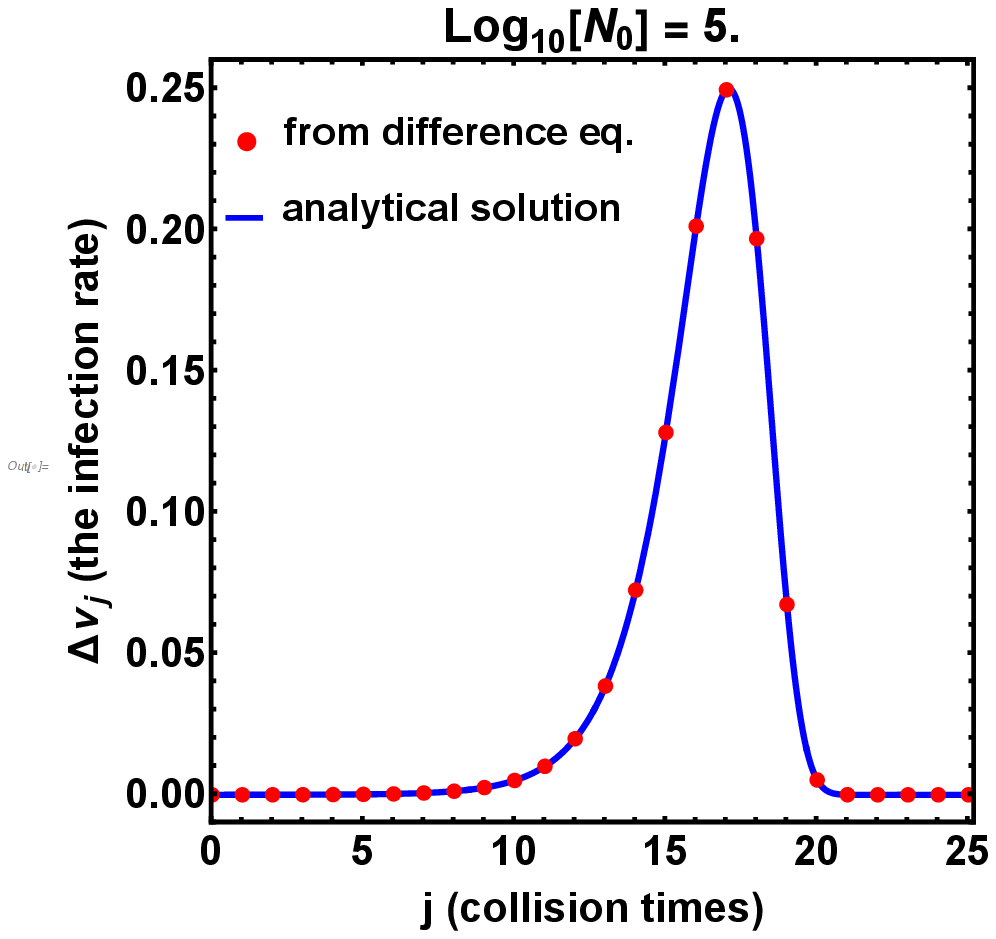}}}
 \caption{(left): ${\nu}_{j}$ from Eq.(3.4) (red points) and Eq.(3.13a) (blue curve); (right):
${\Delta}{\nu}_{j}$ from Eq.(3.11) (red points) and Eq.(3.13b) (blue curve).}
\end{figure}
\noindent
 number
of infections reaches approximately one half of its final value after about 16 collisions,
which is one collision before the growth rate in the right panel reaches its peak
value of 0.25 at $j_{max}~{\approx}~17$.  At this point, 1/4 of the total population is being
infected per collision time.  The full width at half maximum is
approximately 4 collision times, which means that the area under the curve
is about 1, consistent with the final height of the infection curve in the left
panel.  Also, note that each profile is asymmetric.  In the left panel, the trip from
the point of maximum slope to the upper level is shorter than the trip
from the lower level to the point of maximum slope.  Likewise, in the right panel,
the trip from the peak to the final base level is shorter than the trip from the initial
base level to the peak.
\medskip

Although the shape of the growth-rate profile does not vary with $N_{0}$,
its location, $j_{max}$, does vary with $N_{0}$, as shown by the measurements
that are plotted as red dots in Figure~3.
\begin{figure}[h!]
 \centerline{%
 \fbox{\includegraphics[bb=85 235 537 720,clip,width=0.70\textwidth]{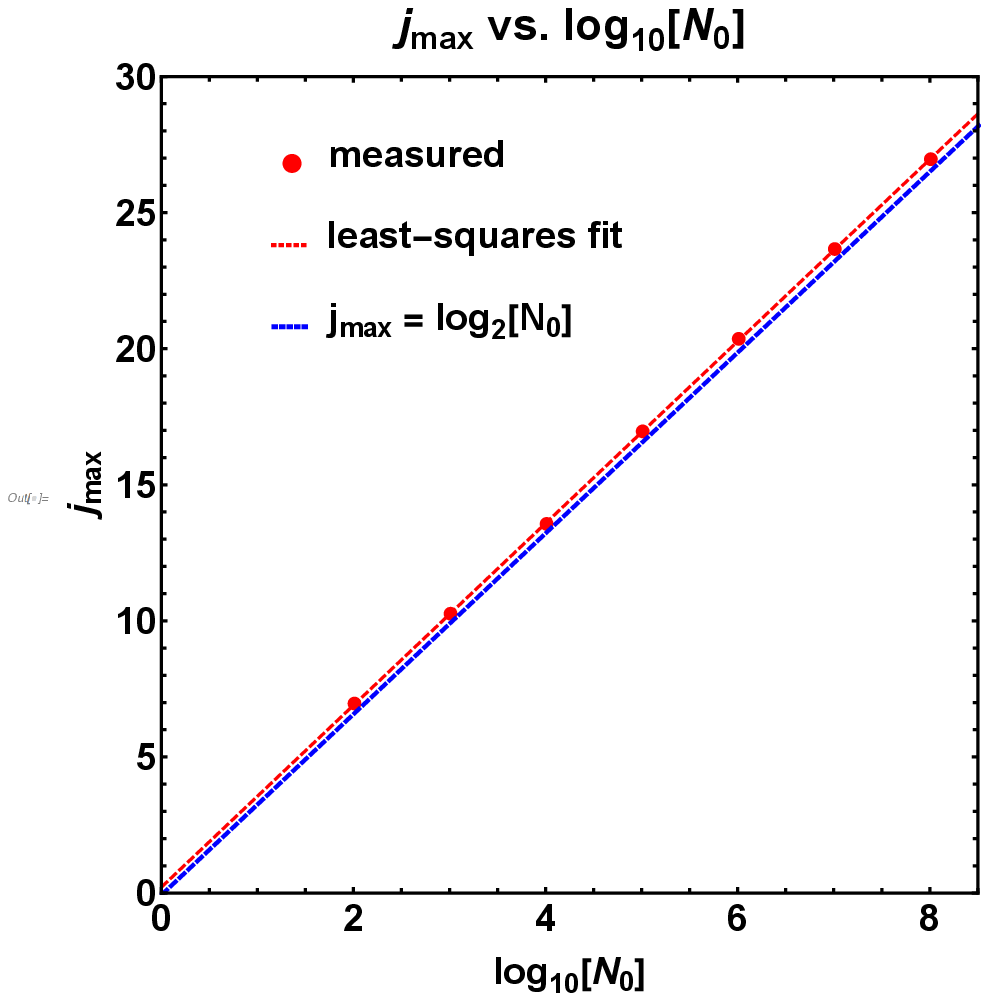}}}
 \caption{Measurements of the peak location, $j_{max}$, plotted versus
$\log_{10}N_{0}$ (red dots), and their least-squares linear fit (dashed red line), compared with
the relation, $j_{max}=\log_{2}N_{0}~{\approx}~
3.322~{\log_{10}N_{0}}$ shown by the dashed blue line.}
\end{figure}
\noindent
The least-squares best fit to these
measurements is shown by the dashed red line, whose equation is
\begin{equation}
j_{max}~=~0.286~+~3.343 \log_{10}N_{0}.
\end{equation}
This red line is nearly coincident with the dashed blue line, which represents the relation
\begin{equation}
j_{max}~=~\log_{2}N_{0}~=~\frac{\log_{10}N_{0}}{\log_{10}2}~{\approx}~3.322~\log_{10}N_{0},
\end{equation}
which is the number of two-factor collisions that it takes to reach the total population, $N_{0}$.

\subsection{Analytical Solution of the Difference Equation}

It is instructive to see if we can find an analytical solution to Eq. (3.4).  Let's
begin with the growth equation in the form
\begin{equation}
{\nu}_{j}~=~{\nu}_{j-1} \left \{1+(\frac{1}{1-{\nu}_{0}})(1-{\nu}_{j-1}) \right \}~
{\approx}~{\nu}_{j-1} (2-{\nu}_{j-1}).
\end{equation}
Initially, ${\nu}_{j-1}<<1$, so that ${\nu}_{j}~{\approx}~2{\nu}_{j-1}=
2^{2}{\nu}_{j-2}=\cdots=2^{n}{\nu}_{j-n}$.  Setting $j=n$, we obtain the solution
${\nu}_{n}=2^{n}{\nu}_{0}$, corresponding to an exponentially increasing number
of infections.  So it is the second term that limits the exponential growth and stops the
infection.
\medskip

We can use a similar approach when the second term is included.  In this case, we
change variables by introducing ${\mu}_{j}=1-{\nu}_{j}$ and obtain
\begin{equation}
1-{\mu}_{j}~=~(1-{\mu}_{j-1})[2-(1-{\mu}_{j-1})]~=~(1-{\mu}_{j-1})(1+{\mu}_{j-1})~=~1-{\mu}_{j-1}^{2}.
\end{equation}
Consequently,
 ${\mu}_{j}={\mu}_{j-1}^{2}={\mu}_{j-2}^{4}={\mu}_{j-3}^{8}={\cdot}{\cdot}{\cdot}={\mu}_{j-n}^{2^n}$.
Setting $j=n$, we obtain ${\mu}_{n}={\mu}_{0}^{2^{n}}$.  Finally, changing back to 
${\nu}_{j}$, we have
\begin{equation}
{\nu}_{j}~=~1-(1-{\nu}_{0})^{2^{j}}
\end{equation}
as the solution to Eq.(3.7).
\medskip

Next, noting that $1-{\nu}_{0}~{\approx}~e^{-{\nu}_{0}}$, we can change Eq.(3.9) to
\begin{equation}
{\nu}_{j}~{\approx}~1-e^{-{{\nu}_{0}}2^{j}}~=~1-2^{-({{\nu}_{0}}/ln2)2^{j}}~=~1-2^{-2^{(j-k)}},
\end{equation}
where $({\nu}_{0}/ln2) = 2^{-k}$.  Because ${\nu}_{0}=1/N_{0}$, it follows that
$k=\log_{2}(N_{0}ln2)$, which is 1 unit less than $j_{max}$, the location of the peak of the
growth rate, ${\Delta}{\nu}_{j}={\nu}_{j}-{\nu}_{j-1}$, as we shall see next.
\medskip

We begin by using Eq.(3.7) to evaluate ${\Delta}{\nu}_{j}={\nu}_{j}-{\nu}_{j-1}$.  The result is
\begin{equation}
{\Delta}{\nu}_{j}~=~(\frac{1}{1-{\nu}_{0}}){\nu}_{j-1}(1-{\nu}_{j-1})~{\approx}~{\nu}_{j-1}(1-{\nu}_{j-1}).
\end{equation} 
The quantity, ${\nu}_{j-1}(1-{\nu}_{j-1})$, has its maximum value of 1/4 when ${\nu}_{j-1}=1/2$.
Thus, ${\Delta}{\nu}_{j}$ has its maximum value when ${\nu}_{j-1}=2^{-2^{j-1-k}}=1/2$.  This
happens when $j-1-k=0$, and therefore
\begin{equation}
j_{max}~=~j~=~k+1~=~1~+~\log_{2}(N_{0}ln2),
\end{equation}
which is $j_{max}=\log_{2}N_{0} + 0.471$ when the numerical value of $1+\log_{2}(ln2)$ is used.
Also, because ${\nu}_{j-1}=(1-{\nu}_{0})^{2^{j-1}}~{\approx}~2^{-2^{(j-j_{max})}}$, it
follows that ${\Delta}{\nu}_{j}=2^{-2^{(j-j_{max})}}\left \{ 1 - 2^{-2^{(j-j_{max})}} \right \}$.
Summarizing the results for both ${\nu}_{j}$ and ${\Delta}{\nu}_{j}$, we have
\begin{subequations}
\begin{align}
{\nu}_{j}~=~1~-~2^{-2^{\left \{j-(j_{max}-1) \right \}}},\\
{\Delta}{\nu}_{j}~=~2^{-2^{(j-j_{max})}} \left \{ 1 - 2^{-2^{(j-j_{max})}} \right \}.
\end{align}
\end{subequations}
\medskip

As we saw in Figure~2, these solutions for ${\nu}_{j}$ and ${\Delta}{\nu}_{j}$ agree very
well with the data points that were obtained from the recursive equations, Eq.(3.4) (or (3.7)) and 
Eq.(3.11).  The point at $j=17$ occurred very slightly to the left of the peak at $j_{max}=17.1$.  Also,
we previously noted in the right panel of Figure~2 that the full-width
at half maximum ($w$) of the growth-rate peak is approximately 4 collision times.   Now, with
Eq.(3.10), we can calculate this width precisely by
setting ${\nu}_{j-1}(1-{\nu}_{j-1})=(1/2){\Delta}{\nu}_{max}=1/8$.  This is a quadratic equation
whose roots are
${\nu}_{j-1}=(1/2)(1~{\pm}~1/{\sqrt{2}})$.   Setting them equal to $2^{-2^{(j-k)}}$, we obtain
$2^{j-k}=1-\log_{2}(1~{\pm}~1/{\sqrt{2}})$, whose numerical values are $0.2284$ and $2.7716$.
The corresponding values of $j-k$ are $-2.130$ and $1.471$, and their difference (the full-width
at half maximum) is $w=3.601$, about 10\% less than our original estimate from Figure~2.
\medskip

It is interesting to compare the width and height of this growth-rate curve with the total area
under it (which we expect to be 1 because
$\displaystyle\sum_{j=0}^{j=25}{\Delta}{\nu}_{j}={\nu}_{25}-{\nu}_{0}=1-1/N_{0}~{\approx}~1$).
For our calculated full-width at half maximum, the product of height and width is
$0.25{\times}3.601=0.900$, which is slightly less than 1.  On the other hand, if we had used
the full-width at $1/e$ maximum instead of the
full-width at half maximum, we would have obtained $0.25{\times}4.398=1.100$, which is
slightly greater than 1.  It is an interesting coincidence that the average of these two widths is almost
exactly 4.000, which gives an area of 1.000.  Now, let us return to the more realistic case in which
the red molecules are contagious for a finite time.

\section{The Spreading When Red Molecules Have a Finite Lifetime}
 \subsection{The Model}
In this section, we shall assume that the infected red molecules do not
remain contagious indefinitely, but lose their ability to infect other
molecules after a finite time, $c$.  We suppose that $c > 1$, so that
the red molecules remain contagious for at least one collision time.
Otherwise, the molecules would lose their ability to infect before they
encounter uninfected blue molecules in the next collision, and the
epidemic would be over.  So, we suppose that after a red molecule suffers
$c$ collisions, it loses is ability to infect others.  At this point, we change
its color from red to green.
\medskip

In the real world, a molecule could become green either because it
recovered from the spreading disease or because it did not recover
and died.  In the latter case, the green molecule ought to be removed
from the gas.  For the current corona virus, this applies to about 3-5\%
of the green molecules (or less depending on how many of the infections
were not reported).  Consequently, in this paper we will neglect this
effect and keep all of the green molecules in the gas.  As we shall see later,
these green molecules provide a kind of `shielding effect' by reducing the
probability that the red molecules will come in contact with the remaining blue
molecules during the next collision.
\medskip

Figure~4 shows the progress for $c=2$.  In the left panel, one collision has
taken place, and the original red molecule from $j=0$ has infected another
molecule at $j=1$.  These two
\begin{figure}[h!]
 \centerline{%
 \fbox{\includegraphics[bb=105 18 525 785,clip,angle=90,width=0.75\textwidth]{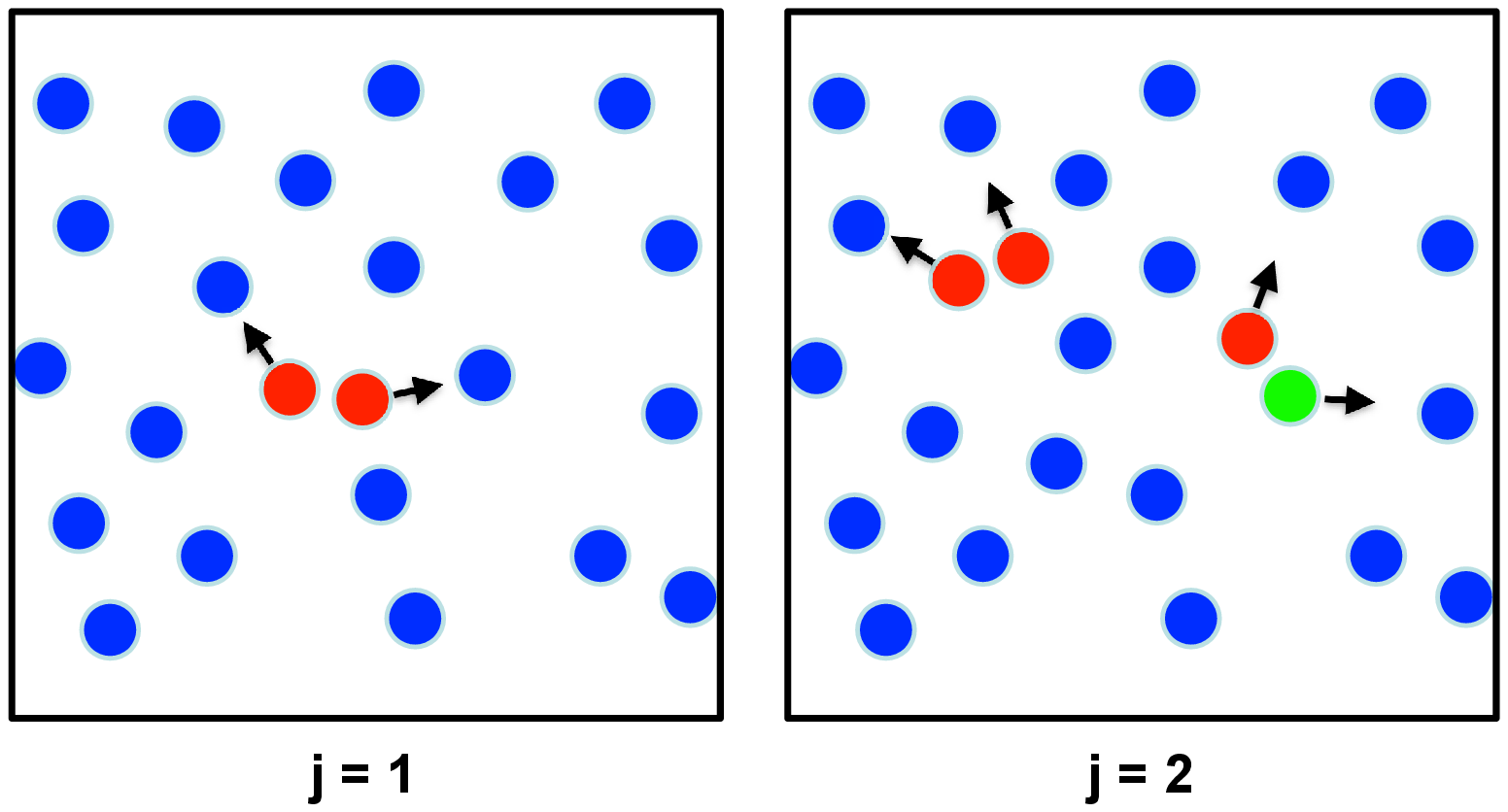}}}
 \caption{Same as Figure~1, but with a finite contagious time of $c=2$
collision times.  After 2 collision times, the initially infected molecule (now green)
is no longer contagious.}
\end{figure}
\noindent
red molecules continue on to infect two
more molecules by $j=2$ (right panel), but by this time, the original red
molecule has lost its contagiousness and has become green.  Like the
three red molecules, the green molecule will go on to collide with
other blue molecules, but it will not infect them or change their color.
\medskip
\subsection{The Equations}
Proceeding as we did in the previous section,  the probability that a red
molecule will collide with a blue molecule is still given by Eq. (3.1).
However, because some of the infected molecules are no longer red (\textit{i.e.}
contagious), Eq. (3.2) must be replaced by
\begin{equation}
n_{inf}^{(j)}~=~n_{inf}^{(j-1)}~+~n_{c}^{(j-1)}p_{j-1},
\end{equation}
where $n_{c}^{(j-1)}$ is the number of contagious molecules at the time $j-1$.
Now, we need to find an equation for the number of contagious molecules.
\medskip

To get a feel for this process, we refer to Figure~5, which is a sketch of
the evolving
\begin{figure}[h!]
 \centerline{%
 \fbox{\includegraphics[bb=70 155 540 640,clip,angle=90,width=0.55\textwidth]{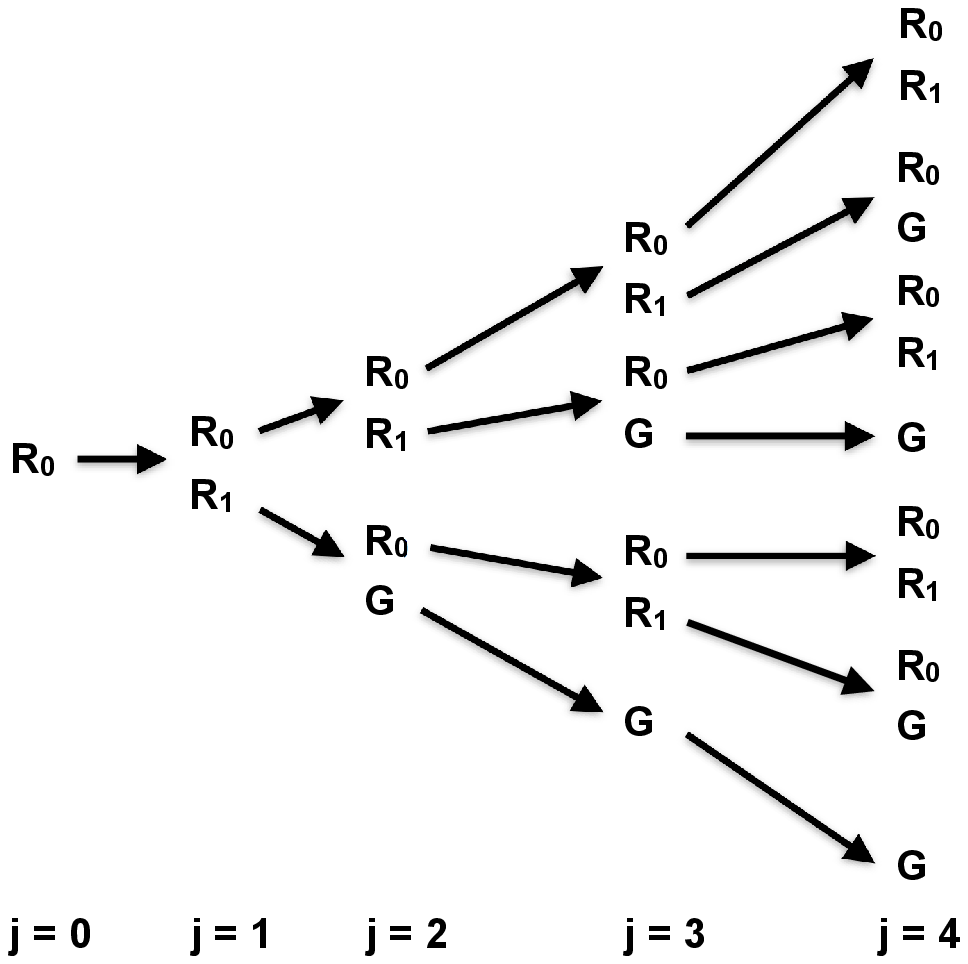}}}
 \caption{The spread of infection during $j=0-4$, assuming that
infected molecules remain contagious for 2 collision times ($c=2$), and are then
colored green ($G$).    At a time, $j$, the number of contagious molecules is equal to the
number of infected molecules at that time minus the number of infected molecules 2 steps
earlier.  The subscript on R refers to the time since that molecule became infected.}
\end{figure}
\noindent
number of red and green molecules as a function of the
collision time, $j$, assuming that there
are no collisions between red molecules
during this time.  At the left end of the array, we begin at $j=0$ with $R_{0}$,
which is a symbol for a single red molecule of `age' 0.  After the first
collision at $j=1$, there are two red molecules - the original red molecule,
$R_{1}$,
whose age is now 1, and the newly infected molecule, $R_{0}$.  One step
later, there are 4 infected molecules - $R_{0}$ and $R_{1}$, derived from $R_{0}$
at time $j=1$, plus $R_{0}$ and $G$, derived from $R_{1}$ at time $j=1$.
Instead of calling the the originally infected molecule by the name $R_{2}$,
we realize that this molecule is not contagious anymore, and we represent it
by the letter $G$.  As time passes to $j=3$ and $j=4$, the pattern expands,
with each $R_{0}$ always producing an $R_{0}$ and an $R_{1}$, with each
$R_{1}$ producing an $R_{0}$ and a $G$, and with each $G$ always
remaining a $G$.
\medskip

From the evolving pattern in Figure~5 with $c=2$, we can see that the
number of contagious molecules at time $j-1$ is equal to the total number of
infected molecules (contagious plus noncontagious) at that time minus the
total number of infected molecules 2 steps earlier.  More generally, we can
appreciate that all of the contagious molecules ($R$'s) at any time, $j-1$, will
lose their contagiousness (and become $G$'s) after an additional time, $c$,
so that
\begin{equation}
n_{c}^{(j-1)}~=~n_{inf}^{(j-1)}~-~n_{inf}^{(j-1-c)},
\end{equation}
even when there are collisions between red molecules.
Combining Eqs. (3.1), (4.1), and (4.2), and normalizing to the total
population, $N_{0}$, we have
\begin{equation}
{\nu}_{j}~=~{\nu}_{j-1}~+~\left ( \frac{1}{1-{\nu}_{0}} \right )
({\nu}_{j-1}-{\nu}_{j-1-c})(1-{\nu}_{j-1}),
\end{equation}
which reduces to Eq. (3.7) when ${\nu}_{j-1-c}=0$.  Likewise, the
difference form of Eq. (4.3) is
\begin{equation}
{\Delta}{\nu}_{j}~=~\left ( \frac{1}{1-{\nu}_{0}} \right )
({\nu}_{j-1}-{\nu}_{j-1-c})(1-{\nu}_{j-1}),
\end{equation}
which reduces to Eq. (3.11) when ${\nu}_{j-1-c}=0$.
\medskip
\subsection{The Plots}
Our next step is to plot ${\nu}_{j}$ and ${\Delta}{\nu}_{j}$ versus $j$
for various values of the parameter, $c$.  To do this, we set ${\nu}_{0}=1/N_{0}$
and use Eqs. (4.3) and (4.4) for $j~\geq~1$.  Also, we add the condition
that ${\nu}_{j}=0$ for $j < 0$.  This means that ${\nu}_{j-1-c}$ will
not contribute until $j-1=c$.
\medskip

Finally, because the influence of the finite contagious time becomes greatest
as that time approaches the collision time (\textit{i.e.} as $c \rightarrow 1$),
we need to find a way to make these equations apply for non-integral values
of $c$ between 1 and 2.  For this purpose, we used a linear interpolation
between values of $c$ in the range $[1,2]$, or more generally in the range
$[c_{0},c_{0}+1]$, where $c_{0}$ is an integer.  Consequently, we replaced
${\nu}_{j-1-c}$ by the expression
${\lambda}{\nu}_{j-1-c_{0}-1}+(1-{\lambda}){\nu}_{j-1-c_{0}}$, where
${\lambda}$ lies in the range $[0,1]$ and $c=c_{0}+{\lambda}$.
\medskip

Figures~6 and 7 show ${\nu}_{j}$ and ${\Delta}{\nu}_{j}$ plotted versus $j$
for parametric values of $c$ from 1.25
\begin{figure}[h!]
 \centerline{%
 \fbox{\includegraphics[bb=85 270 525 720,clip,width=0.90\textwidth]{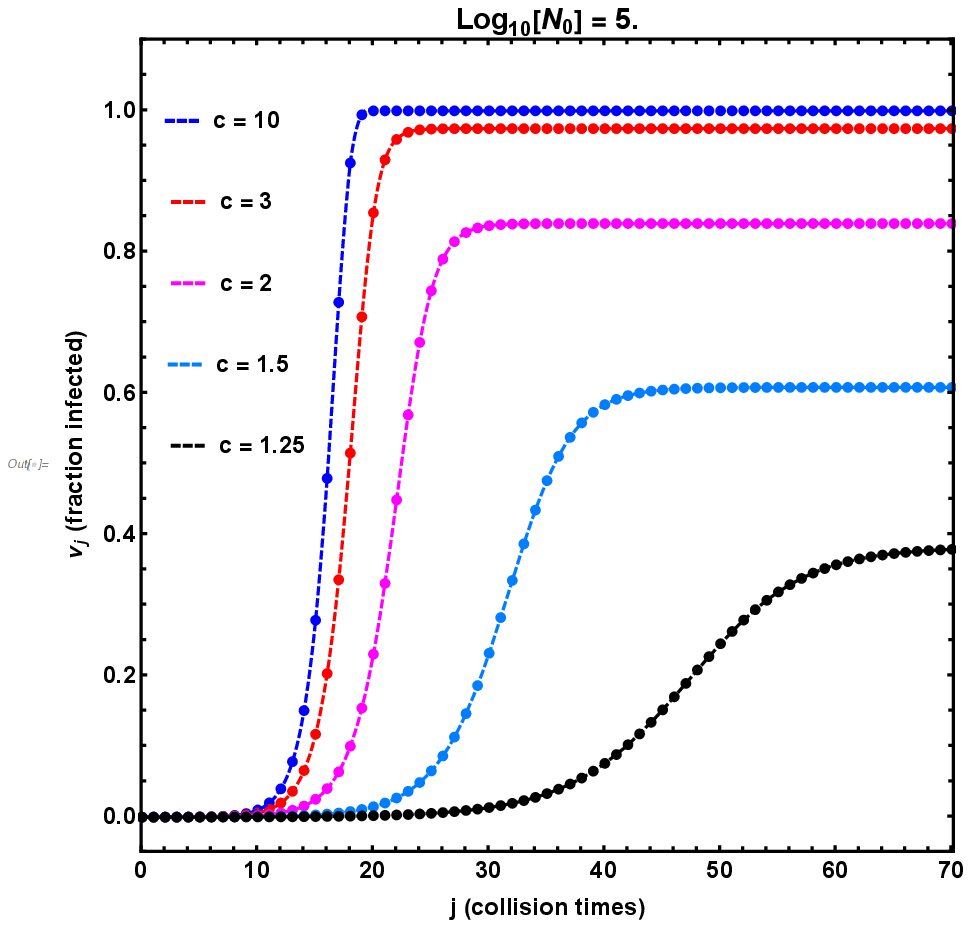}}}
 \caption{The fraction of infected molecules, ${\nu}_{j}$, as a function of time for
values of the contagious time, $c$, from 10 to 1.25, showing the delayed onset and
the reduced final fraction as the contagious time becomes closer to the collision time.}
\end{figure}
\noindent
 to 10.0.  I chose 10 as the largest
contagious time because the curves for $c=10$ and a much larger value of
$c$ like $c=100$ were indistinguishable.  Thus, once the contagious time
reaches 10 collision times, the result is the same as if the red molecules
remain contagious for the duration of the problem.  (The curves were almost
indistinguishable for $c=6$, except that
\begin{figure}[h!]
 \centerline{%
 \fbox{\includegraphics[bb=85 270 530 720,clip,width=0.91\textwidth]{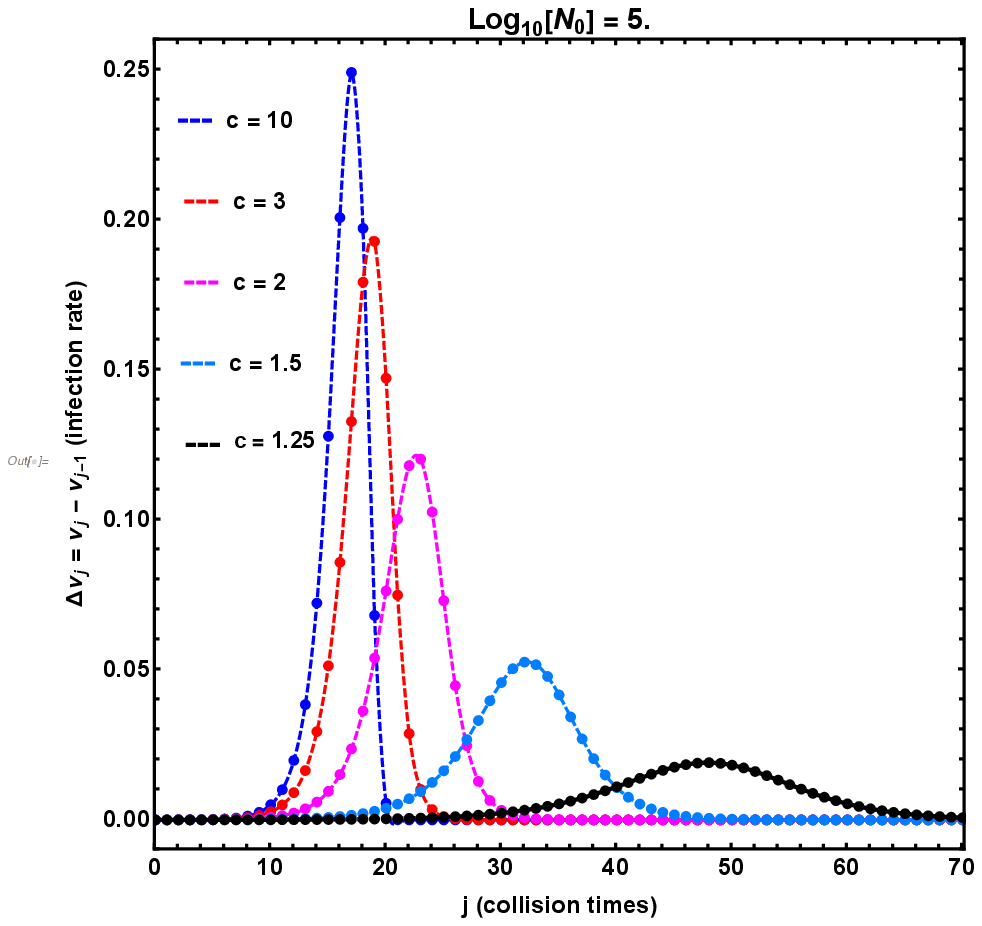}}}
 \caption{The differential fraction of infected molecules, ${\Delta}{\nu}_{j}$,
as a function of time, showing that the profiles become lower, wider, and
shifted to later times as the contagious time, $c$, becomes closer to the
collision time.}
\end{figure}
\noindent
the peak height for the differential
curve was 0.24 instead of 0.25.)  Consequently, the curves with $c=10$ are
an accurate representation of the result when the red molecules remain
contagious for the duration of the problem (\textit{i.e.} the value of ${\nu}_{j}$
eventually reaches 1, and the height of the peak of ${\Delta}{\nu}_{j}$ is 0.25
at the location where $j_{max} \approx 17$, which is approximately
$\log_{2}N_{0}=16.6$.)
\medskip

In retrospect, it is obvious why the growth curve with $c=10$ is indistinguishable from the
corresponding curve for permanently contagious molecules (with $c={\infty}$).  For $c=10$,
the lag is so long that the plot of ${\nu}_{j-1}$ finished its rise from 0 to 1 before the lagged
quantity, ${\nu}_{j-1-c}$, became appreciable (${\nu}_{9}=0.005$).  Equivalently, there are
10 points between ${\nu}_{j}~{\approx}~0$ and ${\nu}_{j}~{\approx}~1$.  For a smaller value
of $c$, the lag is shorter, and ${\nu}_{j-1-c}$ becomes appreciable before ${\nu}_{j}$ reaches
the end of the rising portion of its curve.  In addition, there are more points between the
beginning and end of the rising portion of the curve.  This produces less spacing between
the points of ${\nu}_{j}$ in Figure~6 and lower peaks in the
plots of ${\Delta}{\nu}_{j}$ in Figure~7. 

\medskip

For $c=10$, the fraction of infected molecules reaches 1  at $j~{\approx}~20$ and remains
there forever.  This means that
everyone in the population eventually becomes infected.  However,
as $c$ decreases, the final fraction
of infected molecules decreases, reaching about 0.38 when $c=1.25$
collision times.  If the contagious time were about 14 days, $c=1.25$
would correspond to a collision time of 11.2 days.  This means that to
reduce the number of infected molecules to 38{\%} of the total population,
one would have to increase the average time between collisions to about
11 days.  Such `social distancing' prevents $62{\%}$ of the population from
becoming infected and shows the importance of making the time between
collisions as close as possible to the 14-day contagious time.  But even for
11 days, the end of the epidemic would be delayed appreciably from
$j~{\sim}~20$ to $j~{\sim}~70$, which is roughly from 1 year to 3 years if
the collision time is 2 weeks.

Figure~7 shows that the infection rate weakens, broadens, and
shifts to later times as the contagious time, $c$, approaches the collision time.
I think this is what the media reports for the corona virus are referring to when
they use the term `flattening the curve'.  Corresponding plots for other
values of $N_{0}$ show essentially the same profiles, shifted in time.  In the
next section, we shall analyze the dependences of these profiles on $N_{0}$ and
$c$.

\section{Quantitative Results: Profile Dependence on $\mathbf{N_{0}}$ and $\mathbf{c}$}

In the previous section, we saw that the growth rate gradually rises to
a peak $\sim0.25 N_{0} {\tau_{s}}^{-1}$ when the contagious time, $c$, is large
compared to 1, and that the peak height becomes smaller as $c$ approaches 1.
In this section, we will determine the location, height, and width of this peak
as a function of $N_{0}$ and $c$.  Also, we will determine the dependence
of the final value of ${\nu}_{j}$ on these parameters.

\subsection{The Final Value of the Growth Curve}

We can obtain a clue to the relation between $c$ and ${\nu}_{f}$ by examining the behavior
of ${\nu}_{j-1}$ in the region where the growth curve has its maximum slope.   In this region,
the slope is fairly constant so that it will have approximately the same value
at several neighboring points.  In particular, we would expect that
\begin{equation}
\frac{{\nu}_{j-1}-{\nu}_{j-1-c}}{c}~{\approx}~{\nu}_{j-1}-{\nu}_{j-2}~{\approx}~
{\nu}_{j}-{\nu}_{j-1}~=~{\Delta}{\nu}_{j},
\end{equation}
if $c$ is close to 1.   In this case, Eq. (4.4) reduces to
\begin{equation}
{\Delta}{\nu}_{j}~{\approx}~c {\Delta}{\nu}_{j}(1-{\nu}_{j-1}),
\end{equation}
which would be an identity if $c(1-{\nu}_{j-1})=1$.  Equivalently,
\begin{equation}
{\nu}_{j-1}~{\approx}~1-\frac{1}{c}.
\end{equation}
This point of maximum slope lies somewhere between the starting and ending
values ${\nu}_{0}$ and ${\nu}_{f}$, perhaps becoming centered between them if the growth-rate
curves become symmetric when $c$ approaches 1 (as Figure~7 suggests).   In this case, we would
expect that
\begin{equation}
{\nu}_{f}~{\approx}~{\alpha}(1-\frac{1}{c}),
\end{equation}
where ${\alpha}$ is a number on the order of 2.
\medskip

Although Eq.(5.4) seems to reproduce the calculated values of ${\nu}_{f}$ fairly well as $c$ approaches 1,
it does not satisfy the requirement that ${\nu}_{f}~{\rightarrow}~1$ as $c~{\rightarrow}~{\infty}$.  To satisfy
this requirement and still retain the behavior for $c~{\approx}~1$, we can use an exponential of the form
$1-e^{-{\alpha}(c-1)}$.  This expression approaches 1 as $c-1~{\rightarrow}~{\infty}$, and it approaches
${\alpha}(c-1)$ as $c-1~{\rightarrow}~0$.  In fact, if we retain only the first-order terms in $c-1$, we obtain
the equality:
\begin{equation}
{\alpha}(1-\frac{1}{c})~=~{\alpha}\left \{\frac{c-1}{1+(c-1)} \right \}~={\alpha}(c-1)~=~1-e^{-{\alpha}(c-1)}.
\end{equation}

Figure~8 shows ${\nu}_{f}$, computed from Eq.(4.3) and plotted for values of $c$ in the range
\begin{figure}[h!]
 \centerline{%
 \fbox{\includegraphics[bb=88 245 540 720,clip,width=0.65\textwidth]{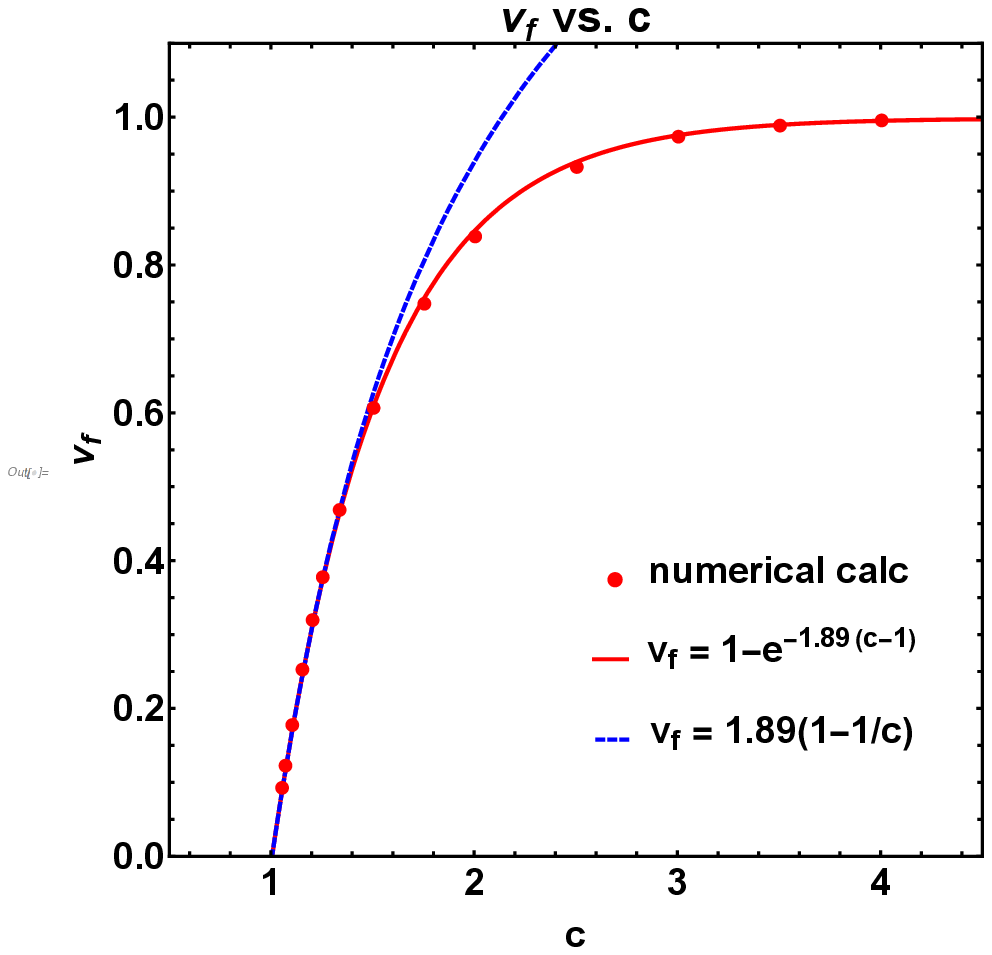}}}
 \caption{The final value, ${\nu}_{f}$, of the growth profile plotted versus the contagious
lifetime, $c$, with the solid red line indicating the root-mean-square best fit given by
Eq. (5.6) and the dashed blue line indicating the approximation given by Eq. (5.4) with
${\alpha}=1.890$.}
\end{figure}
\noindent
$1.05-4$.  The value of $N_{0}=10^5$ used in these calculations determines only the value of $j$
where the final value of ${\nu}_{f}$ occurs, but it does not determine ${\nu}_{f}$ itself.  In other
words, we would obtain the same plot of ${\nu}_{f}$ versus $c$ for other values of $N_{0}$.
The solid red curve is the least-squares-best-fit to these points using the relation
\begin{equation}
{\nu}_{f}~=~1~-~e^{-1.890(c-1)},
\end{equation}
and the dashed blue curve is the approximation given by Eq. (5.4) with ${\alpha}=1.890$.
\medskip

As we can see, Eq. (5.4), represented by the dashed blue line, fits the data very well for
$c$ up to 1.5, but the exponential relation given by Eq. (5.6) is required to extend the fit
to larger values of $c$.  Consequently, Eq. (5.6) provides the desired relation between
the amount of social distancing and the fraction of molecules that eventually become
infected (and therefore the fraction that ultimately escapes the epidemic).  Moreover,
for social distancing with $c<1.5$, a rough estimate of the final fraction
of infected molecules is just ${\nu}_{f}~{\sim}~2~(1-1/c)$.

\subsection{The Height and Width of the Growth-Rate Peak}
In the previous section, we found that Eq. (5.6) provides an accurate expression for the final
fraction of infected molecules, ${\nu}_{f}$, as a function of the parameter, $c$.
Next, we look for analogous expressions for the height and width of the peak in the growth-rate
curve.
\medskip

We begin by referring to Figure~9, which contains plots of the accumulated infection,
\begin{figure}[ht!]
 \centerline{%
 \fbox{\includegraphics[bb=20 117 590 675,clip,angle=90,width=0.47\textwidth]{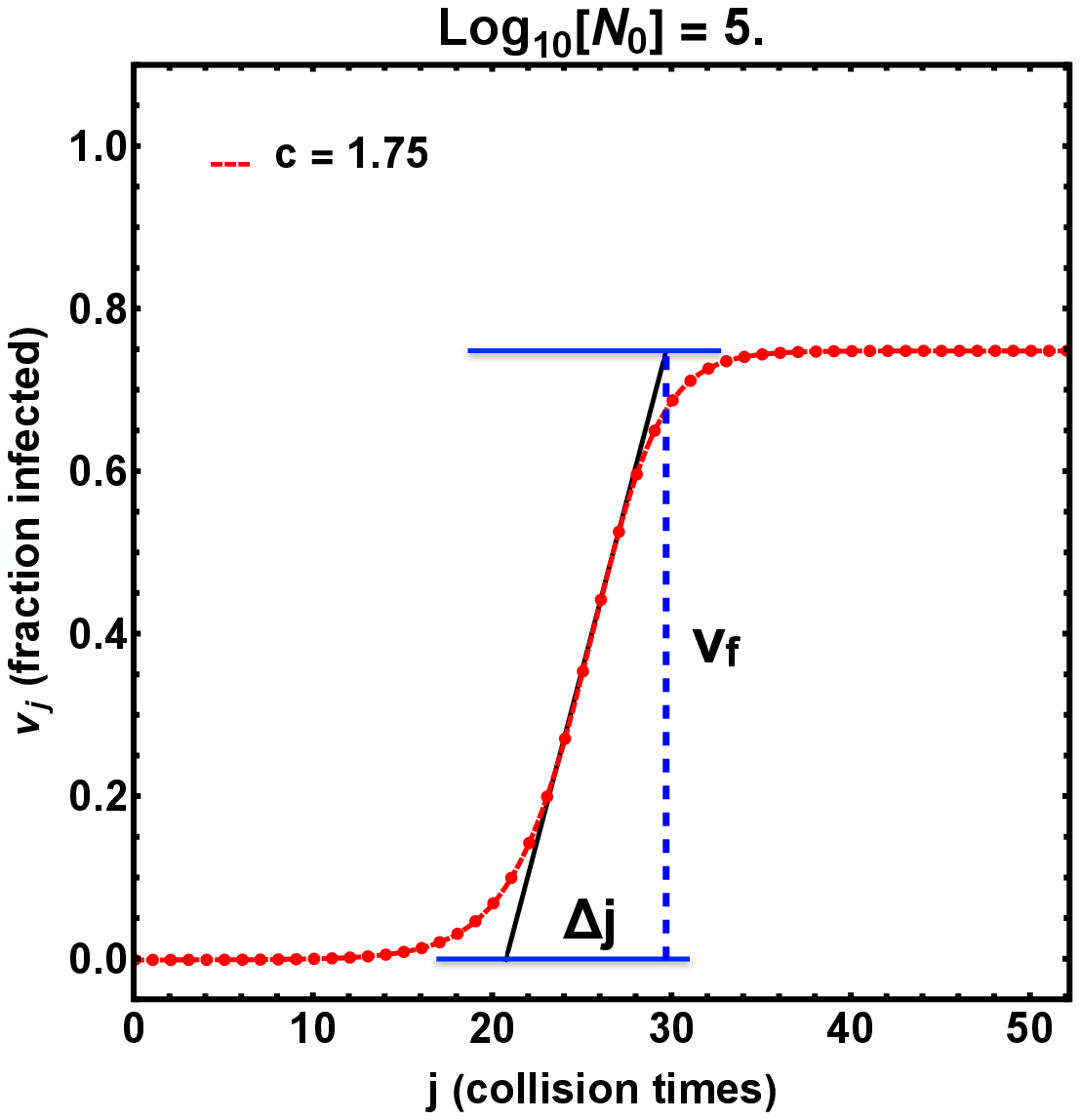}}
 \hspace{0.01in}
 \fbox{\includegraphics[bb=20 117 600 675,clip,angle=90,width=0.47\textwidth]{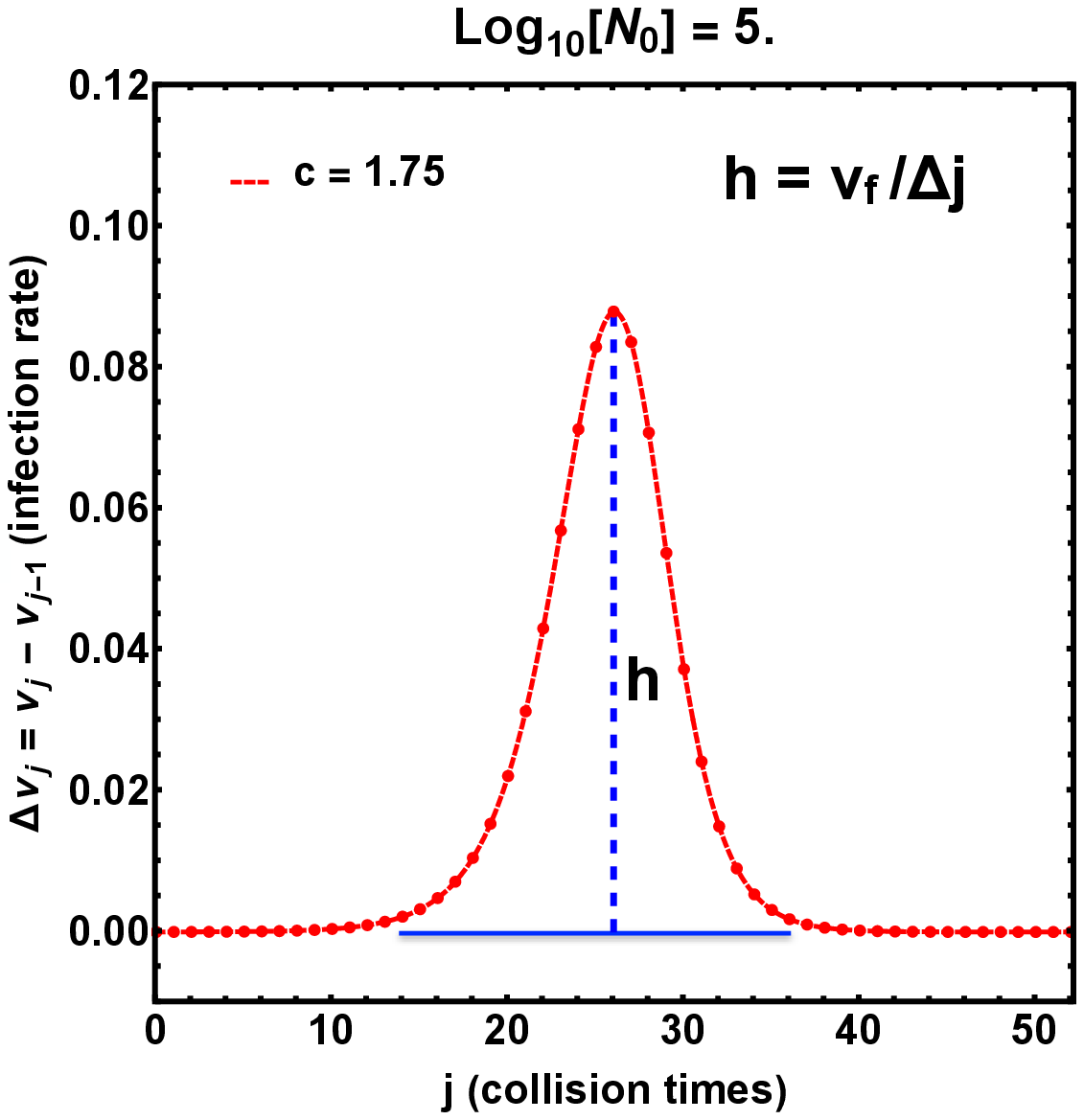}}}
 \caption{Plots of ${\nu}_{j}$ and ${\Delta}{\nu}_{j}$ for $c=1.75$, used to illustrate the relation
between peak height, $h={\Delta}{\nu}_{max}$, and the width, $w_{e}={\Delta}j$.}
\end{figure}
\noindent
${\nu}_{j}$, and the infection rate, ${\Delta}{\nu}_{j}$, as a function of the collision step number,
$j$.  The left panel
shows the accumulated infection curve (in red) and a black line segment drawn tangent to it
at the point of maximum slope.  Actually, this point is really an extended region, which is why we
were able to deduce that ${\nu}_{j-1}~{\approx}~1-1/c$ in the center of this region. (See the discussion
leading to Eq. (5.3).)  The ends of this line segment lie on the horizontal blue lines where
${\nu}_{j}={\nu}_{f}$ and ${\nu}_{j}={\nu}_{0}=1/N_{0}~{\approx}~0$, and are separated by a
horizontal distance, ${\Delta}j$.
Consequently, ${\nu}_{f}/{\Delta}j$ is the maximum slope of the ${\nu}_{j}$ curve, which in turn is equal
to the peak value, ${\Delta}{\nu}_{max}$, of the growth-rate curve in the right panel.  (Here, we ignore
the subtle difference between ${\Delta}{\nu}_{j}$ and $d{\nu}_{j}/dj$ because
${\Delta}{\nu}_{j}={\nu}_{j}-{\nu}_{j-1}~{\approx}~({\nu}_{j}-{\nu}_{j-{\epsilon}})/{\epsilon}~{\rightarrow}~
d{\nu}_{j}/dj$ as ${\epsilon}~{\rightarrow}~0$).  Thus, we can write
\begin{equation}
h~=~\frac{{\nu}_{f}}{{\Delta}j},
\end{equation}
where $h$ is the peak height, ${\Delta}{\nu}_{max}$.
\medskip

Next, in Figure~10, we plot the numerical values of the ratio $h/{\nu}_{f}$ versus the parameter, 
\begin{figure}[ht!]
 \centerline{%
 \fbox{\includegraphics[bb=90 255 545 720,clip,width=0.65\textwidth]{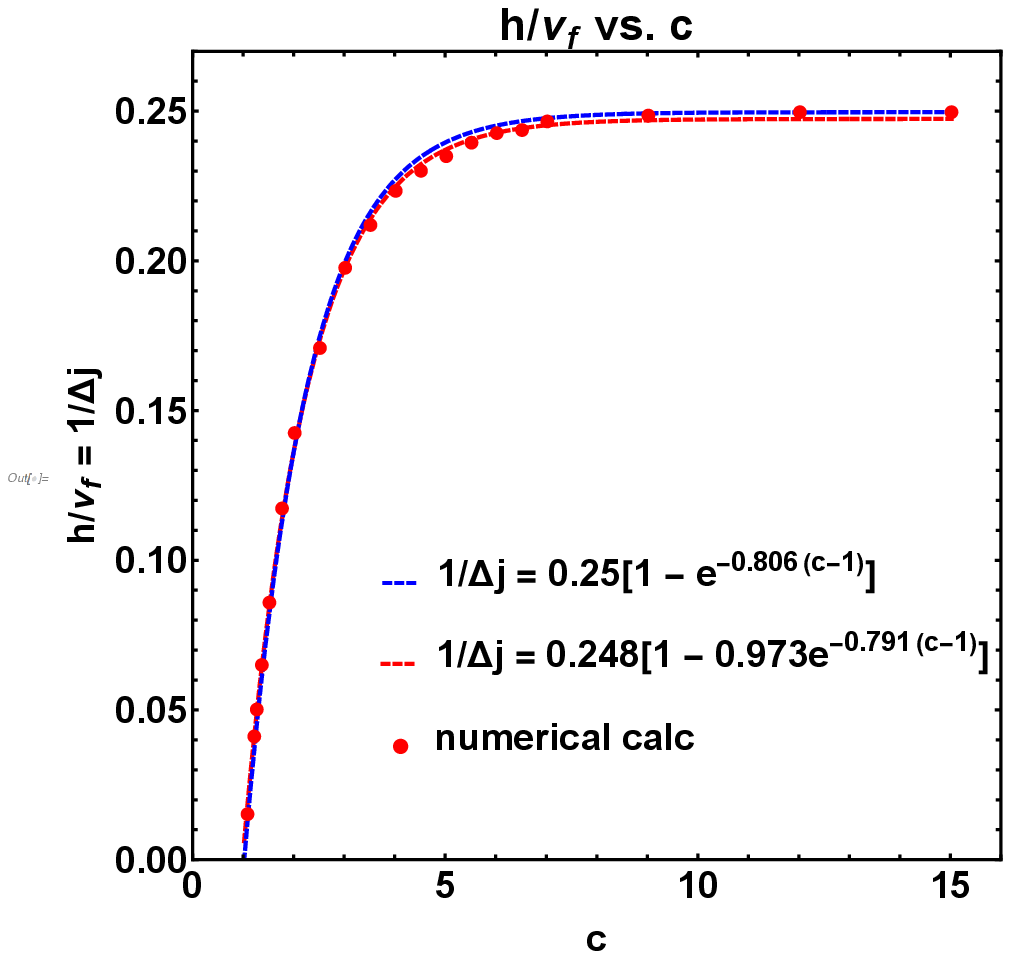}}}
 \caption{Plots of the numerical values of $h/{\nu}_{f}$ versus c (red points) fit by two different
expressions given by Eqs.(5.8a) and (5.8b) (blue and red dashed lines).}
\end{figure}
$c$, and fit the resulting points with two slightly different formulas given by
\begin{subequations}
\begin{align}
\frac{h}{{\nu}_{f}}~=~\frac{1}{{\Delta}j}~=~0.25\left [1-e^{-0.806(c-1)}\right ],\\
\frac{h}{{\nu}_{f}}~=~\frac{1}{{\Delta}j}~=~0.248\left [1-0.973e^{-0.791(c-1)}\right ].
\end{align}
\end{subequations}
\medskip

Because ${\Delta}j$ is a measure of the width of the growth-rate curve, we refer to it as the width,
$w_{e}$, of that curve.  Consequently, that width can be obtained by taking the reciprocal of the
expressions in Eqs.(5.8a) and (5.8b)\footnote{Other measures of the width like the full-width at half
maximum are slightly smaller than $w_{e}$.  For example, the width, $w$, would be given by
$w= \ln(1+{\sqrt{2}})w_{e}=0.881w_{e}$  if the fraction of infected molecules, ${\nu}_{j}$, were fit
by a hyperbolic tangent of the form ${\nu}_{j}=({\nu}_{f}/2)[1+\tanh\{{\gamma}(j-j_{max})\}]$, and the
width would be $w=(2^{2/3}-1)^{1/2}w_{e}=0.766w_{e}$ if ${\nu}_{j}$ were fit by a square-root expression
of the form ${\nu}_{j}=({\nu}_{f}/2)[1+{\gamma}(j-j_{max})\{1+{\gamma}^{2}(j-j_{max})^{2}\}^{-1/2}]$,
where ${\gamma}=2/w_{e}$.}.
As we can see, the simpler formula given by Eq.(5.8a) ends up at 0.25, but misses some of the points
at the bend of the curve.  However, the more complicated formula given by Eq.(5.8b) passes
through the points at the bend of the curve, but ends up slightly lower than 0.25.  Both fits pass through
all of the points where $c~{\leq}3$.
\medskip

The next step is to see how well these expressions combine with Eq.(5.6)
for ${\nu}_{f}$ to fit the calculated points for $h$.  The result is shown in Figure~11.
Here, the product of the two
\begin{figure}[ht!]
 \centerline{%
 \fbox{\includegraphics[bb=90 253 550 720,clip,width=0.65\textwidth]{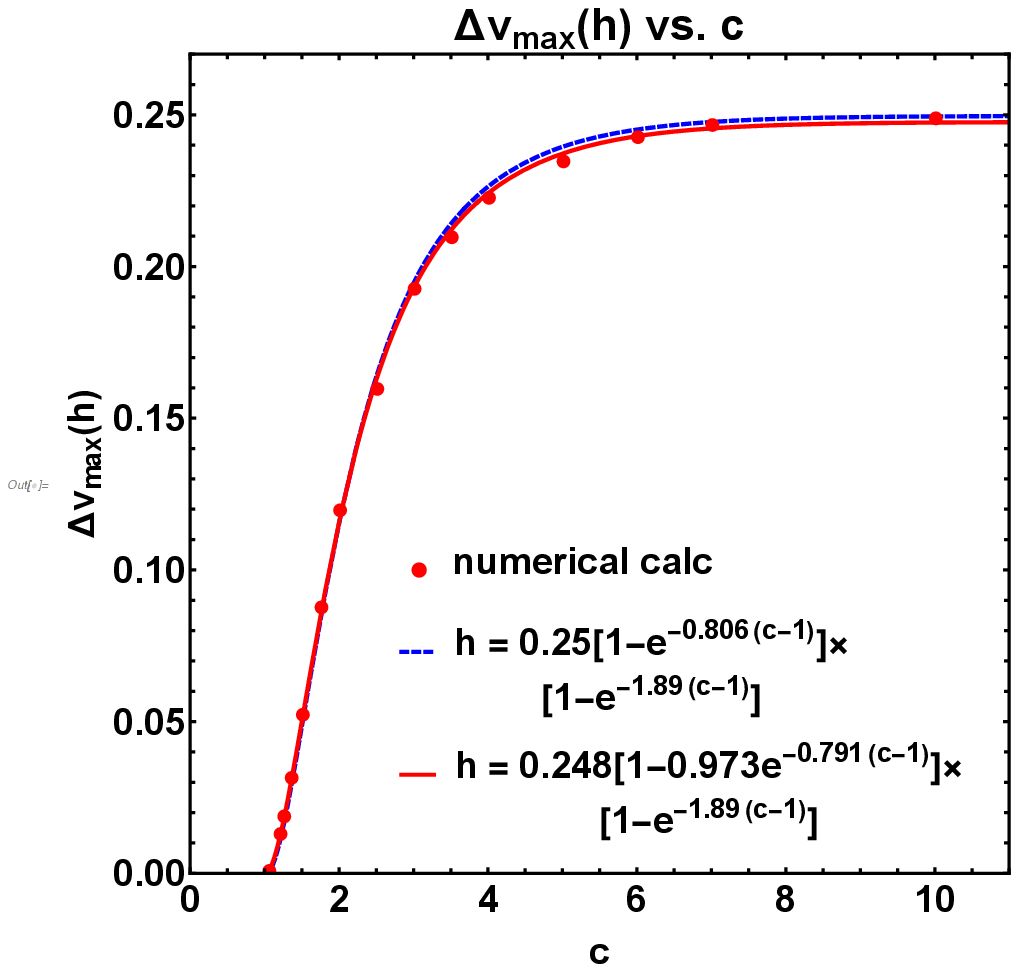}}}
 \caption{Calculated values of ${\Delta}{\nu}_{max}=h$, plotted versus $c$ (red points), fit
by the product of Eqs. (5.6) and (5.8a) (dashed blue line) and (5.8b) (solid red line).}
\end{figure}
\noindent
fits reproduces the initial curvature in the plot of $h$ versus $c$.
Also, as expected, the better fit to $1/{\Delta}j$ in Figure~10 provides the better fit to $h$ in
Figure~11.  However, even the simpler expressions given by Eqs.(5.6) and (5.8a) provide
an essentially perfect fit for $c~{\leq}~3$, which is the region of interest when a substantial
amount of social distancing is maintained.
\medskip

In summary, we can describe the growth curves of Figure~6 and growth-rate curves of
Figure~7  in terms of the contagious lifetime parameter, $c$, using the approximate relations
for ${\nu}_{f}$, $1/w_{e}$, and $h$:
\begin{subequations}
\begin{align}
{\nu}_{f}~=~1~-~e^{-1.890(c-1)},\\
\frac{1}{w_{e}}~=~0.25\left [1-e^{-0.806(c-1)}\right ],\\
h~=~\frac{\nu_{f}}{w_{e}}~=~0.25\left [1-e^{-0.806(c-1)}\right ][1-e^{-1.890(c-1)}].
\end{align}
\end{subequations}

\clearpage
\subsection{Location of the Peak of the Growth Rate}

I have used plots like those shown in Figure~7 to measure the locations, $j_{max}$,
of the peaks for values of $c$ ranging from 1.25 to 10.0, and values of
$\log_{10}N_{0}$ ranging from 2 to 8.  Figure~12  shows the measured values of
$j_{max}$ plotted versus $\log_{10}N_{0}$ for given values of $c$.  The dashed lines
\begin{figure}[h!]
 \centerline{%
 \fbox{\includegraphics[bb=90 250 537 725,clip,width=0.71\textwidth]{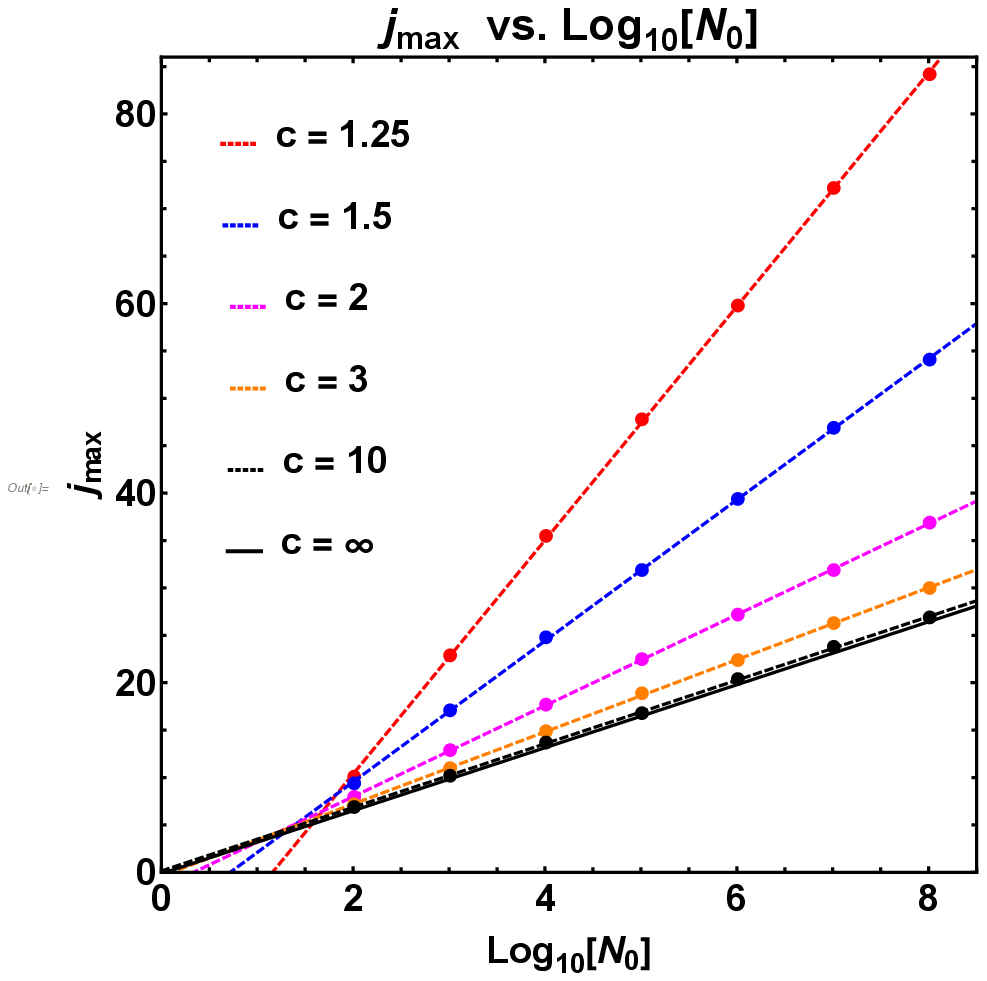}}}
 \caption{Measured $j_{max}$ values plotted  versus $\log_{10}N_{0}$, showing the  linear
relation between these two quantities for values of $c~{\geq}~1.25$.  The
line marked $c=\infty$ corresponds to Eq. (3.6).}
\end{figure}
are the least-squares best fits to these points given by:  
\begin{eqnarray}  j_{max}~=~ \left\{\begin{array}{clcl}
-14.071~+~12.357~\log_{10}N_{0},~~~~c=1.25,\\
\\
-5.154~+~7.439~\log_{10}N_{0},~~~~c=1.5,\\
\\
-1.381~+~4.799~\log_{10}N_{0},~~~~c=2,\\
\\
-0.156~+~3.796~\log_{10}N_{0},~~~~c=3,\\
\\
+0.312~+~3.359~\log_{10}N_{0},~~~~c=10.
\end{array}
\right.
\end{eqnarray}
Also, the black line represents Eq.(3.6) ($j_{max}=\log_{2}N_{0}~{\approx}~3.322\log_{10}N_{0}$) for
permanently contagious molecules with $c={\infty}$.
Its slope (3.322) and intercept (0.000) are very close to the slope (3.359) and intercept (0.312)
of the line with $c=10$.
\medskip

Based on this figure, there can be little doubt that $j_{max} = p_{c}+q_{c}~{\log_{10}N_{0}}$
for each value of the parameter, $c$, where $p_{c}$ and $q_{c}$ are the least squares best-fit
parameters for the dashed lines that pass through these points.  However, an even more
interesting property of Figure~12 is that the spread of the $j_{max}$ values increases with
$\log_{10}N_{0}$, and that this spread tends to a very small minimum value near the point
$(1.5,6)$ where the straight lines nearly intersect.  This means that for a small population,
the maximum growth rates occur at approximately the same time regardless of the value
of $c$ and the corresponding amount of social distancing.  This contrasts with the result
for a large population for which the location of the peaks depends much more strongly on
the amount of social distancing.  Thus, to achieve the same maximum growth rate, the
larger population must wait a much longer time than a smaller population.  As we shall see
in subsection (5.3.2), this property can be used to deduce $j_{max}$ as a function of
$\log_{10}N_{0}$ and $c$.  But first in subsection (5.3.1), we will deduce an expression for
$j_{max}$ using the linearized solution of the growth equation given in Appendix A.

\subsubsection{The slope-intercept method}
Appendix A provides a way to estimate the
values of $p_{c}$ and $q_{c}$ and thereby reproduce the $j_{max}$ expressions in
Eq. (5.10).  Referring to Figure~25 in Appendix A, we see that the linearized solutions
give the initial behavior of ${\nu}_{j}$ for each value of $c$.  Because these linearized
solutions lack the damping effect of the non-linear term, they eventually diverge from the
non-linear solutions whose curves bend over and approach final values of ${\nu}_{f}$.
However, the plots of the linearized solutions in Figure~25 seem to cross the
${\nu}_{j}={\nu}_{f}$ threshold approximately where the non-linear solutions have their
maximum slopes.  Such a relation would provide a way of determining the location of
those maximum slopes and therefore the locations of the peaks of the corresponding
growth-rate curves of ${\Delta}{\nu}_{j}$.  For this purpose, we would simply equate the
linearized solution, ${\sigma}_{c}{\rho}_{c}^{j}{\nu}_{0}$,  and the final value
of the growth curve, ${\nu}_{f}$, and then solve for $j$.
\medskip
The first step gives
\begin{equation}
{\sigma}_{c}~{{\rho}_{c}}^{j}~=~\frac{{\nu}_{f}}{{\nu}_{0}}~=~{\nu_{f}}N_{0},
\end{equation}
where ${\nu}_{f}$ is obtained from Eq.(5.6), and the parameters ${\sigma}_{c}$ and
${\rho}_{c}$ can be obtained from Table 1 of Appendix A (or from Eq. (A9) for values of
$c$ in the range 1-2), or from Eqs. (C1a) and (C1b) of appendix C.  The
resulting value of $j$ is $j_{max}-1$ because this analysis is based
on the growth curves in Figure~6 and Figure~25, whereas the peak value of the growth-rate
curve comes from the difference, ${\Delta}{\nu}_{j}={\nu}_{j}-{\nu}_{j-1}$ shown in Figure~7.
However, because this method is based on a rough estimate, I would be surprised if the
accuracy of the resulting value of $j_{max}$ were closer than 1 unit of collision time.  This
is comparable to the differences between ${\Delta}{\nu}_{j}$ and $d{\nu}_{j}/dj$, which
Figure~7 shows are often about 0.5 of a unit.
\medskip

Solving Eq. (5.11) for $j$ and setting $j_{max}=j+1$, we obtain
\begin{equation}
j_{max}~=~\left \{1+\frac{ \log_{10}({\nu}_{f}/{\sigma}_{c}) }{\log_{10}({\rho}_{c}) } \right \}~+~
\left \{\frac{1}{ \log_{10}({\rho}_{c}) } \right \}\log_{10}N_{0},
\end{equation}
which is in the format $p_{c}+q_{c} \log_{10}N_{0}$ used in Eq. (5.10).  Although, I have used
logarithms to base 10 for comparison with the expressions in Eq. (5.10), because these
logarithms occur as ratios, they can be expressed in any base,
including base $e$, base 2, or even base ${\rho}_{c}$, without affecting the value of $j_{max}$.
\medskip

Next, obtaining ${\nu}_{f}$ from Eq. (5.6) and substituting the values of ${\sigma}_{c}$ and
${\rho}_{c}$ from Table 1 of Appendix A, we find that
\begin{eqnarray}  j_{max}~=~ \left\{\begin{array}{clcl}
-14.934~+~12.239~\log_{10}N_{0},~~~~c=1.25,\\
\\
-4.802~+~7.383~\log_{10}N_{0},~~~~c=1.5,\\
\\
-0.668~+~4.785~\log_{10}N_{0},~~~~c=2,\\
\\
+0.463~+~3.780~\log_{10}N_{0},~~~~c=3,\\
\\
+1.000~+~3.322~\log_{10}N_{0},~~~~c={\infty}.
\end{array}
\right.
\end{eqnarray}
These values are comparable to the least-squares best-fit values in Eq. (5.10).  The coefficients
of $\log_{10}N_{0}$ are essentially the same.  However, the constant terms in Eq. (5.13) are
systematically a little larger than those in Eq. (5.10).  For the larger values of $c$, this discrepancy
is probably due to the intrinsic differences between the digital peak and the peak of the continuous
curve.  For the smaller values of $c$, and in particular for $c=1.25$, the discrepancy may reflect
the weaker and broader profiles whose peak locations are less well defined.
\medskip

Figure~13 shows the $j_{max}$ values calculated using the full non-linear equation (Eq. (4.4))
and the values obtained with the approximate method given by Eq. (5.12), both using the
nominal value of $\log_{10}N_{0}=5$.  The red dots are based on Eq. (4.4), whereas the blue
dots and their interpolated curve are based on Eq. (5.12).  The red and blue points all lie on the
same dashed blue curve, indicating a clear dependence of $j_{max}$ on $c$.  For $c~{\geq}~1.4$,
the corresponding red and blue points of each pair are very close together, often occulting each other
with virtually identical values.  However, for  $c<1.4$, the curve steepens and the
separations systematically increase as $c$ approaches 1.
\begin{figure}[h!]
 \centerline{%
 \fbox{\includegraphics[bb=90 247 545 720,clip,width=0.81\textwidth]{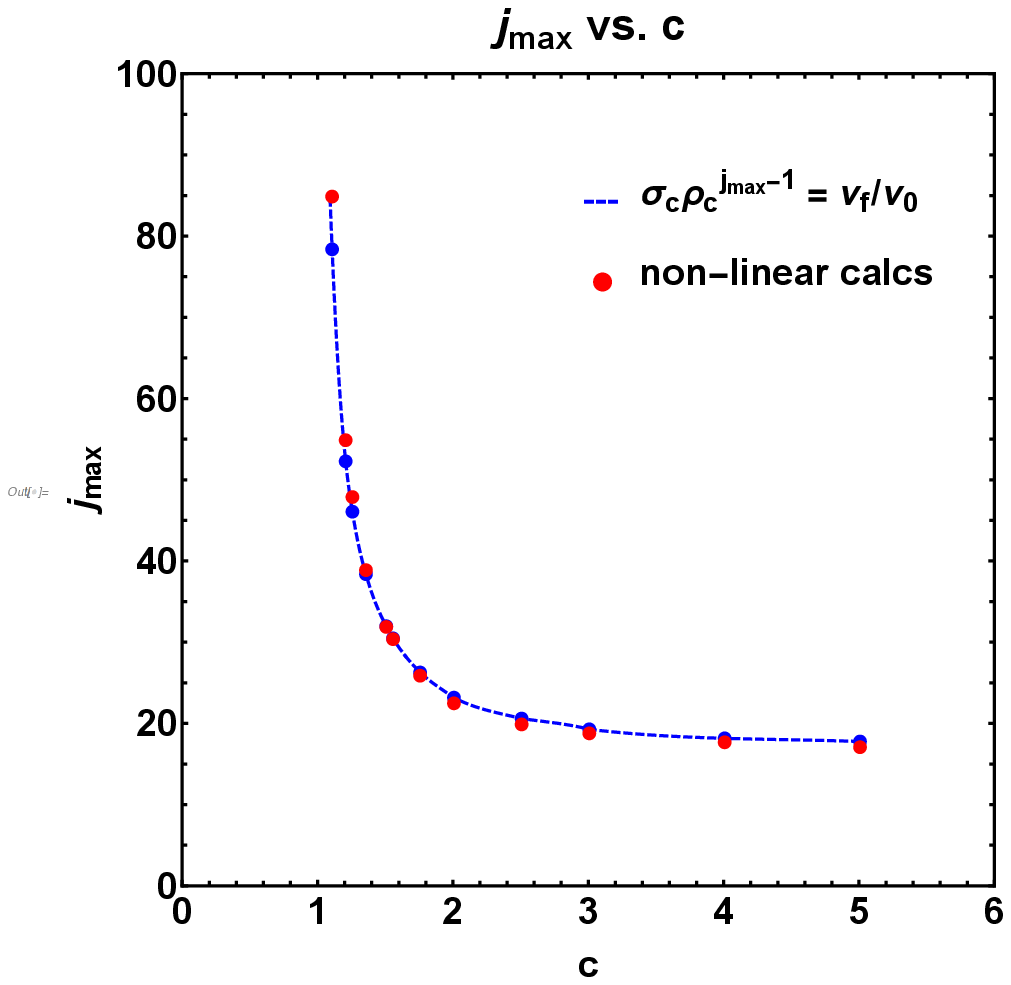}}}
 \caption{$j_{max}$ values, obtained from the non-linear Eq. (4.4) (red points) and calculated
with Eq. (5.12) (blue points and dashed line), plotted versus $c$ for $\log_{10}N_{0}=5$. }
\end{figure}
\medskip

At first, one might suppose that the overall agreement between the calculated values in Eq. (5.13) and
the measured values in Eq.(5.10) supports our ansatz that the point of maximum slope occurs where the
linearized solution for ${\nu}_{j}$ equals the final value ${\nu}_{f}$.  However, the agreement was
better for $q_{c}$ than for $p_{c}$, which contains the entire dependence on ${\nu}_{f}$.
So the discrepancy in the values of $p_{c}$ probably reflects the eventual breakdown of this assumption,
especially as $c$ approaches 1 and the growth-rate curves become increasingly flat.
Next, we will abandon this assumption, and see if the nearly common intersection of the lines in
Figure~12 gives a point-slope formula for $j_{max}$ that is more accurate than the slope-intercept approach.

\subsubsection{The point-slope method}
In this approach, we return to Figure~12 where the measured values of $j_{max}$ give linear fits that
nearly intersect at a common point near $(x_{0},y_{0})=(1.5,6)$.  We begin by precisely determining
the point of closest approach to the lines with $c=1.25$, 1.5, 2, 3, and ${\infty}$, leaving out $c=10$
because it is essentially the same as $c={\infty}$.
\medskip

We do this in the following way.  We recognize that even though these five lines may not intersect
in the same point, lines drawn parallel to them can be adjusted so that they do intersect at a common
point.  This can be done for many such common locations and our objective is to find the one that makes
the root-mean-square distance between all those parallel lines a minimum.  (Because one set of lines
passes through a common point, $(x_{0},y_{0})$, this is equivalent to making the mean-squared
distance from that point to the original lines a minimum.)  Consequently, we derive an expression for the
mean squared distance, $D^{2}(x_{0},y_{0})$, between all of the parallel lines.  Then we minimize
$D^{2}(x_{0},y_{0})$ with respect to $x_{0}$ and $y_{0}$, and obtain two linear equations for $x_{0}$
and $y_{0}$.  When we solve those equations simultaneously, we obtain
$x_{0}=1.586$ and $y_{0}=6.153$.   Using these values to evaluate $D^{2}(x_{0},y_{0})$ and
taking the square root, we find that $D_{rms}=0.054$.
\medskip

Next, we write the equation for the lines that pass through the point $(x_{0},y_{0})$ with the same slope,
$q_{c}=1/\log_{10}{\rho}_{c}$, that we used in the previous subsection.  This point-slope approach gives
the result
\begin{equation}
j_{max}~=~y_{0}~+q_{c}(\log_{10}N_{0}-x_{0})~=~(y_{0}-x_{0}q_{c})~+~q_{c}\log_{10}N_{0},
\end{equation}
where $(x_{0},y_{0})=(1.586,6.153)$ and $q_{c}=1/\log_{10}{\rho}_{c}$.  The quantity, ${\rho}_{c}$, is given
by Table 1 of Appendix A.  It is the largest positive root of the equation $r^{c+1}-2r^{c}+1=0$
when $c$ is an integer greater than 1.  However, when $c$ is not an integer, it is necessary to
interpolate the linearized difference equation to obtain an equation for $r$, as we did in Appendix A
for $c$ between 1 and 2.  An approximate expression is given in Eq.(C1a) of appendix C.
\medskip

Figure~14 shows these lines plotted with the measured points from
Figure~12.  The passage of these lines through the
common point, $(x_{0},y_{0})$, is clearly visible.  These lines fit the measured points almost as
well as the best-fit lines in Figure~12, verifying that $j_{max}$ is well represented by a line of slope
$1/\log_{10}{\rho}_{c}$, passing through a point, $(x_{0},y_{0})$, that does not depend on $c$.
\begin{figure}[h!]
 \centerline{%
 \fbox{\includegraphics[bb=90 250 540 725,clip,width=0.71\textwidth]{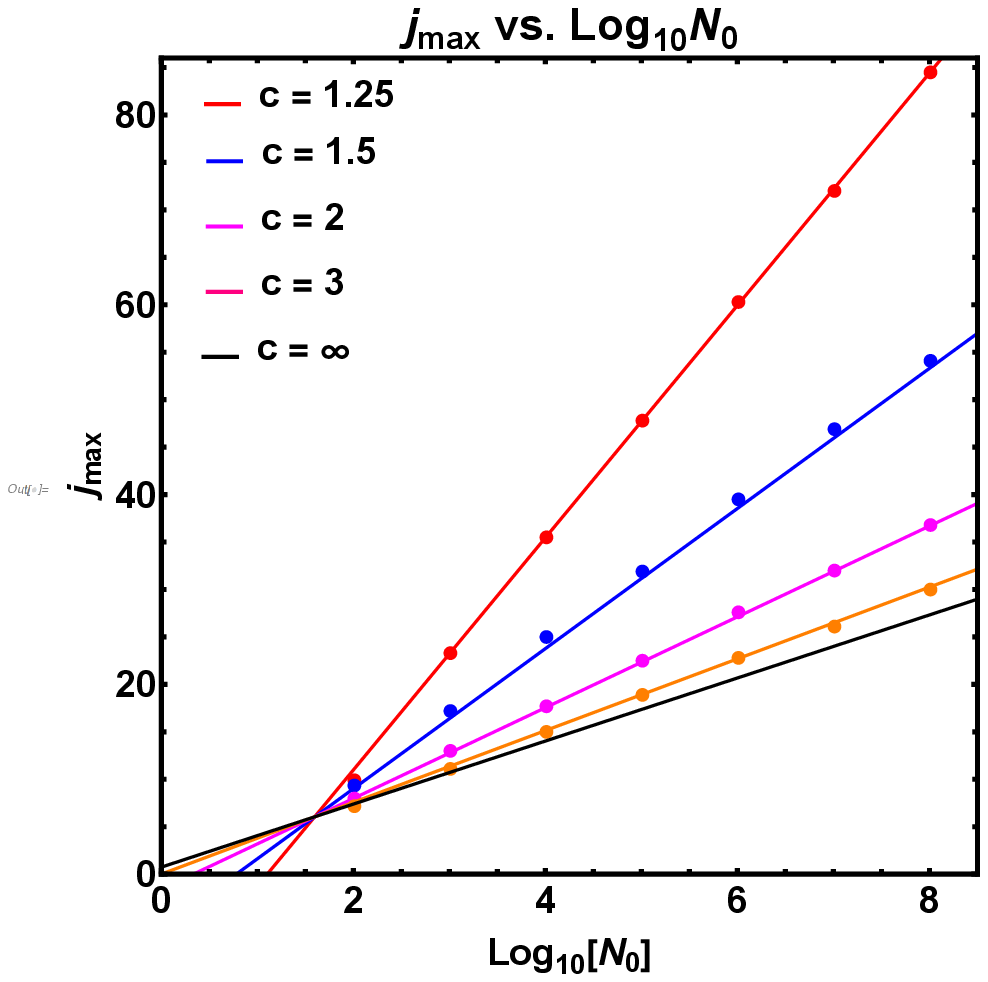}}}
 \caption{$j_{max}$ values calculated using Eq.(5.13) and plotted  versus $\log_{10}N_{0}$
for values of $c~{\geq}~1.25$ for comparison with the straight lines shown in Figure~12.}
\end{figure}
Eq.(5.15) gives numerical values for comparison with Eqs.(5.13) and (5.10).
\begin{eqnarray}  j_{max}~=~ \left\{\begin{array}{clcl}
-13.258~+~12.239~\log_{10}N_{0},~~~~c=1.25,\\
\\
-5.576~+~7.383~\log_{10}N_{0},~~~~c=1.5,\\
\\
-1.436~+~4.785~\log_{10}N_{0},~~~~c=2,\\
\\
+0.158~+~3.780~\log_{10}N_{0},~~~~c=3,\\
\\
+0.884~+~3.322~\log_{10}N_{0},~~~~c={\infty}.
\end{array}
\right.
\end{eqnarray}

\clearpage

\section{Red, Green, and Blue Molecules}
\subsection{Contributions of the red and green molecules}
It is interesting to interpret the growth equation in terms of the fractions of red, green, and blue molecules.
For $j~{\leq}~c$, all of the infected molecules are still contagious (red) and the incremental increase of
infections (normalized to the population, $N_{0}$) from one collision to the next is 
\begin{equation}
{\Delta}{\nu}_{j}~=~{\nu}_{j}~-~{\nu}_{j-1}~=~\overbrace{{\nu}_{j-1}}^{\textstyle{red}}\overbrace{(1-{\nu}_{j-1})}^{\textstyle{blue}}.
\end{equation}
Here, ${\nu}_{j-1}$ is the fraction of infected molecules at time $j-1$ (all of which are red) and $1-{\nu}_{j-1}$
is the fraction of noninfected blue molecules, which is also the probability of colliding with a blue molecule.
Consequently, the incremental increase in infections is just the product, ${\nu}_{j-1}(1-{\nu}_{j-1})$.   However, for
$j~{\geq}~c+1$, the molecules that were infected $c$ steps earlier are no longer contagious.  Consequently,
at step $j-1$, the fraction of infected, but noncontagious, green molecules is ${\nu}_{j-1-c}$, and the
incremental increase becomes
\begin{equation}
{\Delta}{\nu}_{j}~=~\overbrace{({\nu}_{j-1}-\underbrace{{\nu}_{j-1-c}}_{\textstyle{green}})}^{\textstyle{red}}
\overbrace{(1-{\nu}_{j-1})}^{\textstyle{blue}}.
\end{equation}
Thus, after a time, $c$, the increase in the number of infections is still equal to the product of the numbers
of red and blue molecules, but this increase is smaller than it would have been if the red molecules were
permanently contagious and there were no green molecules.  Eq.(6.2) indicates that the increase is smaller
because the number of red molecules is reduced by the current number of green molecules.
\medskip

It is also interesting to consider what happens toward the end of the epidemic when
${\Delta}{\nu}_{j}~{\rightarrow}~0$.  For permanently contagious molecules, Eq.(6.1) tells us that the
factor $(1-{\nu}_{j-1})~{\rightarrow}~0$ and therefore the other factor ${\nu}_{j-1}~{\rightarrow}~1$.
In this case, the epidemic ends because all the blue molecules become infected, and now exist as
contagious red molecules.  However, if the infected molecules remain contagious for a limited time,
$c<10$, then all of the blue molecules do not become infected and the factor $(1-{\nu}_{j-1})$ does
not approach 0.  Instead, the epidemic ends when other factor
$({\nu}_{j-1}-{\nu}_{j-1-c})~{\rightarrow}~0$.  This means that all
of the infected molecules become green and there are no red molecules left to infect blue ones.  As we have seen in Figures~6 and 8 and Eq.(5.6), by decreasing $c$ (and thereby increasing the amount of social distancing), we cause a greater number of blue molecules to avoid infection by the time that the epidemic
is over.
\medskip

Figure~15 shows the separate evolutions of red (${\nu}_{j}-{\nu}_{j-c}$) and green
(${\nu}_{j-c}$) molecules as
\begin{figure}[ht!]
 \centerline{%
 \fbox{\includegraphics[bb=88 247 547 716,clip,width=0.47\textwidth]{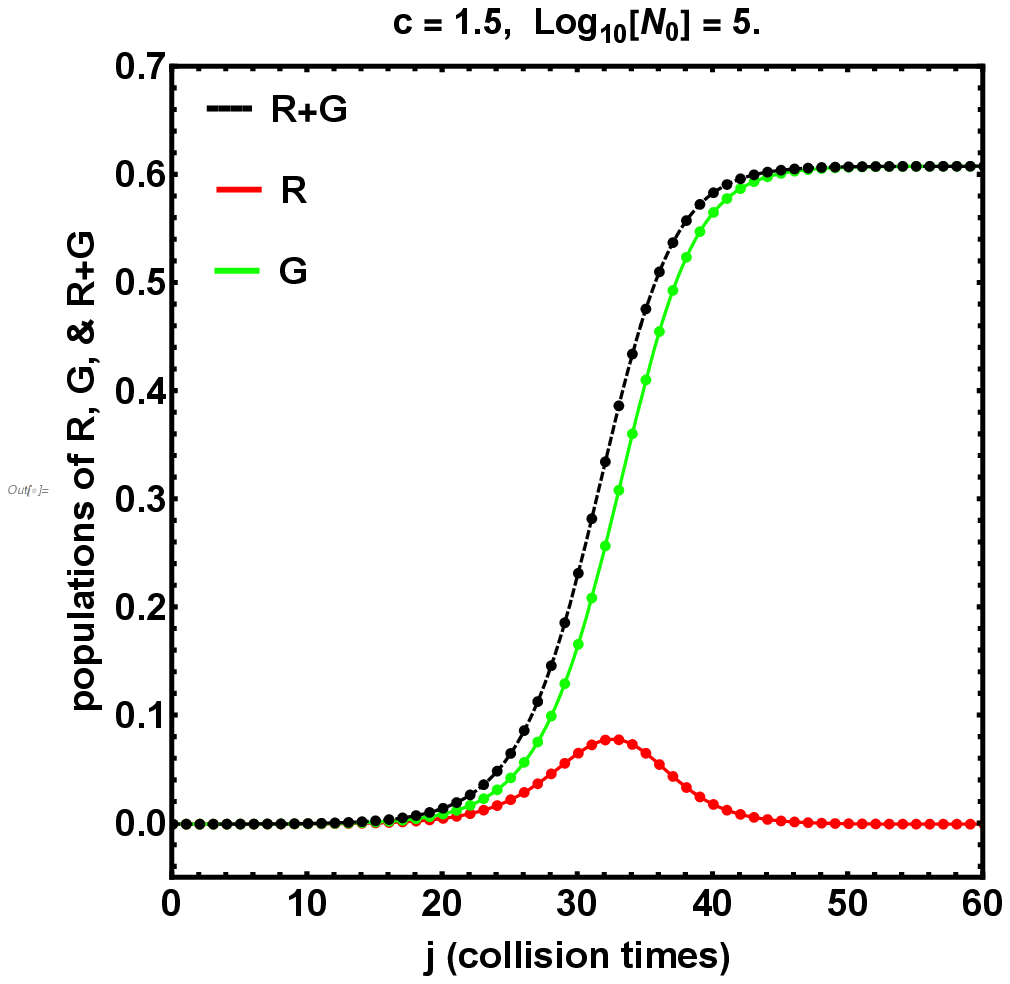}}
\vspace{0.01in}
 \fbox{\includegraphics[bb=88 255 547 716,clip,width=0.47\textwidth]{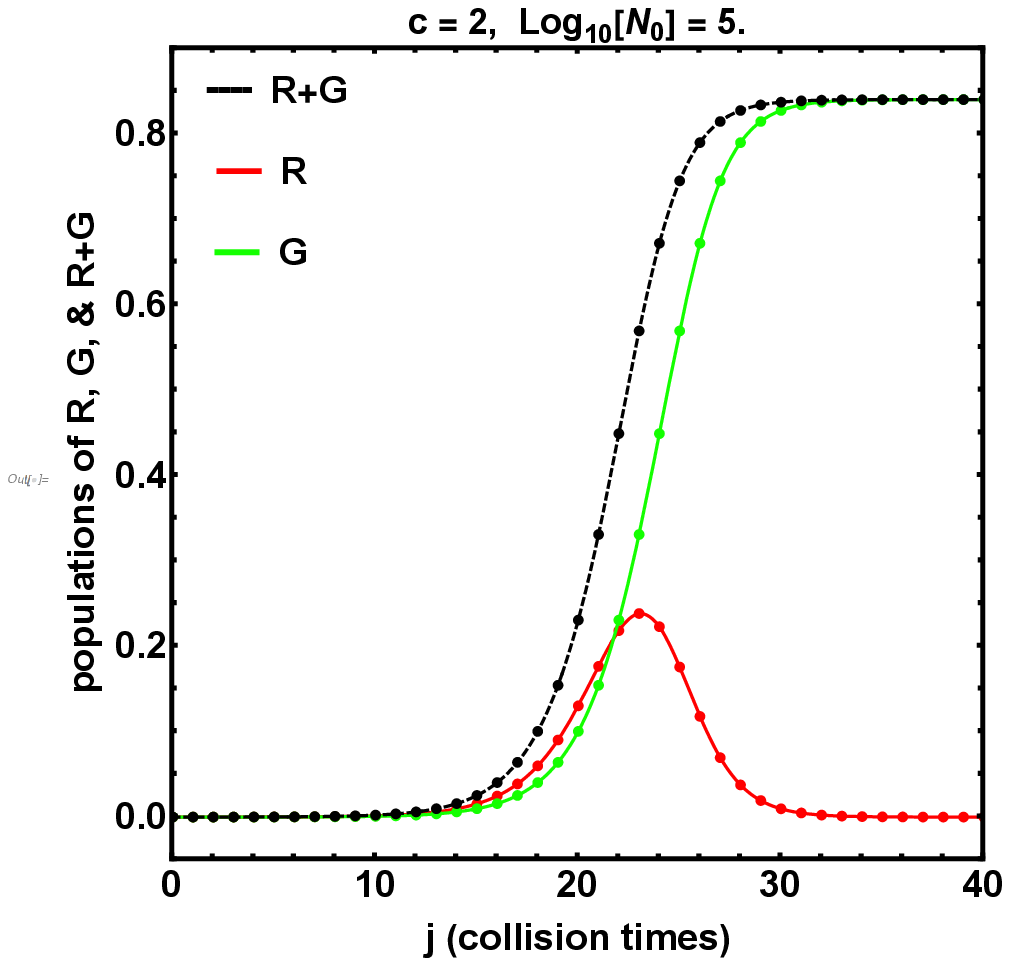}}}
 \hspace{0.01in}
\centerline{%
 \fbox{\includegraphics[bb=88 254 547 716,clip,width=0.47\textwidth]{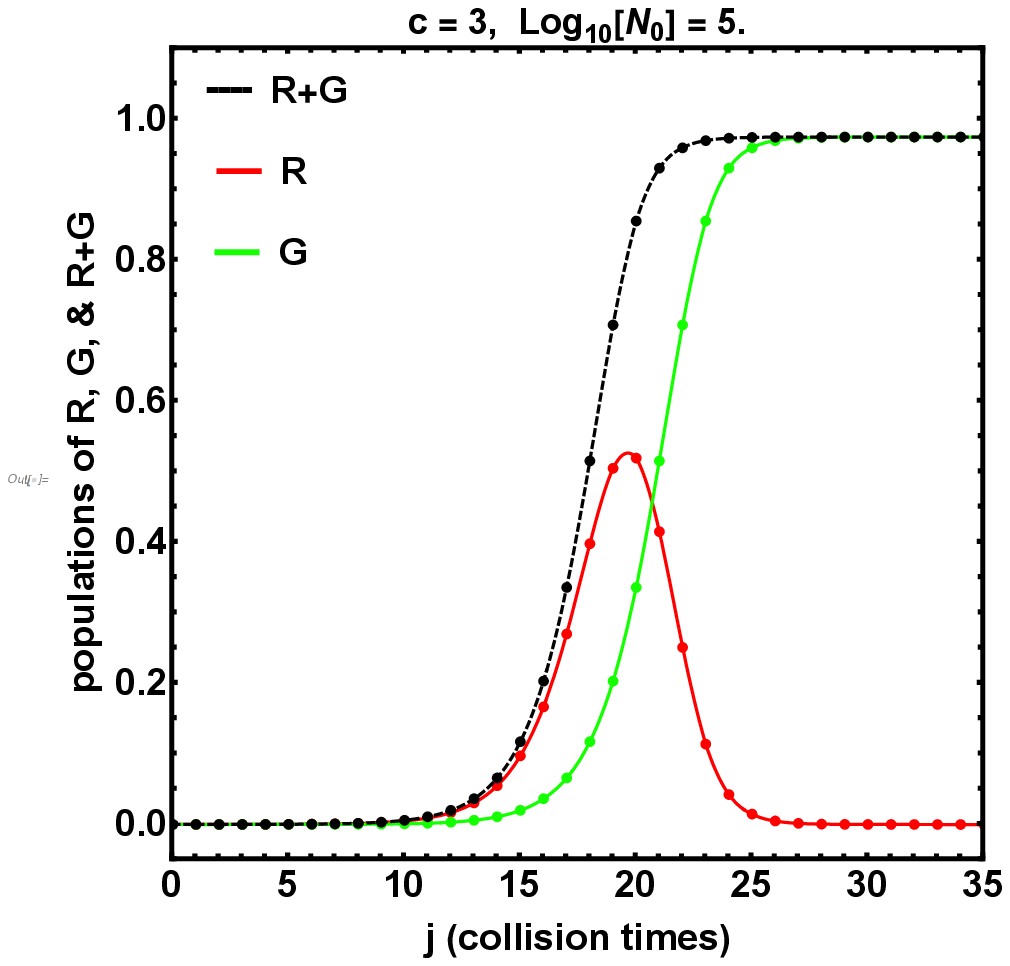}}
\vspace{0.01in}
\fbox{\includegraphics[bb=88 255 547 716,clip,width=0.47\textwidth]{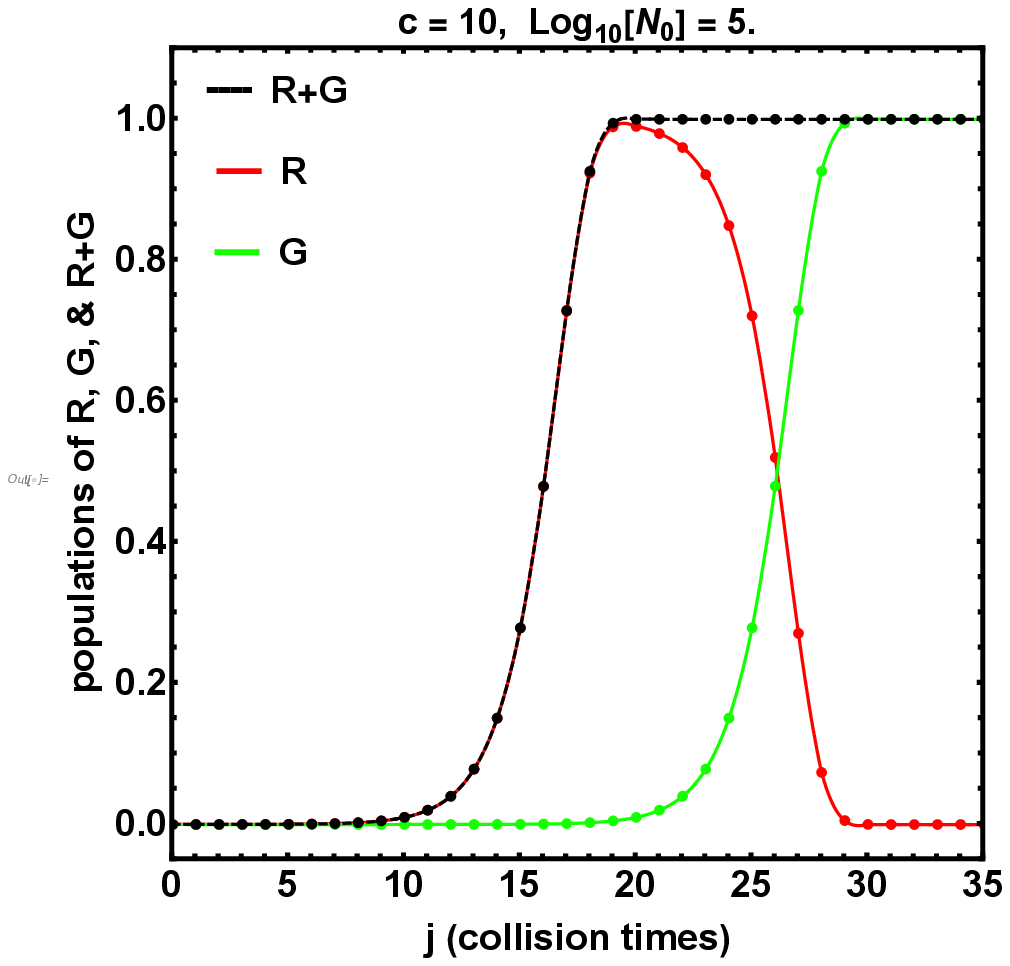}}}
\caption{Plots of ${\nu}_{j}$ (Red+Green), ${\nu}_{j}-{\nu}_{j-c}$ (Red), and ${\nu}_{j-c}$ (Green), showing
their separate evolutions for $c=1.5$ (upper left), $c=2$ (upper right), $c=3$ (lower left),
and $c=10$ (lower right).  Note the scale changes between some of the images.}
\end{figure}
\noindent
well as their sum, ${\nu}_{j}$ (the
combined total of all infections, whether still contagious or not).  (We do not need to provide a
separate plot for blue molecules because their number, $1-{\nu}_{j}$, is just 1 minus the the
combined number of red and green molecules, ${\nu}_{j}$.)  The dashed black curves show the
familiar 3-part variation of ${\nu}_{j}$ as a function of time in units of the step time, ${\tau}_{s}$.
The change begins slowly for a while, then increases rapidly for a time on the order of
$w_{e}={\Delta}j$, and then levels out at a final fraction of infected molecules, whose value, ${\nu}_{f}$, depends on the value of $c$.  Because the number of green molecules is given by the lagged
quantity, ${\nu}_{j-c}$, it shows the same time dependence as the number of infected molecules,
${\nu}_{j}$, but delayed by $c$ collision times.
\medskip

The red molecules do not have this 3-part monotonic increase.  Shown by the red curves in Figure~15,
the number of red molecules have a peaked profile, increasing to a maximum value
and then falling back to zero at the end of the epidemic.   This is what one would expect from their
definition, ${\nu}_{j}-{\nu}_{j-c}$, which is the difference between two terms that eventually become
equal.
\medskip

We can learn more about the time variation of the red molecules by examining the plots in Figure~15 as
a function of $c$.  When $c$ is much smaller than the width of the peak (which is essentially the
rise time, $w_{e}={\Delta}j$, of the black profile of total infections), we see that
\begin{equation}
{\nu}_{j}-{\nu}_{j-c}~=~c({\nu}_{j}-{\nu}_{j-1})~=~c{\Delta}{\nu}_{j}.
\end{equation}
Thus, for $c<<w_{e}$, the plots of the number of red molecules are within a factor of $c$ of
matching the plots of infection rate, ${\Delta}{\nu}_{j}$, in Figure~7.  This near equality is shown
in Figure~16, which compares ${\nu}_{j}-{\nu}_{j-c}$ and  $c{\Delta}{\nu}_{j}$ for $c=1.5$ and 2
and $\log_{10}N_{0}=5$.  In the left panel, the two curves for $c=1.5$ are nearly identical.  In
the right panel where $c=2$,
the peak of the red curve is still nearly equal to the height of the black curve, but is now shifted
to the right by about 0.5 of a collision time.  However, the agreement breaks down as
$c$ becomes larger and the steep increase in the number of green molecules is appreciably
delayed.  For $c~{\geq}~3$, we return to Figure~15 and see that the rising segment of the red
curve coincides with the rising segment of the black curve, until the infection rate is dominated
by green molecules and the red curve reaches its peak.  This leads to the concept of `herd immunity'
which we shall discuss next.
\begin{figure}[ht!]
 \centerline{%
 \fbox{\includegraphics[bb=88 240 560 716,clip,width=0.46\textwidth]{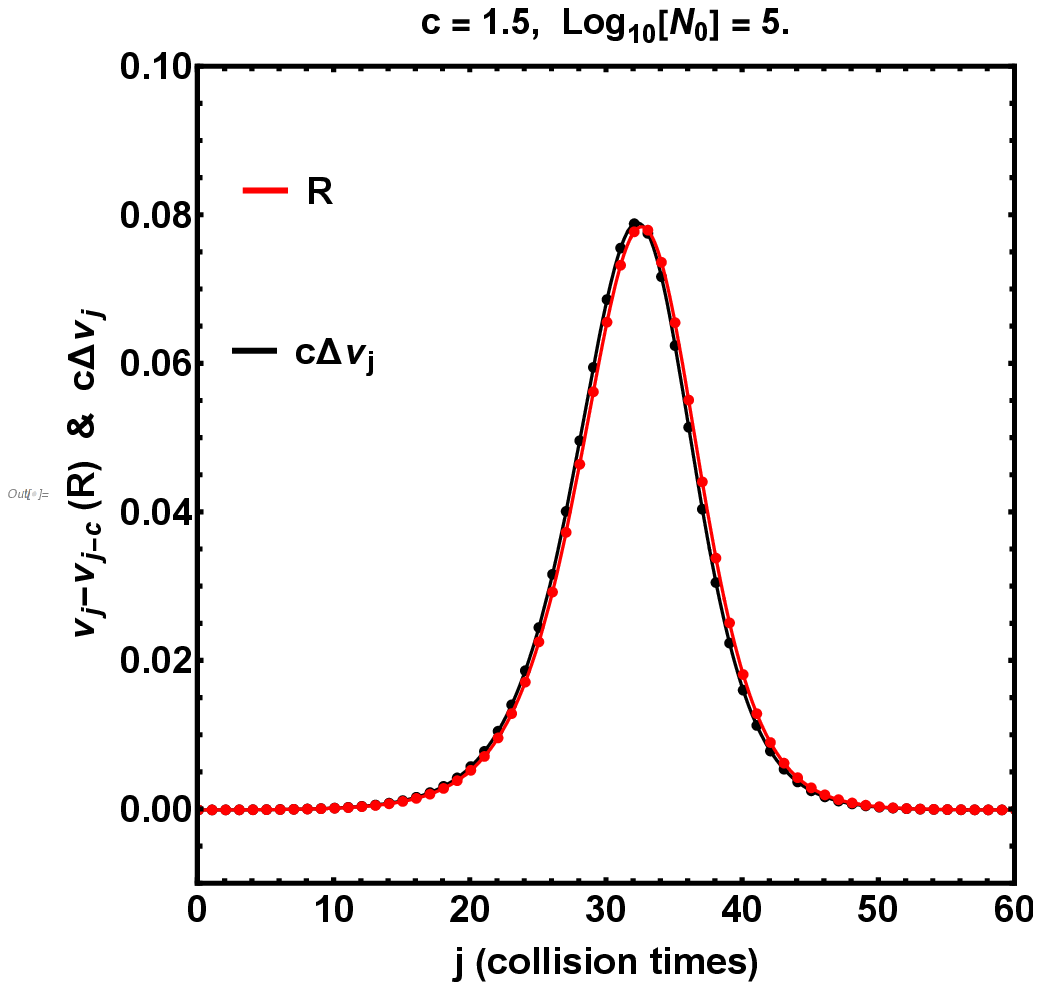}}
 \hspace{0.01in}
 \fbox{\includegraphics[bb=88 240 560 716,clip,width=0.46\textwidth]{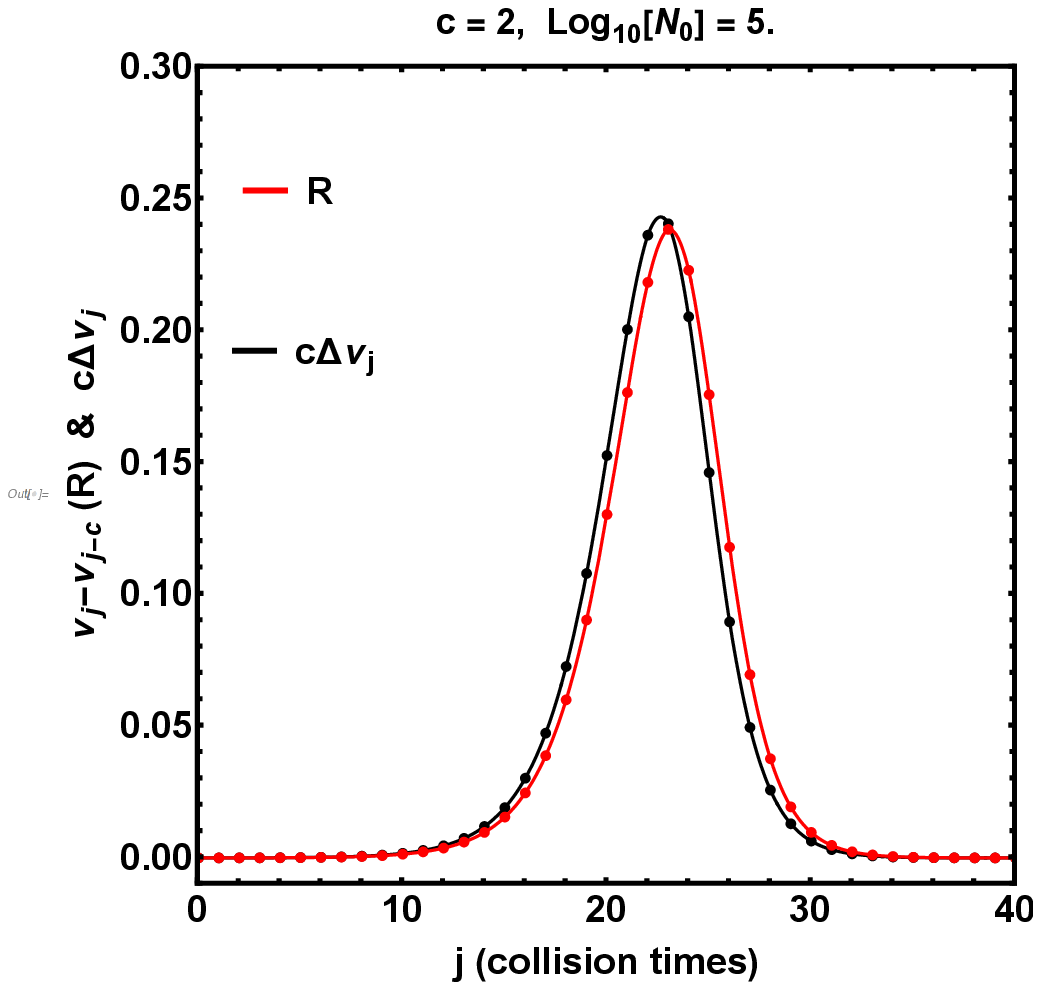}}}
 \caption{Plots of ${\nu}_{j}-{\nu}_{j-c}$  and $c{\Delta}{\nu}_{j}$, comparing the fraction of red
molecules with $c$ times the growth rate of infections as $c$ increases from 1.5 to 2.}
\end{figure}

\subsection{Herd `Immunity'}
An equivalent way of thinking about the relation between the numbers of red and green molecules
is through the so-called `herd immunity' concept that one often encounters in the literature
(see, for example, Fine et al. (2011)).  Here, we think of the green molecules as
shielding blue molecules from the remaining red ones, so that some of these red molecules
cannot find blue molecules before they lose their contagiousness and turn green.  The increased
number of green molecules makes it even more difficult for the surviving red molecules to find
blue ones, and the number of red molecules starts to decrease.   We take the threshold for this
turnaround to be the number of infected molecules (relative to the total population, $N_{0}$)
when the number of red molecules has reached its peak.  However, as Eq.(6.2) indicates, the
remaining blue molecules are still subject to infection as long as there are red molecules in
the population, and the epidemic is not over until all of those red molecules have turned green.
Consequently, even after reaching herd immunity, an appreciable time remains before the red
population is reduced to 0 and a blue molecule is safe from infection.
\medskip

We can think about this process graphically with the help of Figure~15.  The number of
red molecules is given by ${\nu}_{j}-{\nu}_{j-c}$, the difference
between the black and green curves.  This difference is maximum where
${\Delta}({\nu}_{j}-{\nu}_{j-c})=0$, which means that ${\Delta}{\nu}_{j}={\Delta}{\nu}_{j-c}$.
Consequently, the red curve reaches its peak where the profiles of ${\Delta}{\nu}_{j}$ and
${\Delta}{\nu}_{j-c}$ intersect, which is somewhere between the peaks of those
two profiles, depending on the value of $c$.  Another way of interpreting the equality,
${\Delta}{\nu}_{j}={\Delta}{\nu}_{j-c}$, is to say that the red curve reaches its peak
at the value of $j$ where the slopes of the black and green curves are equal.  When $c$ is
small, these points lie slightly above and below the points of maximum slope on the
black and green curves, corresponding to a relatively small peak in the red curve.
As $c$ increases, the peak height of the red curve increases, and the points of equal slope
on the black and green curves move farther apart, so that ${\nu}_{j}$ increases and lies closer
to the `knee' of the black curve and ${\nu}_{j-c}$ decreases and lies closer to the `toe' of the
green curve.  Eventually, for $c=10$, ${\nu}_{j}=1$ and ${\nu}_{j-c}=0$, corresponding to a
common slope of 0.
\medskip

To describe this process more quantitatively, we refer to Figure~17, which compares plots
of the numbers of
red molecules (${\nu}_{j}-{\nu}_{j-c}$) with corresponding plots of the total number of
infections (${\nu}_{j}$) for values of $c$ in the range 1.25 to 20.  In the upper panel,
we find the time that a particular plot of the number of red molecules reaches its peak,
and then in the lower panel, we find the value of ${\nu}_{j}$ for the corresponding plot
of the total number of infections at that time.  Thus, in the upper panel, the blue curve
with $c=1.5$ has a peak of strength 0.08 at $j=33$ which corresponds to the value
${\nu}_{j}=0.38$ in the blue profile in the lower panel.  As $c$ increases, the threshold
value of ${\nu}_{j}$ also increases, eventually approaching 1 as $c$ becomes large.  In the
lower panel, the curves with $c=5$ and $c=20$ are almost indistinguishable, indicating
that ${\nu}_{j}$ is already very close to 1.  This is consistent with our earlier observation
that the growth is essentially the same for $c=10$ as it is when $c~{\rightarrow}~{\infty}$.
\clearpage
\begin{figure}[ht!]
 \centerline{%
 \fbox{\includegraphics[bb=88 265 530 715,clip,width=0.65\textwidth]{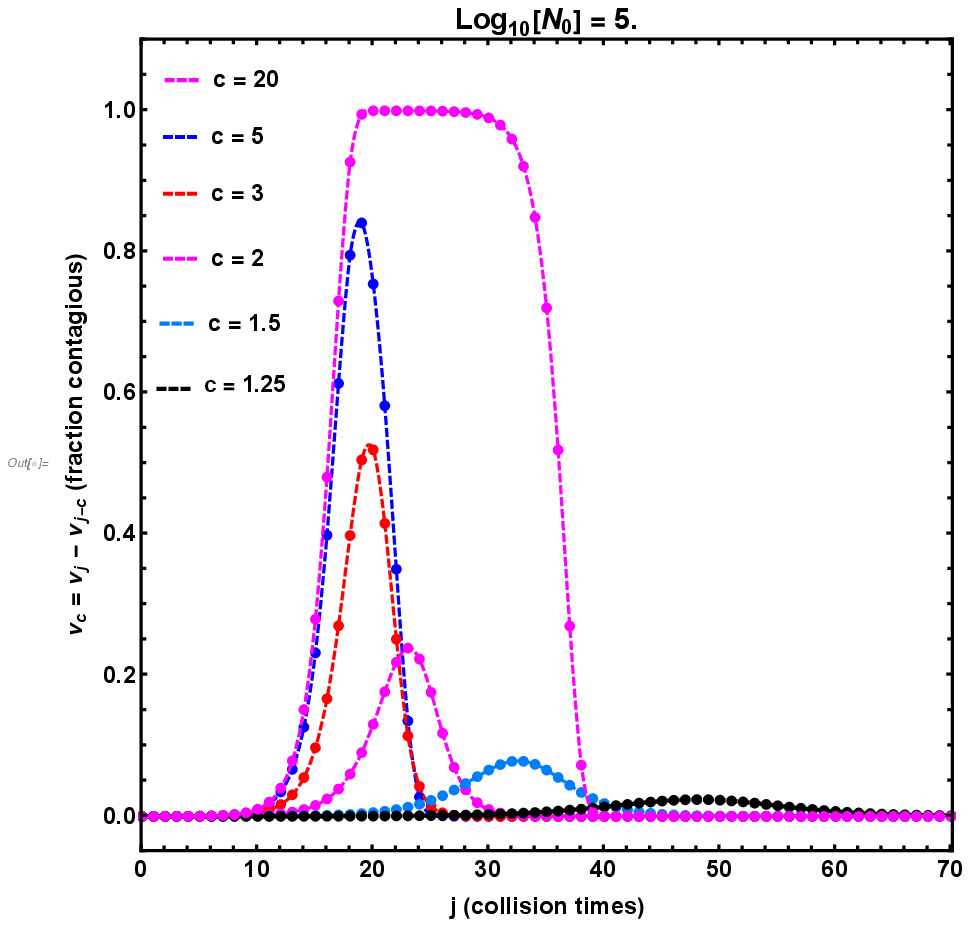}}}
 \vspace{0.01in}
\centerline{%
 \fbox{\includegraphics[bb=88 265 530 715,clip,width=0.65\textwidth]{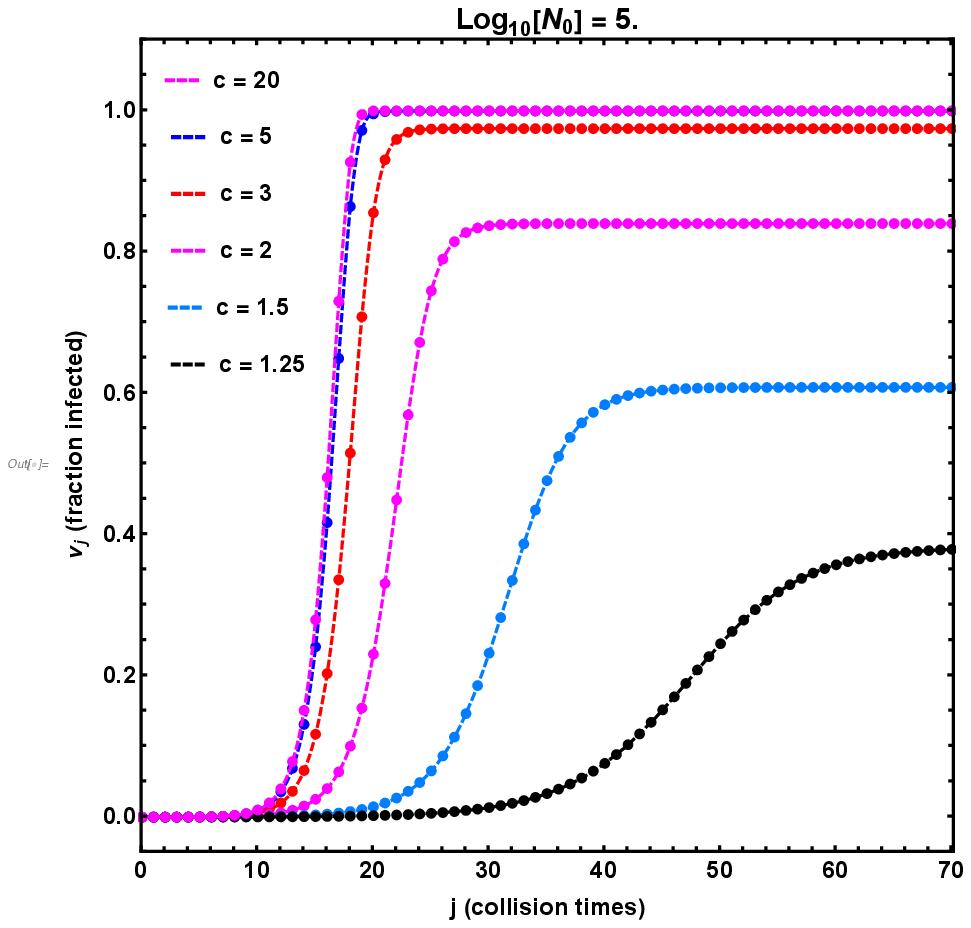}}}
 \caption{The fraction of currently contagious molecules, ${\nu}_{c}^{(j)}$ (top), compared
with the fraction of infected molecules, ${\nu}_{j}$ (bottom) in the total population of
$N_{0}=10^{5}$ molecules.}
\end{figure}
\clearpage

Figure~18 shows a plot of these computed values of ${\nu}_{j}$, which we now refer
to as the herd-immunity threshold, ${\nu}_{herd}$, versus the corresponding values of $c$.
The solid red curve is the root-mean-square best fit to those points using the formula
\begin{equation}
{\nu}_{herd}~=~1~-~e^{-0.860(c-1)}.
\end{equation}
\begin{figure}[h!]
 \centerline{%
 \fbox{\includegraphics[bb=88 246 547 720,clip,width=0.75\textwidth]{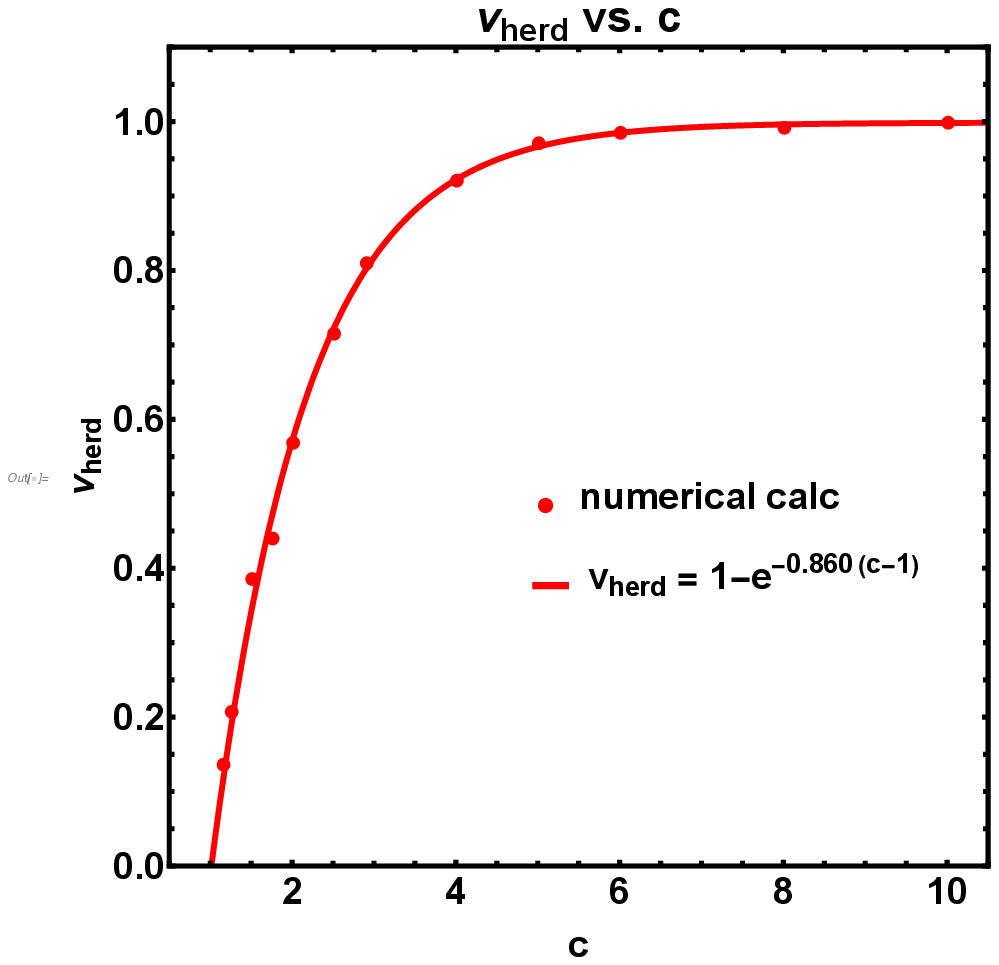}}}
 \caption{A plot of the total number of infections, ${\nu}_{herd}$, when the fraction of
contagious red molecules reaches its peak.  The red curve is the root-mean-square best fit
to the calculated points (red dots). }
\end{figure}
Having learned what the fraction of total infections is when the corresponding fraction of
contagious red molecules reaches its peak, we now ask when the fraction of red molecules
actually reaches its peak.  In particular, after reaching the time, $j_{max}$, that the total
infection rate for red-plus-green molecules reaches its peak, how much longer must we wait until the
number of red molecules reaches its peak?  In Figure~16, the plots of ${\nu}_{j}-{\nu}_{j-c}$ and
$c{\Delta}{\nu}_{j}$ have already shown that the wait time is roughly 0.5 of a collision time
when $c=2$ and even less when $c=1.5$.  So the next step is to make additional
comparisons and plot the results.
\medskip

Figure~19 shows the result of measuring the peak locations of ${\Delta}{\nu}_{j}$ and
${\nu}_{j}-{\nu}_{j-c}$.  The measured lags are relatively small,
\begin{figure}[h!]
 \centerline{%
 \fbox{\includegraphics[bb=88 248 547 720,clip,width=0.75\textwidth]{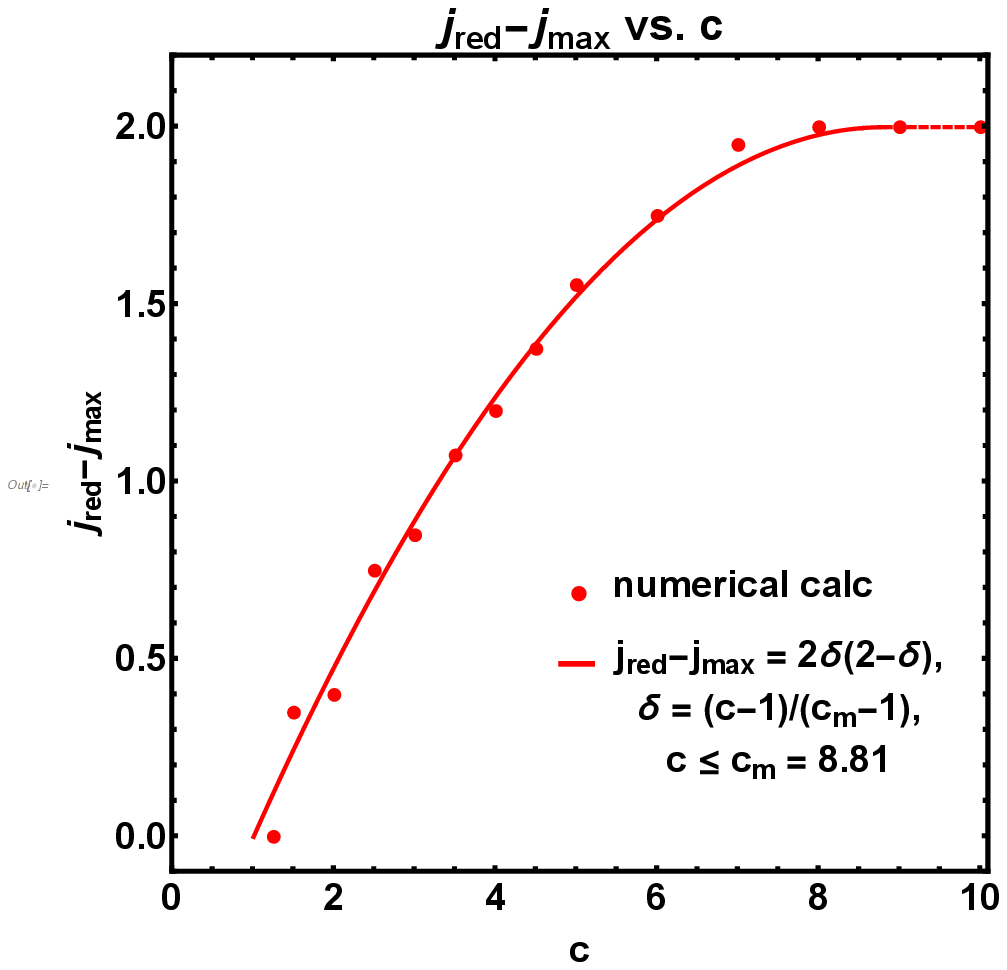}}}
 \caption{The time lag, $j_{red}-j_{max}$, between the peaks of the total number of
infections, ${\nu}_{j}$, and the number of contagious molecules, ${\nu}_{j}-{\nu}_{j-c}$,
plotted versus the contagious time, $c$.  The solid red curve is the root-mean-square best fit
to the calculated points (red dots) given by Eq.(6.5), and the dotted red line is its extension
with a constant lag of 2 after the solid curve reaches its peak at $c_{m}=8.81$.}
\end{figure}
\noindent
never exceeding 2 collision times, which is the difference, $j_{red}-j_{max}$, when
$c~{\rightarrow}~{\infty}$.  (For this calculation, we return to Figure~2 and recognize that
$j_{red}=19$ is the $j$-value when ${\nu}_{j}$ reaches 1, as shown in the left panel, and
that $j_{max}=17$ is the time that ${\Delta}{\nu}_{j}$ reaches its peak, as shown in the right
panel.)  Coming back to Figure~19,  we note that the data points are best fit by the quadratic
function
\begin{equation}
j_{red}-j_{max}~=~2\left (\frac{c-1}{c_{m}-1} \right ) 
\left \{  2 -\left (\frac{c-1}{c_{m}-1} \right ) \right \}~=~0.512(c-1)-0.033(c-1)^2,
\end{equation}
where $c_{m}$ is the value of $c$ where this quadratic function reaches its maximum value
of 2 units of the step time, ${\tau}_{s}$.  In this case, $c_{m}=8.81$.  The numerical form
of this equation shows that the lag starts as a linear function whose slope is nearly 1/2, but
`decelerates' very slightly as $c$ becomes large.    Taking the derivative of Eq.(6.5), we
obtain the slope
\begin{equation}
\frac{d}{dc}(j_{red}-j_{max})~=~\left (\frac{4}{c_{m}-1} \right ) \left (\frac{c_{m}-c}{c_{m}-1} \right )~=~
0.512 \left (\frac{c_{m}-c}{c_{m}-1} \right ),
\end{equation}
which equals 0.51 when $c=1$ and becomes 0 when $c=c_{m}$.  Thus, if $c-1$ is small, then
$j_{red}-j_{max}~{\approx}~0.51(c-1)$, which gives lags of 0.51 for $c=2$ and 0.25
for $c=1.5$, consistent with our measurements from Figure~16.
\medskip

It is interesting to compare the $c$-dependence of this lag with the corresponding formula obtained
for profiles like the hyperbolic
tangent and square-root functions mentioned in the footnote of section 5.2.  For example, we
can use the hyperbolic tangent formula, ${\nu}_{j}=({\nu}_{f}/2)[1+\tanh\{{\gamma}(j-j_{max'})\}]$,
to construct the fraction of red molecules, $R_{j}={\nu}_{j}-{\nu}_{j-c}$, and then determine the
location of its peak value by setting ${\Delta}R_{j}=R_{j}-R_{j-1}=0$.  The result is
$j-j_{max'}=(c+1)/2$, where $j_{max'}$ is the $j$-value for which ${\nu}_{j}$ has its greatest slope.
$j_{max}$ (without the prime) is where ${\Delta}{\nu}_{j}$ has its peak and differs from $j_{max'}$
by 1 unit according to $j_{max}=j_{max'}+1$.  Consequently, $j_{red}-j_{max}=(c+1)/2 -1 = (c-1)/2$,
which is essentially the same as the starting value of $0.51(c-1)$, that we found
in Figure~19 for the RGB-profiles\footnote{Note that to obtain this expression, we had to determine the
red peak from ${\Delta}R_{j}=0$, rather than $dR_{j}/dj=0$, which gives $j=j_{max'}+c/2$, and
we had to distinguish between the place, $j_{max'}$, that ${\nu}_{j}$ has its greatest slope, and
the place, $j_{max}=j_{max'}+1$, that ${\Delta}{\nu}_{j}$ has its peak.  These small differences matter
when the range of $j_{red}-j_{max}$ is only 2 units of ${\tau}_{s}$.}. 
\medskip

Despite this agreement between the lag obtained for the RGB-model and the lag obtained for
the hyperbolic tangent formula, there is an important difference.
Namely, as $c$ increases, the $RGB$-lag gradually deviates from its initially linear
$c$-dependence due to the presence of the small quadratic term in Eq.(6.5).   As shown in
Figure~19, the plot bends over and eventually saturates at 2 units of collision time when $c$
becomes comparable to the width of the ${\nu}_{j}$ profile.  In contrast, the lag for the
hyperbolic tangent model remains linear as $c$ increases until the saturation occurs.  This
distinction reflects different curvatures of the RGB-profile above and below the point that
${\nu}_{j}$ has its maximum slope, especially as $c$ becomes large.  We have already seen
this asymmetry in the plots of ${\nu}_{j}$ and ${\Delta}{\nu}_{j}$ in Figure~2 when the red
molecules were permanently contagious, and in Figures~6 and 7 when $c>2$.  The
hyperbolic tangent formula does not have this asymmetry and therefore gives a lag that
remains linear as $c$ increases until the saturation occurs. 
\medskip

We can understand why this asymmetry causes $j_{red}-j_{max}$ to deviate from its initially linear dependence by referring to plots of $R+G$ (black curve), $G$ (green curve), and $R$ (red curve)
in Figure~15, and considering how the black curve changes around $j_{p}$, its point of maximum
slope.  For $j~{\geq}~j_{p}$, the slope can be approximated by $s_{j}=s_{p}-a_{2}(j-j_{p})$,
where $a_{2}$ is the curvature (\textit{i.e.} the rate of change of the slope with respect to $j$) on
the upper branch of the profile.  Likewise, $s_{j}=s_{p}-a_{1}(j_{p}-j)$ on the lower
branch where $j_{p}~{\geq}~j$.  We assume that both $a_{2}$ and $a_{1}$ are positive so that
the slope decreases on both sides of $j_{p}$, as it ought to do at a point of maximum slope.
\medskip

The essential point here is that for each value of $j$, the red curve is produced by subtracting
contributions from the upper branch of the black curve where $j=j_{p}+{\delta}$ and the lower
branch of the (identical, but shifted) green curve where $j=j_{p}-c+{\delta}$.  Consequently,
if $c$ is small and the black and green curves lie close together, then the peak of the red curve
will lie at the mid-point between the black and green curves where ${\delta}=c/2$ and the slopes
are both very close to the peak value $s_{p}$.  (As noted above, the
condition ${\Delta}R_{j}=0$ adds an extra half step, and the reference to $j_{max}$, rather than
$j_{p}$, subtracts a full step, so that the result is $j_{red}-j_{max}=(c-1)/2$, rather than $c/2$.)
\medskip

However, when $c$ becomes large, the slope will be $s_{p}-a_{2}(j-j_{p}-{\delta})$ on the upper
branch of the black curve and $s_{p}-a_{1}(j_{p}-j-c+{\delta})$ on the lower branch of the green
curve.  To find the value of $j$ for which these slopes are equal, we simply equate those
expressions and solve for $j$, obtaining $j-j_{p}=\{a_{1}/(a_{1}+a_{2})\}c$.  If the accelerations
are equal, then $j=j_{p}+(c/2)$ (or $j_{red}-j_{max}=(c-1)/2$ for the condition ${\Delta}R_{j}=0$).
But if $a_{2}>a_{1}$, then $j-j_{p}<c/2$, and the location of the red curve falls behind the mid-point
by the amount $(c/2)\{(a_{2}-a_{1})/(a_{2}+a_{1})\}$.  This asymmetry causes the lag to deviate
from its initially linear dependence on $c$ that we found in Figure~19.
\medskip

In summary, the `herd immunity' occurs when red molecules are turning green as fast as
they are being produced by the infection of blue molecules.   At this time, the number of
red molecules has reached its peak and the infection rate of blue molecules is slightly
past its peak.  In fact, it is convenient to think in terms of the time that the blue molecules
are being removed most rapidly (and the red-plus-green molecules are being created
most rapidly).  This time occurs approximately half-way up the growth
profile for red-plus-green molecules, which means that the red peak will occur somewhere
in the time remaining between the middle and end of that profile.  As Figure~19 showed,
this time lag, $j_{red}-j_{max}$, depends on $c$, but it is relatively small and always less than
2 collision times, ${\tau}_{s}$.  Thus, for serious social distancing and a value of $c$ close to 1,
the time, $j_{max}$, will be greatly delayed, as will $j_{red}$, which follows closely after it.
However, the good news is that the immunity threshold, ${\nu}_{herd}$ will
be correspondingly small and so will the total number of infected molecules, ${\nu}_{f}$,
when the epidemic is over.
\medskip

Finally, it is important to recognize that the `herd immunity' does not give the blue molecules
any immunity against infection by the red ones.  It is just a way of saying that the rate of
infection has reached its maximum, and will be decreasing for the remainder of the ramp
time.  Although it may be reassuring to think of the green molecules as shielding blue
molecules from red ones, it is not so reassuring to realize that red molecules provide the
shielding when infected molecules are permanently contagious.  Once the infection rate
has reached its peak, the red molecules are so plentiful that
some block others and prevent them from participating in the `feeding frenzy' that will now
continue at a diminishing rate.  In the words of former baseball player, Yogi Berra (1998),
`It ain't over 'til it's over' and all the red molecules are gone.

\clearpage

\section{Changing Social Distancing During an Epidemic}
\subsection{Removing the Social Distancing Entirely}
An interesting application of these calculations is to see what would happen if the
amount of social distancing were relaxed suddenly before the virus has been completely
eliminated.  Figure~20 shows the result of suddenly changing $c$
from $c=1.5$ to 10 after 42 collisions when the fraction of infected molecules had
increased to within $3.3{\%}$ of its final value of 0.61, and the infection rate had decreased
to about $14{\%}$ of its peak height.  This calculation was done for a population of
$N_{0}=10^{5}$,
which means that only $0.033~{\times}~0.61~{\times}~10^{5}~{\approx}~2000$ more
molecules would be infected if the social distancing were to continue at its current rate.
\medskip

\begin{figure}[ht!]
 \centerline{%
 \fbox{\includegraphics[bb=88 265 530 716,clip,width=0.47\textwidth]{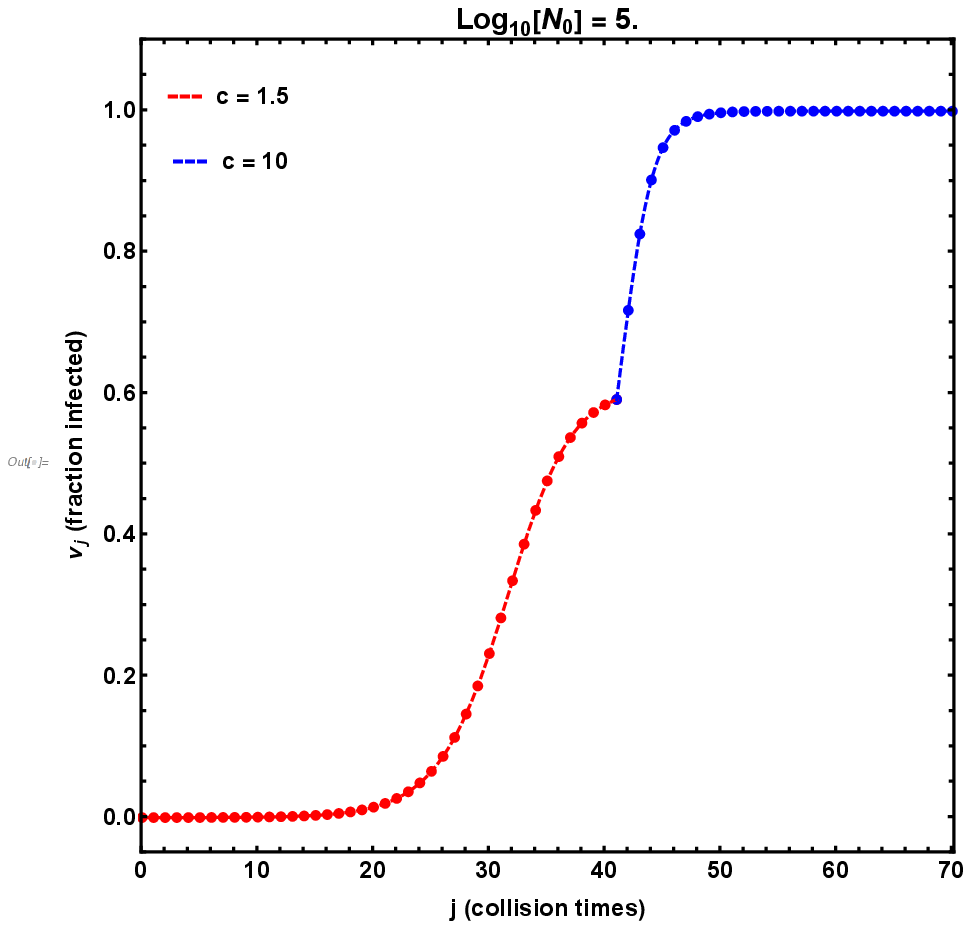}}
 \hspace{0.01in}
 \fbox{\includegraphics[bb=88 265 530 716,clip,width=0.48\textwidth]{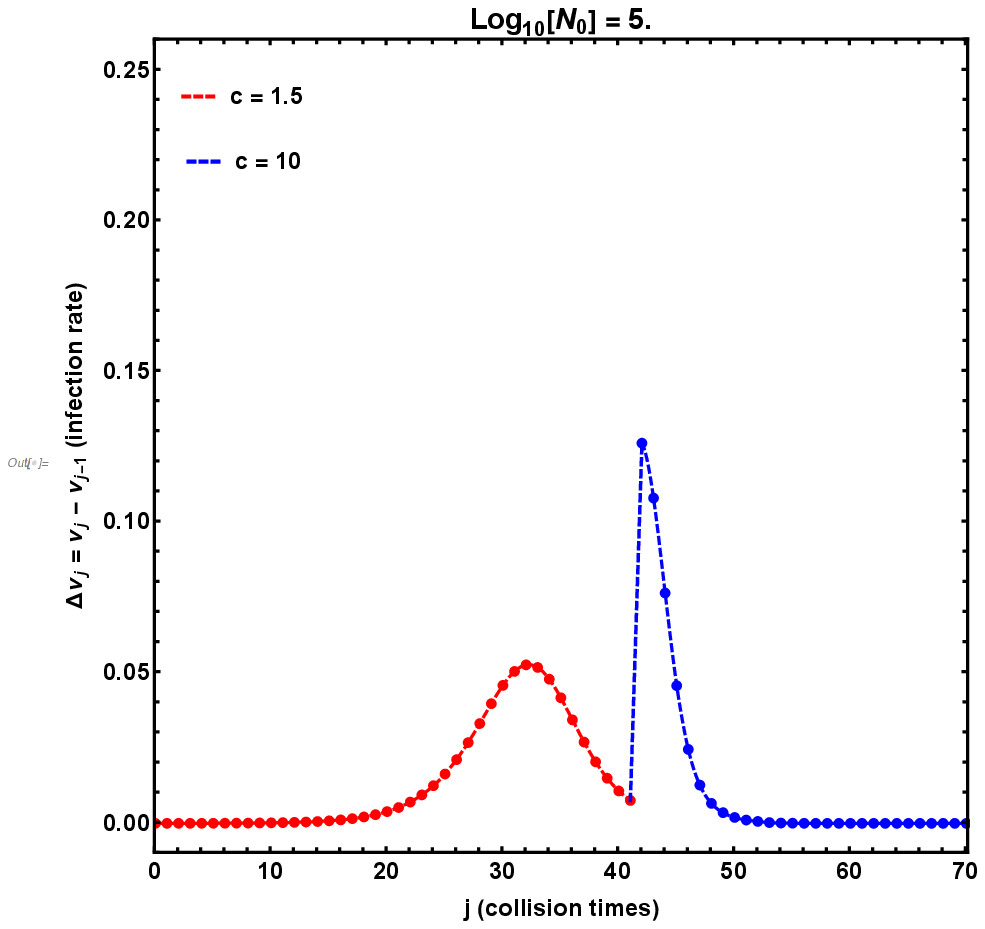}}}
 \caption{The fraction of infected molecules (left), and the infection rate
(right) when social distancing is relaxed suddenly from
$c=1.5$ to $c=10$ after 42 collisions, showing the rapid infection of the
remaining blue molecules.}
\end{figure}

What was the number of contagious molecules at that time?  The fraction of contagious
molecules is so small that we cannot obtain an accurate estimate from Figure~17.  And
in that figure, the
overlap between the curves with $c=1.25$ and $c=1.5$ make the estimate even more
difficult.  However, we can obtain a fairly accurate estimate from the enlarged plot
of ${\nu}_{j}$ in Figure~6 in combination with Eq.(5.1) for ${\Delta}{\nu}_{j}$.  In Figure~6,
the slope of the curve with $c=1.5$ is 0.00657 per step (or collision) length when
$j=40-42$.  For $c=1.5$ (corresponding to 1.5 steps), this means that the
change ${\nu}_{42}-{\nu}_{40.5}=1.5~{\times}~0.00657=0.00986$.  Thus, for a total
population of $N_{0}=10^{5}$ molecules, this corresponds to 986 contagious
molecules after $j=42$ collision times.  This number is relatively small compared
to the $0.6~{\times}~10^{5}$ molecules that had been infected at this time, as we can
see in the bottom panel of Figure~17.  However, as we will see next, those 986 contagious
molecules would soon do great damage.
\medskip

As shown in the right panel of Figure~20, at $j=42$, the infection rate increased suddenly from
${\Delta}{\nu}_{j}~{\approx}~0.007$ to ${\Delta}{\nu}_{j}=0.13$, which is 2.5 times the peak
rate that occurred near $j_{max}=32$.  By comparison, the fraction of infected molecules
increased rapidly from ${\nu}_{j}~{\approx}~0.59$ (near their plateau value of ${\nu}_{j}=0.61$) to
${\nu}_{j}=1$.  Thus, even when the infection rate had decreased to less than $1{\%}$ (which
is about $14{\%}$ of its peak value), the sudden relaxation of social distancing caused a rapid burst
of infection that effectively wiped out the 40,000 surviving blue molecules.  Clearly, 986
contagious (red) molecules were sufficient to restart the infection at a very high rate (2.5 times
the previous peak rate at $j_{max}=32$).

\subsection{Sudden Increase of ${\mathbf{c}}$}
We have seen that $c=10$ is equivalent to a permanently
contagious virus with no social distancing.  Now, we consider a presumably more realistic situation in
which $c$ changes suddenly from $c=1.25$ to $c=2.50$.  Figure~21 shows the first
\begin{figure}[h!]
 \centerline{%
 \fbox{\includegraphics[bb=90 250 547 725,clip,,width=0.47\textwidth]{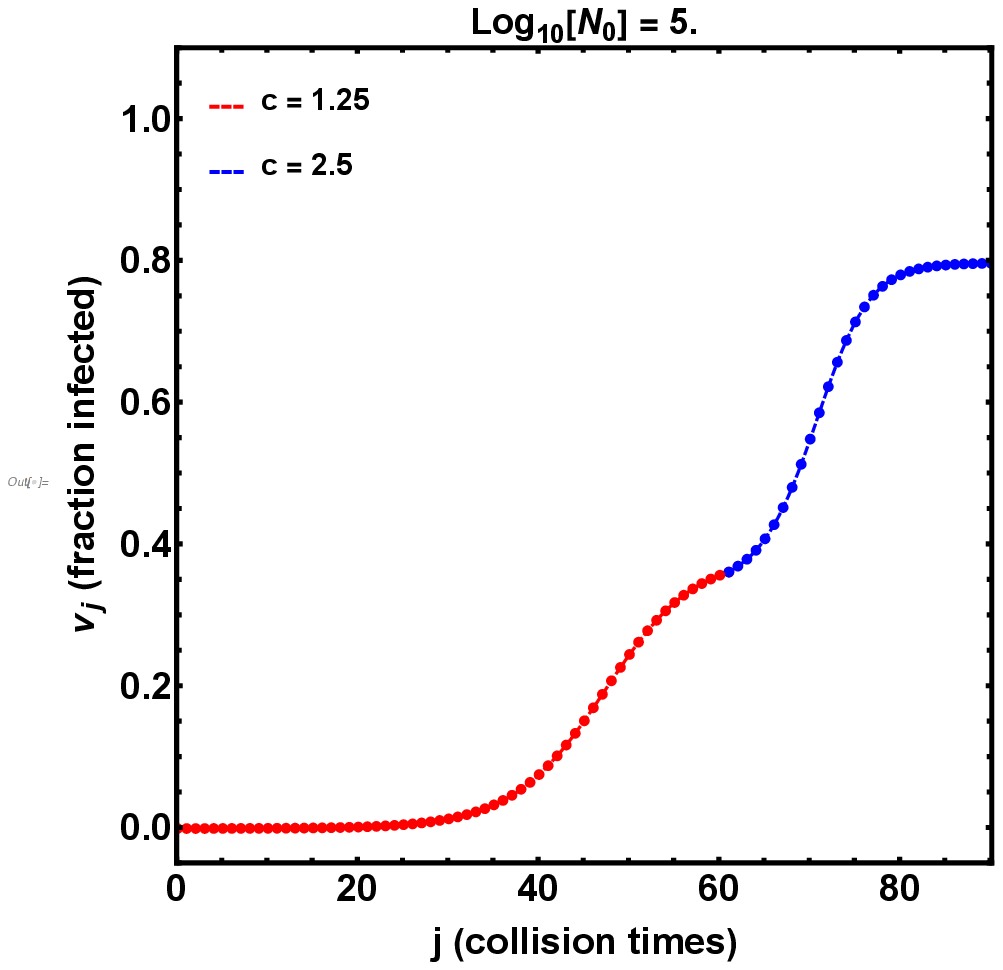}}
\hspace{0.01in}
 \fbox{\includegraphics[bb=90 245 557 721,clip,width=0.48\textwidth]{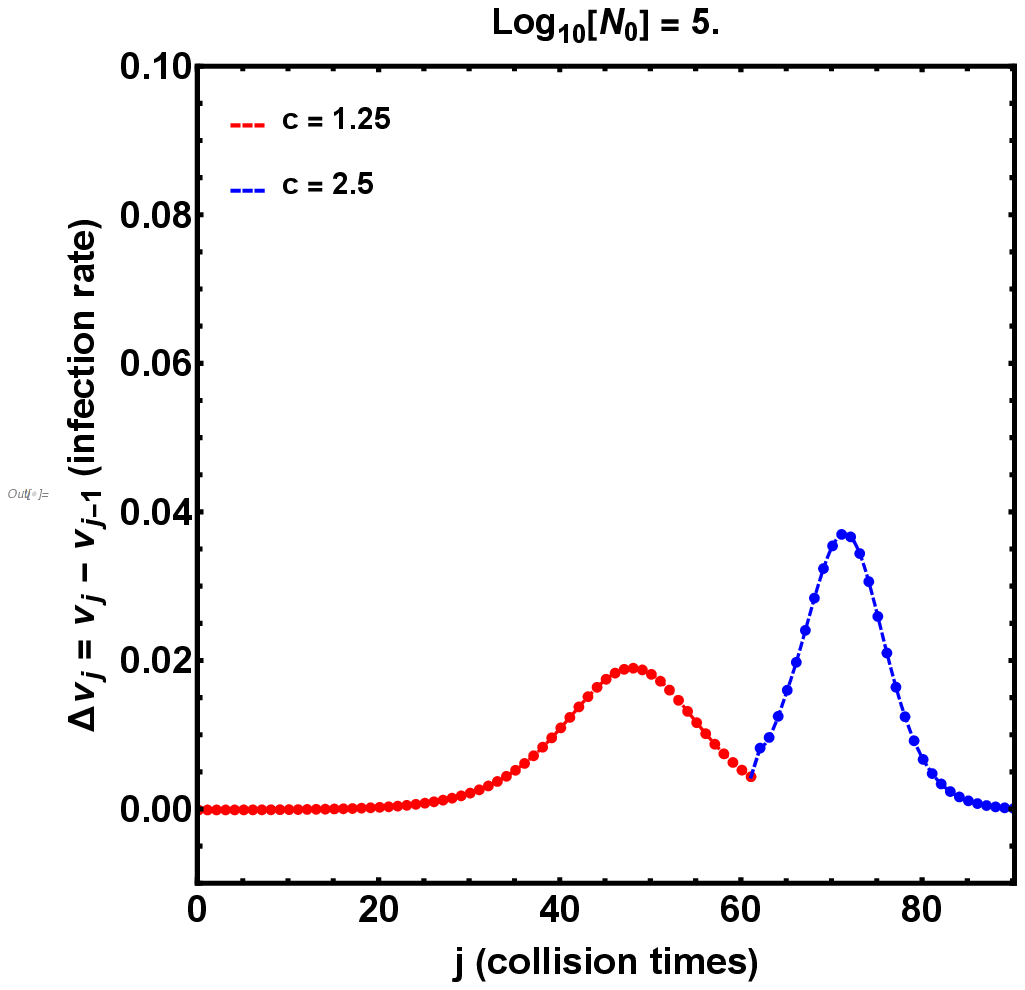}}}
 \caption{The fraction of infected molecules (left), and the infection rate
(right) when social distancing is relaxed suddenly from
$c=1.25$ to $c=2.5$ after 62 collisions when the
infection rate was down to 25$\%$ of its peak value.}
\end{figure}
\noindent
case, in which the
social distancing changes after 62 collisions when the fraction of infections is leveling off near
${\nu}=0.35$ and the infection rate is down to about $25{\%}$ of its peak value.  In this case, there
is a surge in the infection rate, reaching approximately 0.04, which is about 2.0 times the peak height
prior to the change.  Also,the fraction of infected molecules changed
from 0.35 to 0.80, corresponding to infection of 45,000 extra blue molecules in this population
of $N_{0}=1~{\times}~10^{5}$.  Of course, this increase in the number of infections is
proportional to the size of the population and would be higher (or lower) if the population
were higher (or lower).
\medskip

Figure~22 shows the second case, in which $c$ also changes from $c=1.25$ to $c=2.50$, but
after 48 collisions when the epidemic was near its peak.  In this case, there is a much larger
surge in the infection rate, reaching approximately 0.11, which is 5.5 times the peak height
prior to reducing the amount of social distancing.  Also, the fraction of
infected molecules reaches about $0.89$, corresponding to 54,000 more infections than
would have occurred if the social distancing had been maintained at the value of $c=1.25$.
Thus, 9,000 more infections occurred when the change was at $j=48$ near
the peak of the infection rate than when the change was at $j=62$ toward the end of the
epidemic.  As before, this enhancement was for a population of $N_{0}=1~{\times}~10^{5}$
molecules, and would be proportionally more (or less) for a larger (or smaller) population.
\begin{figure}[h!]
 \centerline{%
 \fbox{\includegraphics[bb=90 250 547 725,clip,,width=0.47\textwidth]{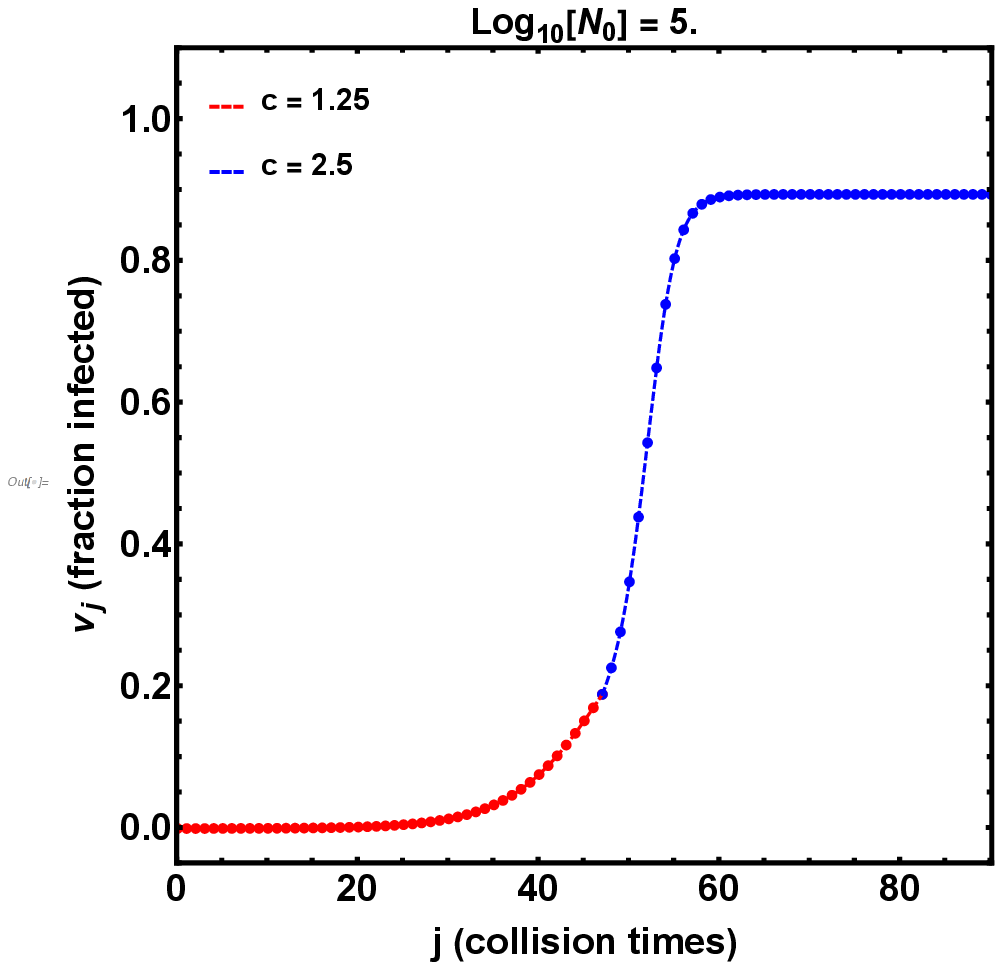}}
\hspace{0.01in}
 \fbox{\includegraphics[bb=90 245 557 721,clip,width=0.48\textwidth]{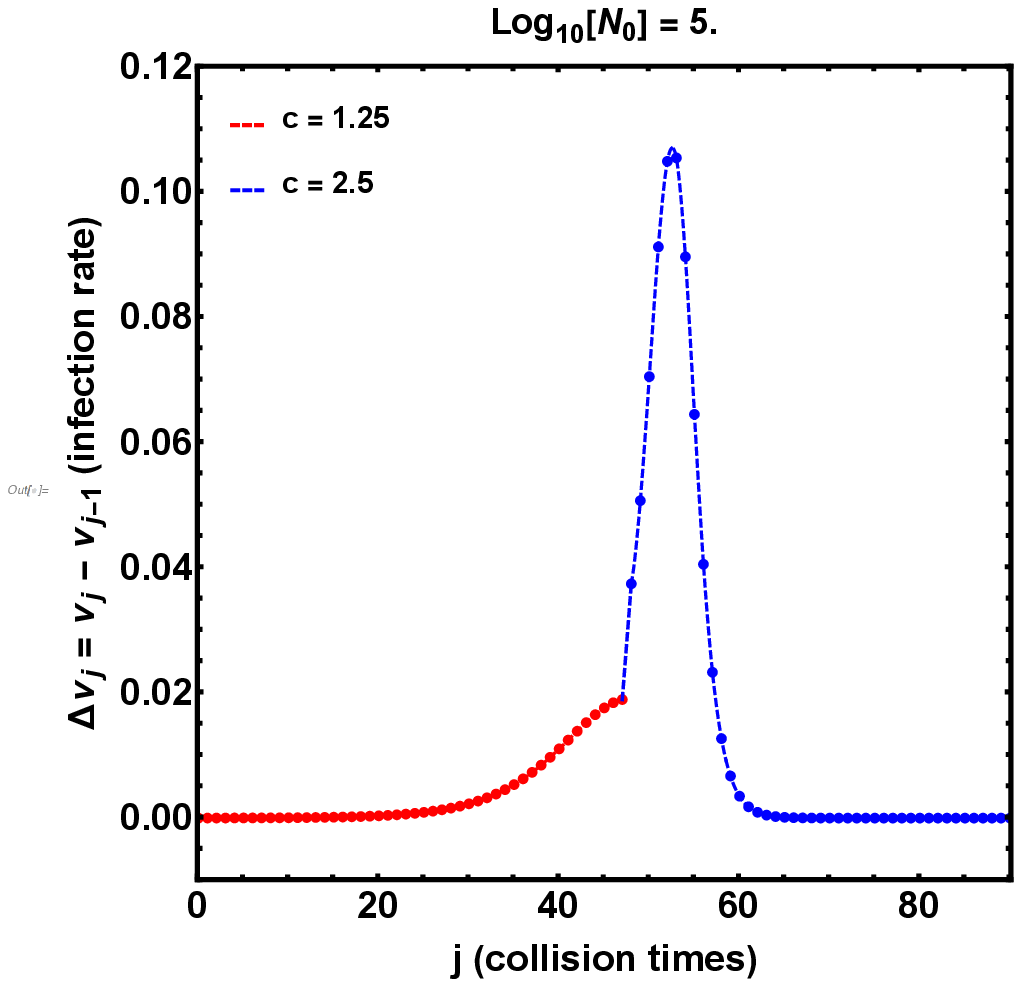}}}
 \caption{The same as Figure~21 except that the social distancing was relaxed after 48 collisions,
when the infection rate was near its peak value.}
\end{figure}
\medskip

Making the change from $c=1.25$ to $c=2.5$ near the peak of the growth rate caused the
final fraction of infected molecules to be ${\nu}_{f}=0.89$.  This value was larger than the
value of ${\nu}_{f}=0.80$ that was obtained in the previous example when the growth rate had
fallen to $25{\%}$
of its peak value.  However, ${\nu}_{f}=0.89$ is still less than the final value of ${\nu}_{f}=0.93$
that would be obtained if the epidemic proceeded from the beginning with $c=2.5$.  Equivalently,
${\nu}_{f}=0.93$ would have been obtained if the transition occurred for $j~{\leq}~35$
toward the start of the epidemic with $c=1.25$.  In fact, the more that the transition is delayed,
the smaller the final fraction, ${\nu}_{f}$, becomes until the limit of ${\nu}_{f}=0.77$ is reached
for $j~{\approx}~75$.  (Nevertheless, ${\nu}_{f}=0.77$ is still larger than the value of
${\nu}_{f}=0.37$ that would have been obtained if the change had not been made and the virus
had progressed to completion with the original value of $c=1.25$.)  Likewise, the maximum growth
rate decreases from ${\Delta}{\nu}_{j}=0.16$ to ${\Delta}{\nu}_{j}=0.027$ when the transition time
moves from $j=35$ to $j=75$.

\subsection{Sudden Decrease of ${\mathbf{c}}$}
Now, let us see what happens when the value of $c$ is decreased suddenly from $c=2.5$ to
$c=1.25$, corresponding to an increase of social distancing.  Figure~23 shows the result when
the change occurs prior to the maximum in the infection rate while $c=2.5$.  Instead of continuing
up to the final value ${\nu}_{f}=0.93$, the fraction of infections levels off at ${\nu}_{f}=0.65$, as shown
in the left panel.  Also, as shown in the right panel, the infection rate abruptly stops its rise and begins
a more rapid decent toward 0.
\medskip

\begin{figure}[h!]
 \centerline{%
 \fbox{\includegraphics[bb=90 250 547 725,clip,,width=0.47\textwidth]{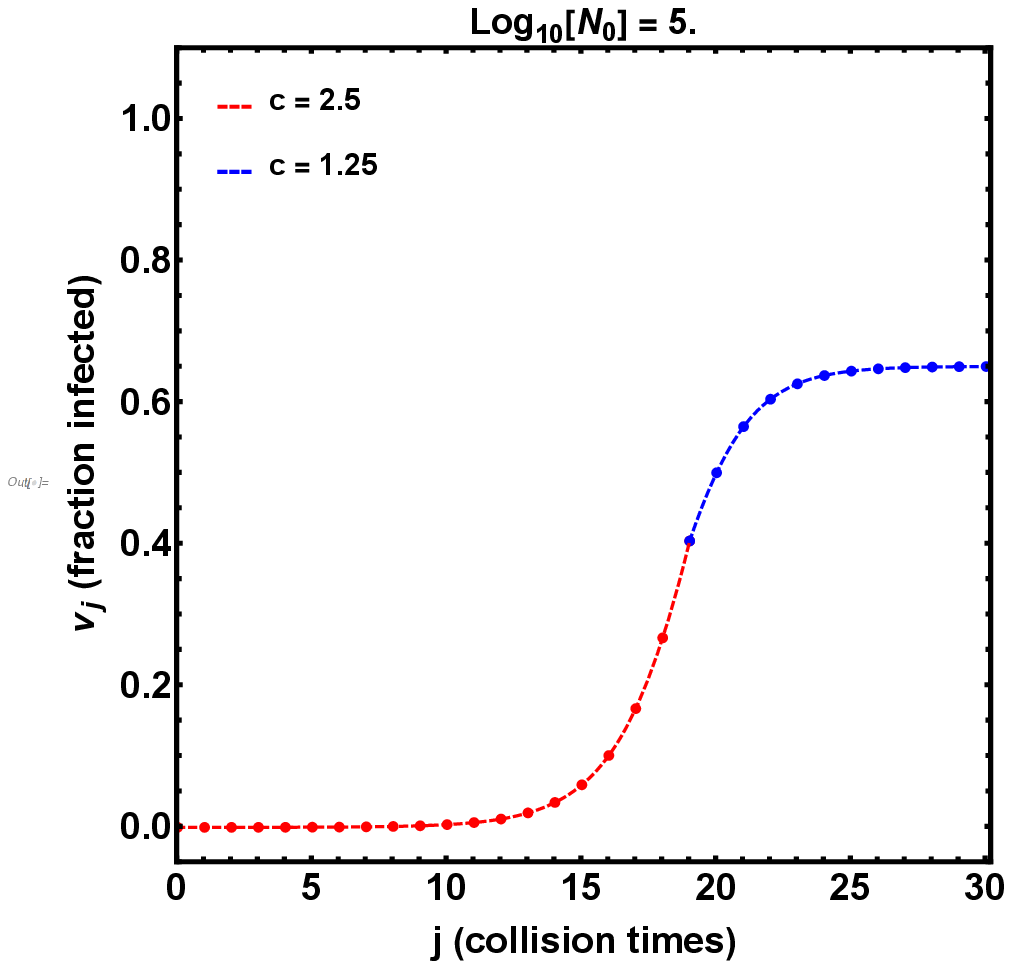}}
\hspace{0.01in}
 \fbox{\includegraphics[bb=90 245 557 721,clip,width=0.48\textwidth]{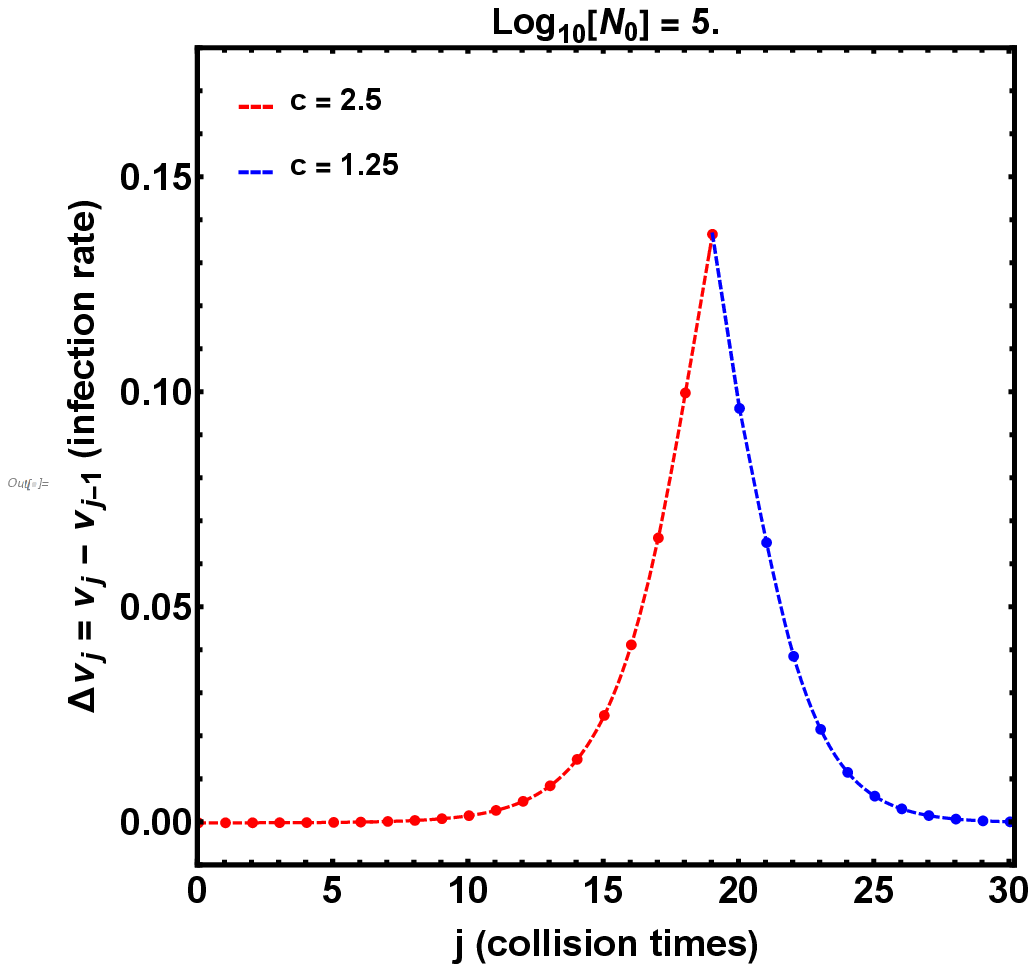}}}
 \caption{Similar to Figure~22 except that the change is done in reverse order with $c=2.5$ changing
to $c=1.25$ at $j=20$ before the infection rate reaches its maximum.}
\end{figure}
Figure~24 shows the effect of making the change at $j=22$, after the peak in the infection rate.  Again,
the fraction of infections levels off quickly, but this time at ${\nu}_{f}=0.80$, which is
\begin{figure}[h!]
 \centerline{%
 \fbox{\includegraphics[bb=90 250 547 725,clip,,width=0.47\textwidth]{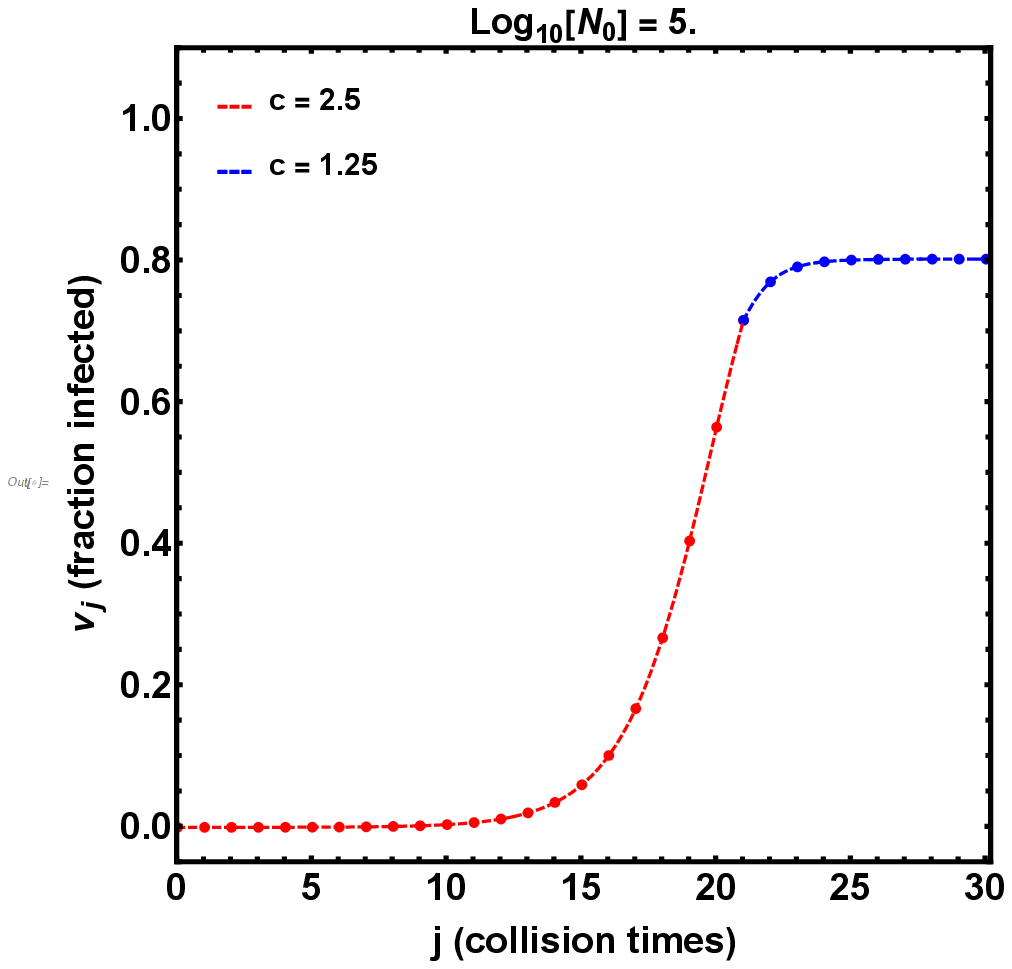}}
\hspace{0.01in}
 \fbox{\includegraphics[bb=90 245 557 721,clip,width=0.48\textwidth]{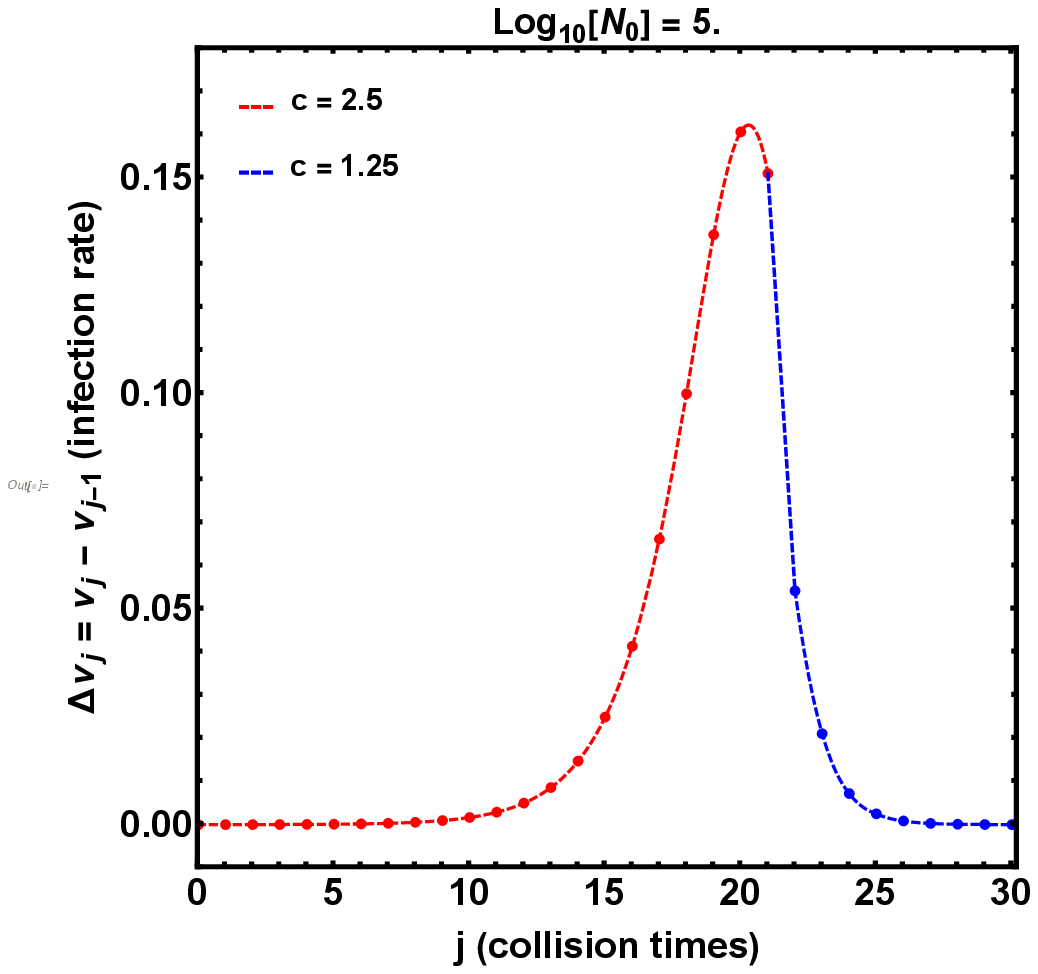}}}
 \caption{Similar to Figure~23 except that the transition occurs at $j=22$, after the maximum of
the infection rate.}
\end{figure}
\noindent
 a larger value than we
obtained by making the change earlier.  Also, the infection rate falls suddenly, but too late to influence
the height of the peak.  So if we increase the amount of social distancing, we will obtain fewer total
infections and a lower peak infection rate if we make the change early before the infection rate reaches its
peak than if we wait until after the peak has occurred.  This is opposite of what we found in the previous
section where the amount of social distancing was suddenly decreased.  In that case, it was necessary to
wait until well after the peak to keep the final number of infections and the infection rate low.  Thus, to
minimize the total number of infections, social distancing should be applied as soon as possible when
entering an epidemic and it should be removed as late as possible when coming out of the epidemic. 

\section{Summary}
This paper describes a mathematical model for the spread of a virus through a population
represented by the colliding molecules of a gas.  In this model, uninfected
molecules are colored blue, contagious molecules are red, and molecules that were infected,
but have lost their ability to infect are colored green.  The epidemic  starts with a single red
molecule entering a gas of $N_{0}-1$ blue molecules.  The red molecule collides with a blue
molecule, which becomes infected and turns red.  Then the two red molecules collide with
two more blue molecules making them red, and the process continues, rapidly increasing the
number of infected molecules in the gas.
\medskip

If infected molecules remained contagious forever, the fraction of infected molecules would
continue to increase exponentially until about half of the molecules are infected.  Then, with
fewer blue molecules left, the infection rate would decrease and the epidemic would end
when all of the remaining blue molecules are infected.  No one has escaped.  However, if
an infected molecule loses its contagiousness after a finite time, ${\tau}_{c}$, then the spread
of the infection is delayed and weakened, and the epidemic ends without infecting all of the
blue molecules.  The spreading rates, times, and magnitudes all depend on the ratio of the
contagious lifetime, ${\tau}_{c}$, to the average time between collisions, ${\tau}_{s}$, the step
rate used in the calculation.  Thus, the key parameter is $c={\tau}_{c}/{\tau}_{s}$.
\medskip

In principle, we could fix ${\tau}_{s}$ and let ${\tau}_{c}$ vary, as if we were studying a variety of
different diseases.  However, I have regarded ${\tau}_{c}$ to be fixed, and imagined that the
collision time, ${\tau}_{s}$, varies with $c$ due to different amounts of social distancing for
a single disease.
\medskip

After setting up the model with permanently contagious molecules, I performed the
numerical computations and graphed the solutions, which included the evolution of the
number of infected molecules as well as the infection rate.  Also, for this case, it was
possible to solve the difference equation for the spreading analytically, exactly reproducing
the ramped transition between the initial and final states of infection as well as the peaked
profile of the infection rate.  For permanently contagious molecules, the final fraction of
infections was ${\nu}_{f}=1$, as expected.  Also, the width of the infection-rate was about
4 units of ${\tau}_{s}$, the peak of the infection-rate profile was $0.25/{\tau}_{s}$, and the
location of the peak was at $j~{\approx}~\log_{2}N_{0}+0.471$, which is 17.08 for a population
of $10^{5}$ molecules.
\medskip

In section 4, I dropped the assumption that the molecules were permanently contagious
and derived the spreading equation for the fraction of infected molecules, ${\nu}_{j}$, for
several values of the contagious lifetime parameter, $c$.  Immediately, the plots showed
that this more general solution reduced to the previous one if $c$ was greater than about 10.
So if the molecules can stay red for at least 10 collision times, they will wipe out the entire
population of blue molecules, as if these red molecules remained contagious forever.
\medskip

However, for smaller values of $c$, and especially values less than 2 and approaching 1, the
results differed quantitatively from the previous solution for $c>10$.  The trends were the same,
showing the same ramped transition from the initial state of uninfected blue molecules to the final
state when the epidemic was over.  However, the difference was that the ramp of rising infection
had a lower slope and a longer duration than when the red molecules had an unlimited lifetime.
Also, the ramp started at a later time than when the red molecules were permanently contagious.
Most important, the decrease in slope was greater than the increase in duration, causing their
product (and the final level of infection) to be less than 1.  At the end of the epidemic, all of the
red molecules were gone.  The population consisted entirely of green molecules that had lost
their contagiousness and blue molecules that had escaped infection.   These changes became
more pronounced as $c$ fell below 2 and approached 1, corresponding to the so-called
`flattening of the curve` that happens with increased social distancing.
\medskip

In the section on red, green, and blue molecules, we learned that the number of green molecules
at a given time was equal to the total number of infected molecules at a time that is $c$ steps
earlier.  This meant that the time history of green molecules is the same as that for all of the
infected molecules, but delayed by $c$ time steps.  Both groups have the same three-phase
evolution with an initially flat distribution, followed by a steep ramp of rapidly rising infections,
which levels off at the final fraction of infected molecules.  The difference between these two
groups gives the fraction of red molecules.  When the shift, $c$, is relatively small, this difference
is just $c$ times the derivative of the profile of infected molecules.  Consequently, for small $c$,
the fraction of red molecules has a profile that is just $c$ times the profile for the growth rate of
all the infected molecules (expressed mathematically, ${\nu}_{j-1}-{\nu}_{j-1-c}=c{\Delta}{\nu}_{j}$).
For $c~{\geq}~3$, this relation breaks down, and the rising phase of the profile for red molecules
matches the ramped profile for all of the infected molecules.  However, unlike the profile for
the total number of infections (red plus green), the profile for red molecules reaches a peak and
then returns to zero, leaving only green and uninfected blue molecules at the end.
\medskip

Finally, in section 7, we examined the effect of changing the amount of social distancing during the
epidemic.  By suddenly removing all of the social distancing (\textit{i.e.} increasing $c$ to ${\infty}$,
or, even to $c=10$), all of the accomplishments disappeared and the fraction of infections rapidly
shot up to the final value of ${\nu}_{f}=1$.  All of the the blue molecules became infected and the
epidemic ended.  For smaller changes either increases or decreases, the result was not quite so
dramatic, but the trend was clear.  To minimize the total number of infections, social distancing
should be applied as soon as possible when entering the epidemic and it should be removed as
late as possible when coming out of the epidemic.

The quantitative results, obtained from section 5 and appendices A-C, are:
\medskip

1. The final fraction of infected molecules, ${\nu}_{f}$ was well fit by the exponential expression
\begin{equation}
{\nu}_{f}~=~1~-~e^{-1.89(c-1)}.
\end{equation}

2.  The `width', $w_{e}$, defined by ${\nu}_{f}/h$ (where $h$ is the maximum value of the infection rate),
could be represented by the relation
\begin{equation}
\frac{1}{w_{e}}~=~0.25[1-e^{-0.806(c-1)}].
\end{equation}
The full-width at half maximum (FWHM) is somewhat less, ${\sim}0.8 w_{e}$, depending on the detailed
shape of the profile.

3. The peak height of the infection rate was given by
\begin{equation}
{\Delta}{\nu}_{max}~=~h~=\frac{{\nu}_{f}}{w_{e}}~=~0.25 \left \{1-e^{-0.806(c-1)} \right \}\left \{ 1~-~e^{-1.89(c-1)}\right \}.
\end{equation}

4. The location, $j_{max}$, of the peak height can be obtained from the empirical relation
${\nu}_{j}~{\approx}~{\nu}_{f}$, where ${\nu}_{j}$ is the dominant term in the solution of
the linearized form of Eq.(4.4) given by
\begin{equation}
{\nu}_{j}~{\approx}~{\sigma}_{c}{\rho}_{c}^{j}~ {\nu}_{0}.
\end{equation}
Here, ${\sigma}_{c}$ and ${\rho}_{c}$ can be obtained from Table 1 of Appendix A or from the
best-fit approximations given in Eqs.(8.7a) and (8.7b) below.  Substituting ${\nu}_{j}={\nu}_{f}$,
and setting $j=j_{max}-1$, we obtain
\begin{equation}
j_{max}-1~=~\frac{ \log{\nu}_{f} } { \log{\rho}_{c} }~-~\frac{\log{\sigma}_{c}}{\log{\rho}_{c}}~+~
\frac{ \log{N_{0}} } { \log{\rho}_{c} }.
\end{equation}
For $c$ in the range (1,2), ${\sigma}_{c}$ and ${\rho}_{c}$ are given by
\begin{subequations}
\begin{align}
{\sigma}_{c}~=~\left (\frac{1}{c-1} \right ) \left ( 1+\frac{c}{s} \right ),\\
{\rho}_{c}~=~\left (\frac{1+s}{2} \right )
\end{align}
\end{subequations}
where $s=\sqrt{1+4(c-1)}$.  In addition, the best-fit approximations can be used over the full
range of $c$:
\begin{subequations}
\begin{align}
{\rho}_{c}~=~2-e^{-0.920(c-1)},\\
{\sigma}_{c}~=~\left (\frac{2}{c-1} \right ) e^{-0.372(c-1)}~+~\left \{ 1-e^{-1.019(c-1) } \right \}.
\end{align}
\end{subequations}
For $c>3$, ${\nu}_{f}~{\approx}~1$, ${\sigma}_{c}~{\approx}~1$, and ${\rho}_{c}~{\approx}~2$,
so that Eq.(8.5) reduces to
\begin{equation}
j_{max}-1~=~\frac{ \log{N_{0}} } { \log{2}}~=~\log_{2}{N_{0}},
\end{equation}
which is the result when the red molecules are permanently contagious.

5. A point-slope approach gave an alternate expression for $j_{max}$.  In this approach,
all of the linear plots of $j_{max}$ versus $\log_{10}{N_{0}}$ were assumed to pass through
the same point, $(\log_{10}{N_{0}},j_{max})=(1.586, 6.153)$, so that the expression for $j_{max}$ became
\begin{equation}
j_{max}~=~6.153~+~\left (\frac{1}{\log_{10}{\rho}_{c}} \right )(\log_{10}{N_{0}}-1.586),
\end{equation}
which reduces to
\begin{equation}
j_{max}-0.884~=~\frac{ \log{N_{0}} } { \log{2}}~=~\log_{2}{N_{0}},
\end{equation}
when $c$ is large and ${\rho}_{c}~{\approx}~2$.

6. The starting position or threshold value, $j_{th}$, of the ramp of rapidly rising infection was defined by
${\nu}_{j_{th}}=0.1{\nu}_{f}$, where ${\nu}_{j_{th}}$ is obtained from Eq.(8.4) with $j=j_{th}$, which is
valid during the initial phase of the variation when ${\nu}_{j}<<1$.  Consequently, $j_{th}$ becomes
\begin{equation}
j_{th}~=~\frac{ \log{\nu}_{f} } { \log{\rho}_{c} }~-~\frac{\log{\sigma}_{c}}{\log{\rho}_{c}}~+~
\frac{ \log{N_{0}} } { \log{\rho}_{c} }~-~\frac{1}{\log_{10}{\rho}_{c}}~=~(j_{max}-1)-\frac{1}{\log_{10}{\rho}_{c}}.
\end{equation}
Thus, $1/\log_{10}{\rho}_{c}$ gives the time interval between $j_{th}$ and $j_{max}-1$, the place
that the curve of accumulated infections has its maximum slope.  (By comparison, $j_{max}$
refers to the location of the maximum growth rate in the plot of ${\Delta}{\nu}_{j}$ versus $j$.)

7. The best-fit relation for the total fraction of infected molecules at the `herd immunity' threshold
was found to be
\begin{equation}
{\nu}_{herd}~=~1~-~e^{-0.860(c-1)},
\end{equation}
and the corresponding expression for the lag between the peak of the total infection rate,
${\Delta}{\nu}_{j}$ and the peak in the fraction of contagious (red) molecules, ${\nu}_{j}-{\nu}_{j-c}$,
was found to be
\begin{equation}
j_{red}~-~j_{max}~=~0.512(c-1)~-~0.033(c-1)^2,
\end{equation} 
until the lag saturated at $c_{m}=8.81$ and $j_{red}-j_{max}=2$.

\clearpage
\section{Discussion}
Now, we will use these calculations to understand the propagation of a
virus.  Remember that they are for an idealized model of a gas of molecules and not
for a realistic population of people in our society.  Also, this model and its calculations
are only a first step in the experiment.  A second step might include more complicated
interactions including multiple collisions, clumps of molecules, and exchanges of
molecules between separate populations.  On the other hand,
this simple model is general and ought to apply to any virus and population for specified
choices of the parameters, $c$ (equivalently, ${\tau}_{c}$), collision time ${\tau}_{s}$,
and $N_{0}$.
\medskip

We found that when the infected molecules retained their contagiousness indefinitely and
there was no `social distancing', the number of infections increased exponentially until
about half of the molecules were infected.  After that, the number of infections increased
more slowly as the remaining molecules of the gas became infected.  The infection rate, which we
represented by ${\Delta}{\nu}_{j}$, was essentially the derivative of the growth curve.  It rose
to its maximum value of $0.25 N_{0}/{\tau}_{s}$ in a time on the order of $\log_{2}N_{0}$,
which is the number of powers of 2 in the number $N_{0}$ (equivalently, $3.322\log_{10}N_{0}$).
\medskip

This means that the time to infect half the population would depend on the size of the population,
and therefore would be longer for a metropolis than for a small community.  In particular, for a small
town of $10^{4}$ molecules, it would take 13.3 collisions to infect half the population, which would be
2-4 weeks if the average time between collisions were 1 or 2 days.  On the other hand, for a large
city of $10^{6}$ molecules, it would take 19.9 collisions and therefore about 3-6 weeks.
\medskip

On the other hand, we found that the shape of the infection-rate curve was independent of $N_{0}$,
so that it had the same width and height for a small town and a large metropolis.
In particular, its width (full width at half maximum) was about 4 collision times, which would be
4-8 days.  Because $0.25 N_{0}/{\tau}_{s}~{\times}~4{\tau}_{s}=N_{0}$, this means that most of
the molecules would be infected during this 4-step interval (of 4-8 days) around the peak of the
distribution.
\medskip

These results are a consequence of using Eq.(3.13a) to calculate ${\nu}_{j}$, the number of
infected molecules expressed as a fraction of the population, $N_{0}$.  In this case, the
final approach from ${\nu}=0.5$ to ${\nu}=1$ was slightly shorter than the initial rise from
${\nu}={\nu}_{0}$ to ${\nu}=0.5$.
Equivalently, the infection-rate curve was asymmetric around its peak with a steeper
fall than rise, which means that the epidemic ends faster than it starts.  
\medskip

Next, I discarded the assumption that the infected molecules remain contagious indefinitely,
and solved the problem for a range of contagious lifetimes, ${\tau}_{c}$, where
${\tau}_{c}$ is $c$ times the collision time (or step time), ${\tau}_{s}$.  The result was a
range of solutions that depend on the parameter, $c$.  One way to think of this is to
suppose that ${\tau}_{s}$ is a constant and ${\tau}_{c}$ varies,
as if we were considering a variety of different diseases.  Another approach is to suppose
that ${\tau}_{c}$ is constant, as it might be for a specific disease, and to consider what happens for
a range of step times, ${\tau}_{s}$ that would occur for different amounts of `social distancing'.
So, for the current pandemic, we can think of the $c$-dependence as an
indication of the influence of `social distancing' on the spread of the disease with
$c=1$ corresponding to perfect distancing and $c={\infty}$ corresponding to the normal
society with no social distancing. 
\medskip

The first result of these new calculations was to find that the plots of the number of infected
molecules
(expressed as a fraction of the total population, $N_{0}$) were similar to the original plot with
no social distancing, showing a relatively long base level followed by a steep ramp
to the final level where the infection stops.  However, for increased amounts of
social distancing, the curves took longer to reach the steep ramp, the slope of the ramp
was lower, and the final level of infected molecules was smaller than in the absence of
social distancing.  Likewise the growth-rate profiles had smaller
heights, larger widths, and were shifted to later times as the amount of social distancing
increased.  This corresponds to the popular term, `flattening the curve'.
\medskip

But how much flattening and delaying occurred?  Let us assume that the contagious time is
${\tau}_{c}=14$ days based on the time that an infected person is required to be quarantined for
the current corona virus.   As shown in
Figure~7, for $c=10$ (corresponding to ${\tau}_{s}={\tau}_{c}/c~{\approx}~14/10 = 1.4$ days
between collisions), the curve peaks at $j_{max}=17$, corresponding
to $1.4~{\times}~17=23.8$ days after the first contagious molecule entered the population of
${\sim} 10^5$ molecules.  For $c=1.25$ (corresponding to $14/1.25=11.2$ days between collisions),
the curve peaks at $j_{max}=48$, corresponding to $11.2~{\times}~48=537.6$ days, or approximately
18 months.  Referring to Figure~6, we found that the fraction of infected molecules dropped from
1.0 to 0.38 (100,000 dropped to 38,000).  So this social distancing saved 62,000 blue
molecules, but required more than 1.5 years to do it.  And this refers to the peak height, not the
final infection rate, which reaches 0 around $j_{max}=70$, corresponding to
$11.2~{\times}~70=784$ days, which is about 26 months.  So the cost of those 62,000
blue molecules was more than 2 years of keeping apart and avoiding collisions for intervals
of about 11 days.
\medskip

The times are shorter for smaller populations and longer for larger populations,
as we found in Eq. (5.10) and Figure~12.  For a small town of population
$N_{0}=10^{4}$, the peaks occur after times of $1.4~{\times}~13.6=19$ days for $c=10$
and $11.2~{\times}~35.5=397.6$ days (about 13 months) for $c=1.25$.  For a metropolis with
$N_{0}=10^{6}$, the corresponding times are $1.4~{\times}~20.2=28.3$ days and 
$11.2~{\times}~60.2=674$ days (about 22 months).  And these are only the times
to reach the peak infection rates.  The time required to bring those rates back to 0
is even larger.  The number of blue molecules saved by social distancing is proportional
to the populations, so for the smaller population of $10^{4}$, the number saved is 6,200,
and for the metropolis of $10^{6}$, the number saved is 620,000.
\medskip

As we have seen in Figure~6, the growth of the numbers of infected molecules seems to
go through three phases.  The evolution begins with a long interval of low values, eventually
changing to an upward ramp that ends in a plateau of final values.  The width of the
infection-rate profile is a rough measure of the duration of this upward ramp, and
therefore an indication of how long the infection-rate remains at high values.  As shown in
Figure~7 and plotted in Figure~10, the width, $w_{e}={\Delta}j$, decreases as $c$ increases.
Therefore, the width increases as the social distancing increases.  In fact, from Eq. (5.8a),
one can show that the width, $w_{e}$, is given approximately by
$w_{e}=5/(c-1)$ for a large amount of social distancing with $c~{\lesssim}~1.5$.
\medskip

Thus, for the example with $c=1.25$ (${\tau}_{s}~{\approx}~11.2$ days), the width, $w_{e}$, is
$20~{\times}~11.2=224$ days, or about 7 months.  Whereas, for our previous example
with $c=10$ (${\tau}_{s}=1.4$ days), the width was approximately $4~{\times}~1.4=5.6$ days.
So when social distancing reduces the fraction of infected molecules
from 1 to 0.38, it increases the duration of the active phase of the disease from about 1 week to
7 months.  By social distancing, we trade a short duration of infection at a high rate
for a long duration of infection at a low rate.  This is the so-called `flattening of the curve'.
However, as Eq.(5.9a) shows, the benefit is to substantially reduce the final number of
infected molecules.
\medskip

It is also interesting to understand this process in terms of the numbers of contagious
molecules (expressed as a fraction of the total population, $N_{0}$).  For example, if
a particular molecule managed to escape the virus and remain blue for a while, what would
its chances of encountering a contagious molecule be at that time?  Recall that the number
of contagious molecules at a given step time, $j$, is equal to the number of infected
molecules at that time minus the number of infected molecules $c$ steps earlier at the time
$j-c$.  This difference depends on the size of  the shift, $c$, relative to the width, $w$, of the
infection-rate profile.  For social distancing with $c~{\leq}~2$, the peak number of
contagious molecules is approximately $c$ times the maximum slope of the growth curve
(which is the peak value of the infection-rate curve).  So the resulting number of contagious
molecules in the top panel of Figure~17 is just $c$ times the corresponding peak heights in
Figure~7.  Because $c>1$ for an uncontained virus, this means that the peak heights for
the numbers of contagious molecules are always greater than the corresponding peak heights
for the infection-rate profiles.  On the other hand, for a nominal amount of social distancing,
the widths and locations of the peaks are the same for the contagious (red) molecules as they are
for the infected (red plus green) molecules.  Based on the numbers that we found in
the examples above, this means that the surviving blue molecules must wait a few years
before re-entering the pool if they wish to avoid collisions with contagious red molecules.

It is important to remember that this molecular model and its numerical calculations began as
an attempt to determine the mathematical properties of the growth curve and growth-rate
curve.  The model was not intended to be a realistic description of the
spread of the Covid-19 disease.  Even the choice of 14 days for ${\tau}_{c}$
was done arbitrarily to permit the calculation of some specific delay times
that might occur in the evolution of the disease.  The resulting delays of months
and years might be different if another value of ${\tau}_{c}$ were used
in the calculation.  Nevertheless, the reader should be pleased to gain some insight into
how the process works, and not be too alarmed or complacent about the resulting numbers.
Remember, these numbers refer to molecules, not people.

Like many people, I was interested in how a virus might spread.  Rather than looking for the
answer in the published literature, I created the molecular model and did the calculations as if I were
solving a puzzle in math or physics.   Only after finishing these calculations, did I perform a Google
search and find the three-component model of Kermack and McKendrick (1927) and a summary of
that model by Weisstein (2004).  Of course, this was `the tip of the iceberg', which led to
more recent references, including those of Anderson and May (1979),
Jones and Sleeman (1983), Smith and Moore (1996), and others.  The three components of
the `SIR-model' referred to individuals who were
susceptible (S), infected and contagious (I), and recovered and not contagious (R) - the same
components that I have called blue (B), red (R), and green (G), respectively.   Therefore,
we should probably call our molecular model the $RGB$ model, after the
$RGB$ color coding used in electronic imaging and photography.  Kermack and
McKendrick described these populations with three differential equations, of which only two
are independent because the total number of molecules is conserved.  In our $RGB$ terminology,
$R+B+G=N_{0}$, and the $SIR$ equations are: 
\begin{subequations}
\begin{align}
\frac{d(R+G)}{dt}~=~+{\beta}BR,\\
\frac{dR}{dt}~=~{\beta}BR-{\gamma}R,\\
\frac{dG}{dt}~=~{\gamma}R.
\end{align}
\end{subequations}
Here, $1/{\beta}$ corresponds to the collision time, ${\tau}_{s}$, and $1/{\gamma}$ is the
$(1/e)$-lifetime of the red molecules in the absence of a source, as one
can see by setting $B=0$ in Eq.(9.1b).  (Equivalently,  $1/{\gamma}$ is the average lifetime of
the red molecules.)  In the RGB-model, the infected
molecules remain contagious for a finite time ${\tau}_{c}=c{\tau}_{s}$ before suddenly losing their
ability to infect.  Thus, $c$ is equivalent to the average lifetime,  $1/{\gamma}$, in units of the
collision time, $1/{\beta}$.
\medskip

As Weisstein pointed out, the key value determining the propagation or damping of
these equations is the `epidemiological parameter', $R_{0} = ({\beta}/{\gamma})B$, where $B$
stands for the fraction of uninfected blue molecules, and ${\beta}/{\gamma}$ is analogous to
the quantity that we call $c$ in the $RGB$-model.  Thus, in Eq.(9.1b), $dR/dt$ changes sign
from positive to negative when $({\beta}/{\gamma})B$ falls below 1, and the number of red molecules
starts to decrease.  Setting $B=1$, we obtain $R_{0}={\beta}/{\gamma}$, which is the fraction of
molecules that a single red molecule at the start of the epidemic will infect before it turns green.
However, in the $RGB$-model, that red molecule would be responsible for $2^{c}-1$ infections
because each infected molecule goes on to infect other blue molecules before the original red
molecule suddenly turns green.  With this distinction, $c$ is analogous to $R_{0}$, and the
difference equations of the $RGB$-model are analogous to the differential equations of the
$SIR$-model, as will be described in detail in Appendix D.

\begin{appendix}

\section{Solution For $\mathbf{{\nu}_{j}<<1}$}
We have seen that when $c~{\geq}~10$, the fraction of infected molecules, ${\nu}_{j}$, can be obtained from
Eq. (3.7), which has an exact solution given by ${\nu}_{j}=1-2^{-2^{\{j-(j_{max}-1)\}}}$, where $j_{max}$ is where
the growth rate ${\Delta}{\nu}_{j}={\nu}_{j}-{\nu}_{j-1}=1/4$ and ${\nu}_{j-1}=1/2$.  This is the solution when
the infected molecules remain contagious for the duration of the epidemic, and all of the blue molecules
eventually become red.  Also, recall that when the non-linear term in Eq. (3.7) is neglected, the
linearized equation becomes ${\nu}_{j}=2{\nu}_{j-1}$ and its solution is ${\nu}_{j}=2^{j}{\nu}_{0}$.  This
raises the question of what the initial-phase of the solution is for other values of $c$.
\medskip

To find out, we linearize Eq. (4.3) by replacing the factor $1-{\nu}_{j-1}$ by 1 and writing
\begin{equation}
{\nu}_{j}~=~2{\nu}_{j-1}~-~{\nu}_{j-1-c}.
\end{equation}
This linear equation has an exact solution composed of terms of the form, $r^{j}$, where $r$ refers to the
roots of the equation, $r^{c+1}-2r^{c}+1=0$, obtained by substituting $r^j$ into Eq. (A1).  When $c=2$,
this equation is $r^3-2r^2+1=0$, whose three roots are $r=1, (1~{\pm}~{\sqrt{5}})/2$.  Consequently,
when $c=2$, the exact solution to Eq. (A1) is
\begin{equation}
\frac{{\nu}_{j}}{{\nu}_{0}}~=~
-1~+\left (1+\frac{2}{{\sqrt{5}}} \right ) \left \{ \frac{1+{\sqrt{5}}}{2} \right \}^{j}
~+~\left (1-\frac{2}{{\sqrt{5}}} \right ) \left \{ \frac{1-{\sqrt{5}}}{2} \right \}^{j},
\end{equation}
where the coefficients, $-1, 1~{\pm}~2/{\sqrt{5}}$ were determined by setting
${\nu}_{j}/{\nu}_{0}~{\approx}~1, 2, 4$ for $j=0,1,2$ and solving the three equations simultaneously.
The approximate numerical form of Eq. (A2) is
\begin{equation}
\frac{{\nu}_{j}}{{\nu}_{0}}~=~-1~+1.894 (1.618)^{j}~-~0.106 (-1)^{j} (0.618)^{j},
\end{equation}
which is quickly dominated by $1.894 (1.618)^{j}$ when $j$ becomes larger than about 6.  This is to be
compared with $1.355(1.839)^{j}$ when $c=3$ and $2^{j}$ when $c$ is greater than 10.  Also, as we shall
see in the next few paragraphs, the dominant terms are $3.732(1.366)^{j}$ for $c=1.5$ and
$7.536(1.207)^{j}$ for $c=1.25$.  These results are given in Table 1 for an expression
of the form
\begin{equation}
\frac{ {\nu}_{j} } { {\nu}_{0} }~=~{\sigma}_{c}~{{\rho}_{c}}^{j},
\end{equation}

\begin{table}[h!]
\caption{${\sigma}_{c}$~and~${\rho}_{c}$}
\begin{tabular}{c c c c}
\hline\hline
$c$ &  ${\sigma}_{c}$ &  ${\rho}_{c}$  \\[1.5ex]
\hline\\
1.25	&	7.536	&	1.207	 \\[1.5ex]
1.50	&	3.732	&	1.366	 \\[1.5ex]
1.75	&	2.500	&	1.500	 \\[1.5ex]
2.00	&	1.894	&	1.618	 \\[1.5ex]
2.50      &    1.694    &    1.740      \\[1.5ex]
3.00	&	1.355	&	1.839     \\[1.5ex]
4.00	&	1.177	&	1.928     \\[1.5ex]
5.00	&	1.095	&	1.966	 \\[1.5ex]
${\infty}$	&	1.000	&	2.000     \\[1.5ex]
\hline 
\end{tabular}
\end{table}

\noindent
where ${\sigma}_{c}$ is the coefficient of the dominant $r^{j}$-term, and ${\rho}_{c}$
is the corresponding value of $r$.  As we would expect, this dominant
term approaches $2^{j}$ as $c$ becomes large.  And as we will show next, it approaches
$\{2/(c-1)\}c^{j}$ as $c~{\rightarrow}~1$.
\medskip

We are especially interested in the solution when $c$ lies between 1 and 2.
Recall that we used a linear interpolation for ${\nu}_{j-1-c}$ in section 4.  For $c$ in the range (1,2),
this linear interpolation is ${\nu}_{j-1-c}=(2-c){\nu}_{j-2}+(c-1){\nu}_{j-3}$.  Consequently, Eq. (A1)
becomes
\begin{equation}
{\nu}_{j}~-~2{\nu}_{j-1}~+~(2-c){\nu}_{j-2}~+~(c-1){\nu}_{j-3}~=~0,
\end{equation} 
whose associated equation for $r$ is
\begin{equation}
r^{3}~-~2r^{2}~+~(2-c)r~+~(c-1)~=~0.
\end{equation}
The three roots of this equation are $r=1, \{1~{\pm}~{\sqrt{1+4(c-1)}}\}/2$.  As before, we
set ${\nu}_{j}/{\nu}_{0}=$ 1, 2, and 4 for $j=$ 0, 1, and 2, and solve for the coefficients of
each $r^{j}$ term..   Then, with the substitution
$s={\sqrt{1+4(c-1)}}$, we can write the general solution for ${\nu}_{j}/{\nu}_{0}$ as
\begin{equation}
\frac{{\nu}_{j}}{{\nu}_{0}}~=~\left (\frac{c-3}{c-1} \right )~+~
\left (\frac{1}{c-1} \right ) \left ( 1+\frac{c}{s} \right ) \left \{\frac{1+s}{2} \right \}^{j}~+~
\left (\frac{1}{c-1} \right ) \left ( 1-\frac{c}{s} \right ) \left \{\frac{1-s}{2} \right \}^{j},
\end{equation}
valid for $1<c~{\leq}~2$.  It is reassuring to see that when $c=2$, $s={\sqrt{5}}$, and
Eq. (A7) reduces to Eq. (A2).  At the other end of the range where $c-1<<1$, Eq. (A7)
approaches the relation
\begin{equation}
\frac{{\nu}_{j}}{{\nu}_{0}}~{\approx}~-\left (\frac{2}{c-1} \right )~+~\left (\frac{2}{c-1} \right )c^{j}~+~
(-1)^{j} (c-1)^{j},
\end{equation}
which allows us to see that ${\nu}_{j}/{\nu}_{0}$ is dominated by $\{2/(c-1)\}c^{j}$ when $c$ is
close to 1 and $c^{j}>>1$.  More generally, we can approximate Eq. (A7) by
\begin{equation}
\frac{{\nu}_{j}}{{\nu}_{0}}~{\approx}~\left (\frac{1}{c-1} \right )
 \left ( 1+\frac{c}{s} \right ) \left \{\frac{1+s}{2} \right \}^{j},
\end{equation}
where $s={\sqrt{1+4(c-1)}}$ and $1<c~{\leq}~2$.
\medskip

We can use Eq.(A4) to compare this analytical solution of the linearized Eq.(A1)
with the numerical solution of the non-linear Eq.(4.3).  The results are plotted in Figure~25 for
$c=10, 2, 1.5, 1.25$ and $N_{0}=1~{\times}~10^5$.  (When $c=10$, the linearized equation is
${\nu}_{j}=2{\nu}_{j-1}$, whose solution is ${\nu}_{j}=2^{j}{\nu}_{0}$.)  Recall that the linearized
equations involve
the approximation $1-{\nu}_{j-1}<<1$, which means that we expect the approximate solutions to be valid
only for small ${\nu}_{j}$.  In Figure~25, we see that the solutions for $c=10$ and $c=2$ are
fairly accurate for values of ${\nu}_{j}$ less than about 0.2.  However, for progressively smaller values of
$c$, the dashed and solid curves begin to separate at smaller values of ${\nu}_{j}$, below 0.1 for $c=1.5$
and then below 0.05 for $c=1.25$.  However, the values of ${\nu}_{f}$ also decrease as $c$ becomes
closer to 1, so the discrepancies will be smaller when normalized to the values of ${\nu}_{f}$.
\begin{figure}[h!]
 \centerline{%
 \fbox{\includegraphics[bb=88 269 527 720,clip,width=0.85\textwidth]{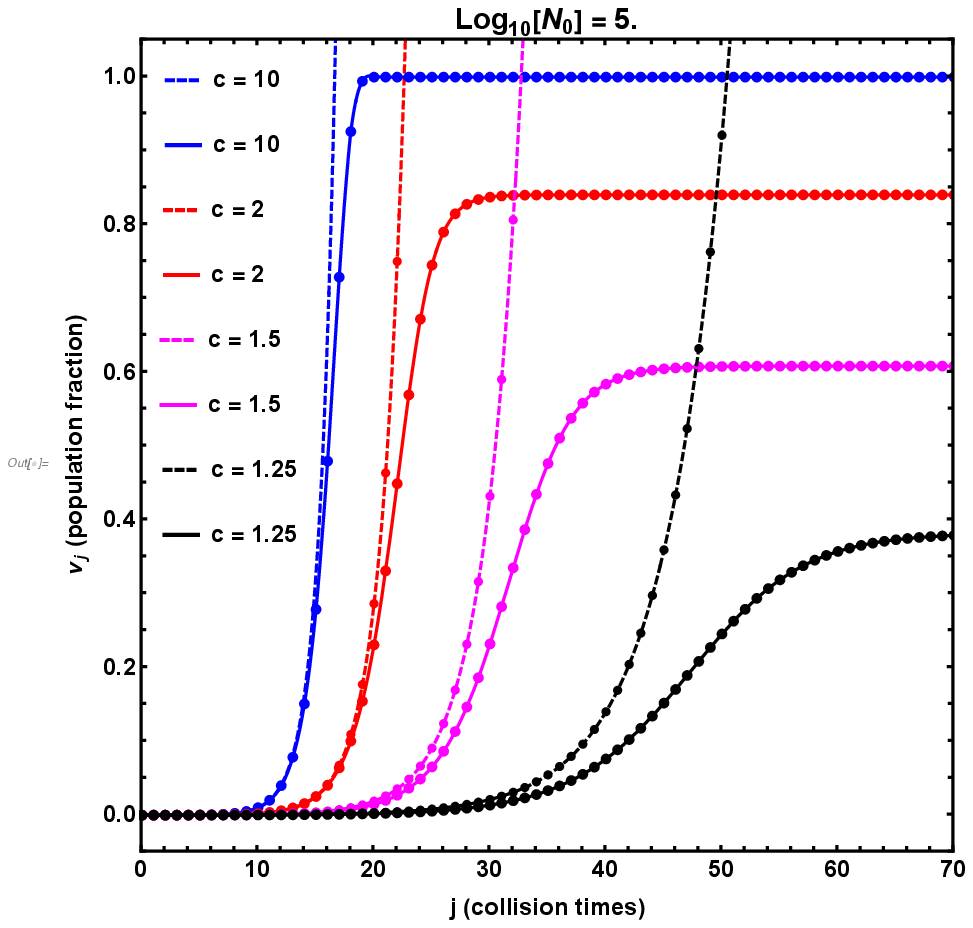}}}
 \caption{Comparing the solutions of the linearized equation (dashed) with the solutions of the
non-linear equation (solid) for the growth of the virus when the values of $c$ are 10, 2, 1.5, and 1.25,
and when $\log_{10}N_{0}=5$.}
\end{figure}
\medskip

\section{Estimating the Starting Point of the Region of Rapid Increase}

Although the solution of the linearized version of Eq. (4.3) diverges from the solution of the complete
non-linear equation, it remains accurate near the initial rising phase of the evolution where
${\nu}_{j}<<1$.  Consequently, we can use the solution given by Eq.(A4) to estimate where the
rapid increase of ${\nu}_{j}$ begins.  We begin by supposing that the starting point can be defined
by a threshold value, arbitrarily taken to be ${\nu}_{j}=0.1{\nu}_{f}$.  The corresponding value
of $j$, say $j_{th}$, is determined by the equation
\begin{equation}
{\sigma}_{c}{\rho}_{c}^{j_{th}}~=~0.1(\frac{{\nu}_{f}}{{\nu}_{0}})~=~0.1~{\nu}_{f}N_{0}.
\end{equation}  
Next, we set $j=j_{max}-1$ in Eq.(5.11) and divide that equation by Eq.(B1).  This division eliminates
${\sigma}_{c}$, ${\nu}_{f}$, and $N_{0}$, and gives
\begin{equation}
{\rho}_{c}^{j_{max}-1-j_{th}}~=~10,
\end{equation}
from which we deduce that
\begin{equation}
(j_{max}-1)-j_{th}~=~\frac{1}{\log_{10}{\rho}_{c}}.
\end{equation}
Thus, we have an expression for the difference, $(j_{max}-1)-j_{th}$, as a function of ${\rho}_{c}$,
which is given by Table 1 of Appendix A and the solution to the equation $r^{c+1}-2r^{c}+1=0$
when $c$ is an integer greater than 1.  (Recall that $j_{max}-1$ and $j_{max}$ refer to different
profiles.  $j_{max}-1$ is the location of the maximum slope of ${\nu}_{j}$, which concerns us here,
whereas $j_{max}$ is the location of the peak of ${\Delta}{\nu}_{j}$.)  
\medskip

Note that our arbitrary threshold of ${\nu}_{th}=0.1{\nu}_{f}$ led to the shift
$(j_{max}-1)-j_{th}=-\log_{10}0.1/\log_{10}{\rho}_{c}=1/\log_{10}{\rho}_{c}$
because the logarithms were taken to base 10.  But we could have used any
base, including ${\rho}_{c}$, in which case, the shift would have been
$-\log_{{\rho}_{c}}0.1/\log_{{\rho}_{c}}{\rho}_{c}=\log_{{\rho}_{c}}10$.  For permanently
contagious red molecules, ${\rho}_{c}=2$ and the shift becomes $\log_{2}10=3.322$.
For convenience, we could have chosen a threshold using a power of 2, like
${\nu}_{th}=2^{-3}{\nu}_{f}$, for example, so that the shift is just $\log_{2}2^{3}=3$ collision times.
Thus, for permanently contagious red molecules, the threshold $j_{th}$ would be just 3 steps back
from the point of greatest slope, which itself is given approximately by
$j_{max}-1~{\sim}~\log_{2}N_{0}$.
\medskip

Also, note that because the shift is $(j_{max}-1)-j_{th}=1/\log_{10}{\rho}_{c}$, the shift will double
from 3 to 6 collision times when ${\rho}_{c}$ is changed from 2 to ${\sqrt2}=1.414$ (corresponding
to $c=1.581$ using Eq. (C1a)).  And it will double again to 12 collision times on going to
${\rho}_{c}=2^{1/4}=1.189$ ($c=1.228$), as one can verify in Figure~6 (but remembering that the
threshold value is not just 0.1 (or 1/8 for the simpler case), but is $0.1{\nu}_{f}$ (or $(1/8){\nu}_{f}$).
Likewise, for the nearly symmetric profiles, the end of the ramp
occurs about the same number of collision times after the point of greatest slope.

\section{Approximating the Linearized-Solution Parameters,
$\mathbf{ {\sigma}_{c} }$ and $\mathbf{ {\rho}_{c} }$}
Because the parameters, ${\sigma}_{c}$ and ${\rho}_{c}$ are important for understanding the initial
behavior of ${\nu}_{j}$, I thought it would be useful to obtain approximate analytical expressions for
those parameters, expressed as a function of $c$.  Consequently, in Figures~26 and 27, I have
plotted the values of ${\rho}_{c}$ and ${\sigma}_{c}$, respectively, obtained from
Table 1 of Appendix A, supplemented by additional values chosen to fill in the range of $c$.
In addition, I have superimposed dashed curves corresponding to the best-fit exponential solutions
given in Eqs. (C1a) and (C1b).
\begin{subequations}
\begin{align}
{\rho}_{c}~=~2-e^{-0.920(c-1)},\\
{\sigma}_{c}~=~\left \{\frac{2}{c-1} \right \} e^{-0.372(c-1)}~+~\left \{ 1-e^{-1.019(c-1) } \right \}.
\end{align}
\end{subequations}

\begin{figure}[ht!]
 \centerline{%
 \fbox{\includegraphics[bb=88 255 540 715,clip,width=0.75\textwidth]{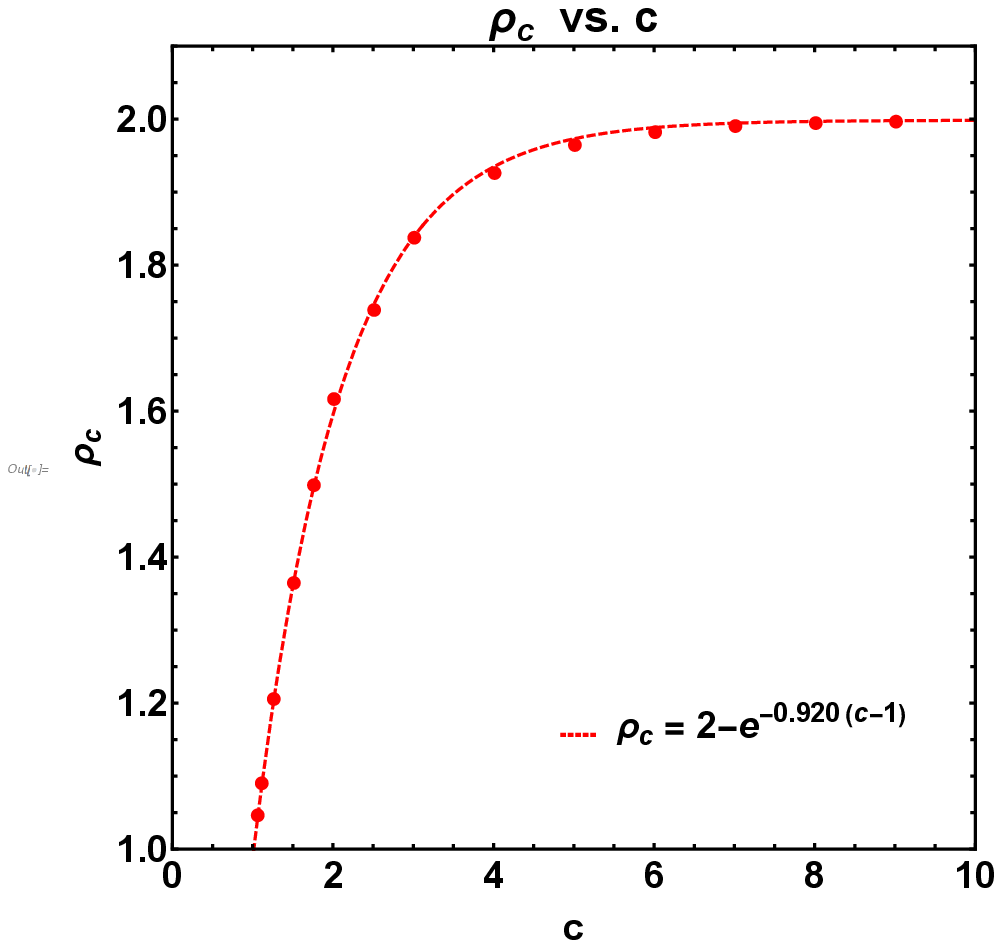}}}
 \caption{Calculated values of ${\rho}_{c}$ (red points), superimposed by the best-fit solution
given by Eq.(C1a) (dashed red curve).  ${\rho}_{c}$ is defined by
${\nu}_{j}={\sigma}_{c} ~{\rho}_{c}^{j}~ {\nu}_{0}$,  the dominant term in the solution to the
linearized equation of spreading.}
\end{figure}
The general form of Eq.(C1a) was chosen so that it would give 1 when $c=1$ and give 2 as
$c~{\rightarrow}~{\infty}$.  That was much easier than choosing Eq.(C1b) so that it would
reduce to $2/(c-1)$ as $c~{\rightarrow}~1$ and approach 1 as $c~{\rightarrow}~{\infty}$.  An additional
exponential factor was necessary to weaken the $2/(c-1)$-term so that ${\sigma}_{c}$ would approach 1
more rapidly than $2/(c-1)$ as $c~{\rightarrow}~{\infty}$. 
\begin{figure}[ht!]
\centerline{%
 \fbox{\includegraphics[bb=88 255 540 715,clip,width=0.75\textwidth]{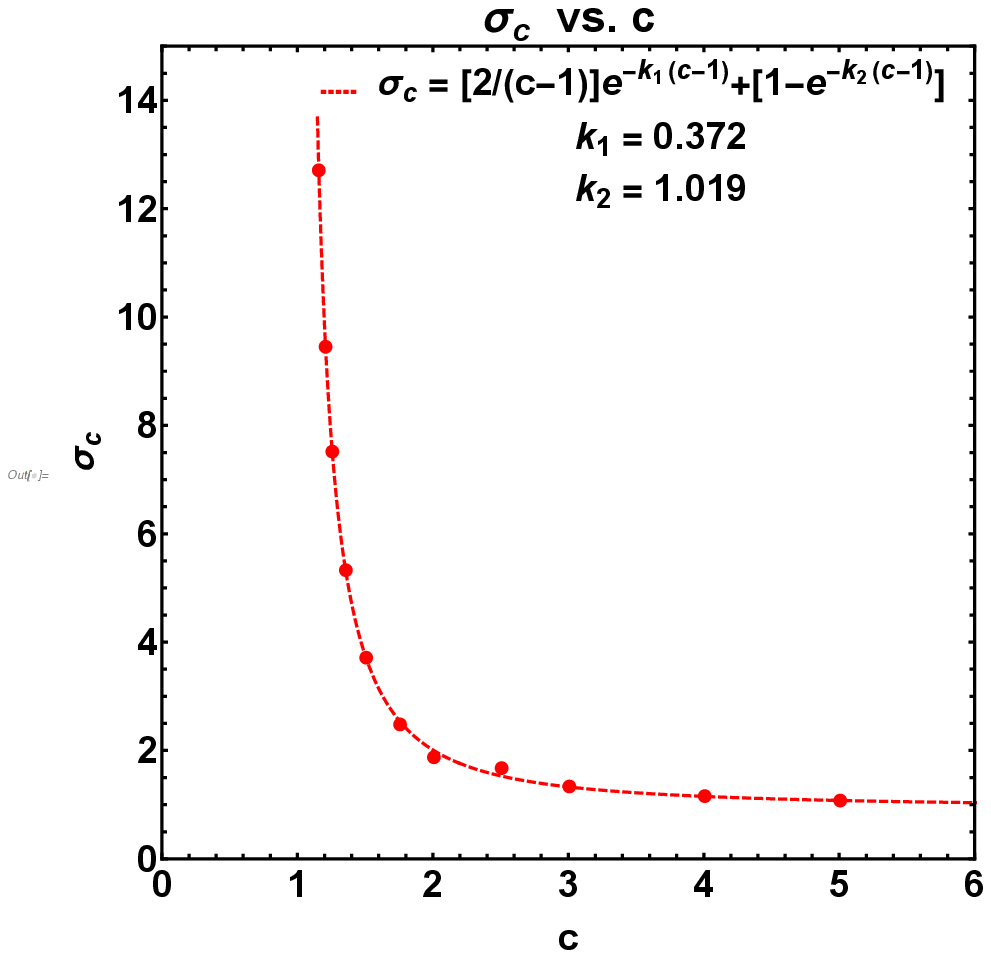}}}
 \caption{Calculated values of ${\sigma}_{c}$ (red points), superimposed by the best-fit solution given
by Eq.(C1b) (dashed red curve).   ${\sigma}_{c}$ is defined by
${\nu}_{j}={\sigma}_{c} ~{\rho}_{c}^{j}~ {\nu}_{0}$, the dominant term in the solution to the linearized
equation of spreading.}
\end{figure}
\medskip

Finally, note that ${\sigma}_{c}$ contributes to the initial behavior of ${\nu}_{j}$ as a shift in the value
of $j$.  We can see this by writing
\begin{equation}
{\nu}_{j}~=~{\sigma}_{c}~{\rho}_{c}^{j}~=~{\rho}_{c}^{j+j_{sh}},
\end{equation}
where $j_{sh}=\log{\sigma}_{c}/\log{\rho}_{c}$ whose logarithms can be taken to any base.  In this case,
Eq. (5.12) of the main text can be rewritten as
\begin{equation}
j_{max}-1~=~\frac{ \log{\nu}_{f} } { \log{\rho}_{c} }~+~\frac{ \log{N_{0}} } { \log{\rho}_{c} }~-~j_{sh},
\end{equation}
Thus, if $c$ were greater than about 2, ${\nu}_{f}$ and ${\sigma}_{c}$ would be approximately 1,
${\rho}_{c}$ would be approximately 2, and $j_{sh}$ would be negligible.   Eq.(C3) would then
become
\begin{equation}
j_{max}-1~=~\frac{ \log{N_{0}} }{\log2}~=~\log_{2}N_{0},
\end{equation}
which is the result for molecules that remain contagious indefinitely.
\medskip

Figure~28 shows values of $j_{sh}$ plotted as a function of $c$.  For comparison,
the dashed blue curve indicates $j_{sh}$, calculated from Eq.(A9) and
$j_{sh}=\log{\sigma}_{c}/\log{\rho}_{c}$.  Eq.(A9) is the approximate solution of
Eq.(A5), which only applies when $c$ lies in the interval (1,2).  Nevertheless, the
extension of this curve fits the data points for $c>2$ reasonably well.  A horizontal
line of the form $j_{sh}=0$ would also be a good fit for $c>3$.  Consequently,
we could also fit the data points using Eq.(A9) inside (1,2) and $j_{sh}=0$ for $c>3$
outside.  
\begin{figure}[ht!]
\centerline{%
 \fbox{\includegraphics[bb=88 255 538 715,clip,width=0.75\textwidth]{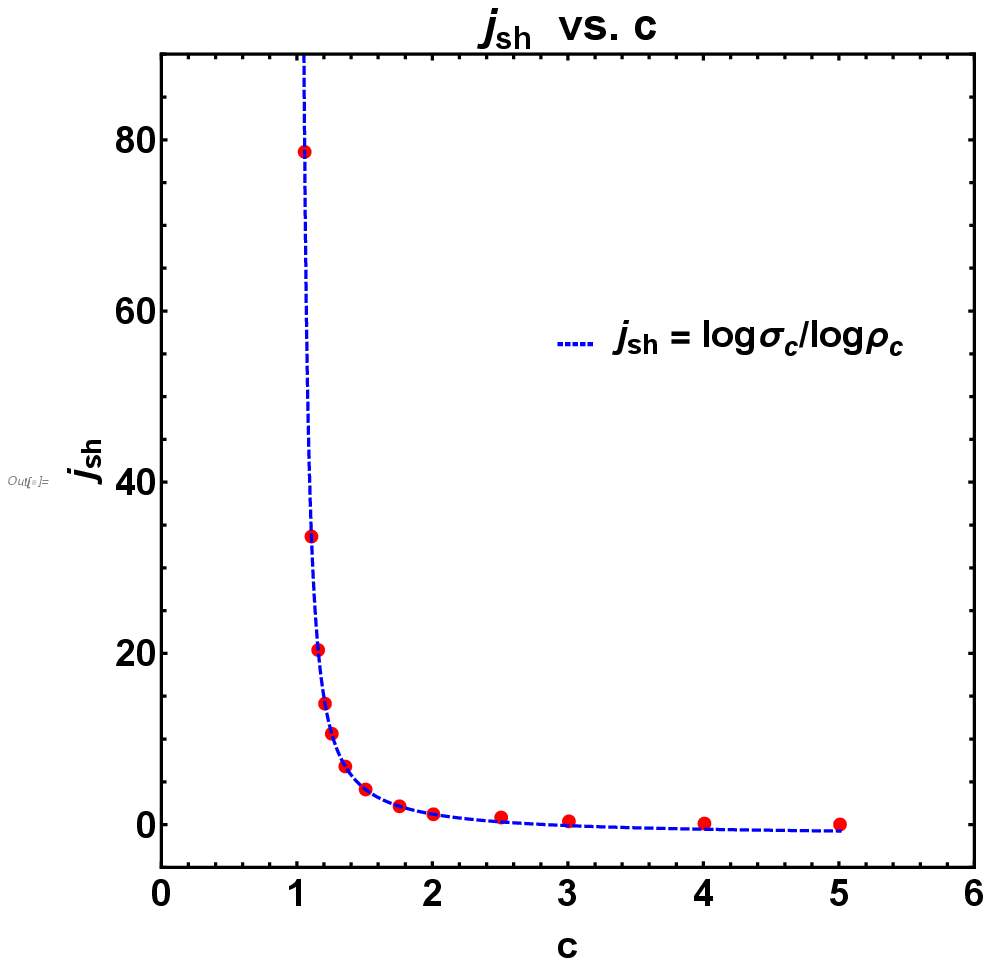}}}
 \caption{Calculated values of $j_{sh}$ (red points) superimposed by a plot of Eq.(A9),
the approximate solution of Eq.(A5) (dashed blue line).  Although Eq.(A5) only applies
for $c$ in the interval (1,2), its extension fits the nearly constant values of $j_{sh}=0$
for $c>3$ very well.}
\end{figure}

\section{Comparison with the Solutions of the SIR Equations}
As I mentioned in the discussion section of this article, I constructed the RGB-model and
solved its equations without any reference to the published literature.  I just wanted
to see if I could figure it out on my own, as if one of my scientific friends had presented
it to me as a mathematics puzzle.  It was only afterward, when I started looking through
the published literature that I found the well-known SIR-model of
Kermack and McKendrick (1927).
\medskip

Both models describe the spread of a virus through
a large population of initially uninfected individuals, and track the population of the same
three types of individuals, which I have called red (infected and contagious), blue 
(uninfected), and green (infected, but no longer contagious).  However, the models take
different approaches.  Whereas the RGB-model is based on individual collisions and discrete
equations, the SIR-model is continuous and is controlled by differential equations.  The contrast
is like comparing the discrete spreading by a random walk with the continuous spreading by
an effective diffusion, reminiscent of the two ways of simulating the transport of magnetic
flux on the Sun's surface (Leighton, 1964; Wang and Sheeley, 1994).
\medskip

Although both models suppose that the infected individuals do not remain contagious indefinitely,
the RGB-model turns off the contagiousness abruptly after a specific time, ${\tau}_{c}$, whereas
the SIR-model allows the contagiousness to decay exponentially at a rate ${\gamma}$.  The models
have two other relevant parameters, ${\tau}_{s}$, the step rate or average time between collisions
of the molecules in the RGB-model, and ${\beta}$, the rate at which the contagious red molecules
and uninfected blue molecules come together.  In each case, the important quantity is the ratio of
these respective parameters -- ${\tau}_{c}/{\tau}_{s}$ for the RGB-model and ${\beta}/{\gamma}$
for the SIR-model.  I have used $c$ for ${\tau}_{c}/{\tau}_{s}$, and in this section, I will also use
$c$ for the ratio, ${\beta}/{\gamma}$.
\medskip

We begin with the SIR equations Eqs.(9.1a), (9.1b), and (9.1c) in the discussion section of this
paper.  They are subject to the initial conditions: $R(0)={\epsilon}=1/N_{0}$, $G(0)=0$, and
$B(0)=1-{\epsilon}$.  This corresponds to our starting condition of one red molecule and
$N_{0}-1$ blue molecules in the RGB-model.  In each case, $R+G+B=1$ in our normalized units.
\medskip

Our first step will be to relate the quantities, R, R+G (which corresponds to ${\nu}_{j}$ in the
RGB-model), and B, to the variable, G.  We do this by
dividing Eq.(9.1a) by Eq.(9.1c) to obtain
\begin{equation}
\frac{d(R+G)}{dG}~=~\frac{{\beta}}{{\gamma}}\left [1-(R+G) \right ].
\end{equation}
Inserting $c={\beta}/{\gamma}$ and solving this differential equation, we obtain
\begin{equation}
R+G~=~1~-~(1-{\epsilon})e^{-cG},
\end{equation}
where we have chosen the constant of integration to satisfy the initial conditions of the
previous paragraph.  And, of course, once $R+G$ is known, $B$ is also determined because
their sum is 1:
\begin{equation}
B~=~(1-{\epsilon})e^{-cG}.
\end{equation}
It takes a little more algebra to obtain $R$.  We begin by rewriting Eq.(D1) as
\begin{equation}
\frac{dR}{dG}~+~cR~=~(c-1)~-~cG.
\end{equation}
The solution to this differential equation is
\begin{equation}
R~=~1~-~G~-~(1-{\epsilon})e^{-cG}
\end{equation}
where we have again determined the constant of integration so that the resulting equation
satisfies the initial conditions of the first paragraph.  Now, we have $R$, $R+G$, and $B$
expressed in terms of $G$.  Unfortunately, that is the end of our analytical solutions.  Next,
we use numerical integration to evaluate $G$ as a function of time, $t$.
\medskip

We begin by combining Eq.(9.1c) and Eq.(D5) as
\begin{equation}
\frac{dG}{dt}~=~{\gamma}R~=~{\gamma} \left [1~-~G~-~(1-{\epsilon})e^{-cG} \right ],
\end{equation}
from which we can obtain an implicit expression for $G(t)$:
\begin{equation}
t~=~c\int_{0}^{G(t)}\frac{dG}{1-G-(1-{\epsilon})e^{-cG}},
\end{equation}
where again we have adjusted parameters, this time by setting ${\gamma}t=({\gamma}/{\beta})({\beta}t)=
(1/c)(t/{\tau}_{s})$ and expressed $t$ in units of ${\tau}_{s}$.
\medskip

As a practical point, we note that
the presence of ${\epsilon}$ prevents the denominator of the integrand from vanishing at
the lower limit of $G=0$.  It also provides the dependence on the population, $N_{0}=1/{\epsilon}$.
As a second point, we note that the upper limit, $G(t)$, is bounded by a maximum value, which
is 1 when $c={\infty}$, but less than 1 for finite values of $c$.  Because $R=0$ at the end of the
epidemic, it follows that the maximum value of $G$ is also the maximum value of $R+G$, which
we have called ${\nu}_{f}$ in the previous sections.  In this case, the maximum value of $G$ is
determined by setting the denominator of the integrand equal to zero.
\medskip

This provides a method for calculating the value of ${\nu}_{f}$ in this SIR-model.  We just set
$1-G-(1-{\epsilon})e^{-cG}=0$, neglect ${\epsilon}$ compared to 1, and solve for $c$ as a
function of $G$.  The result is
\begin{equation}
c~=~-\frac{ln(1-G_{f})}{G_{f}}~=~-\frac{ln(1-{\nu}_{f})}{{\nu}_{f}},
\end{equation}
where $G_{f}$ is the final value of $G$ and $G+R$, which corresponds to ${\nu}_{j}$ in the RGB-model.
Figure~29 shows this value of ${\nu}_{f}(c)$ plotted as the black, dashed curve with the
corresponding values obtained from our RGB-model, already shown in Figure~8.  The
agreement is fairly good overall, and it is nearly perfect in the region where $c<1.4$.  As we shall see
next, the agreement for small values of $c$ is a general property of the two models.  Equivalently,
the disagreement increases as the exponential decay time for R increases in the SIR-model.  
\begin{figure}[h!]
 \centerline{%
 \fbox{\includegraphics[bb=88 250 540 720,clip,width=0.75\textwidth]{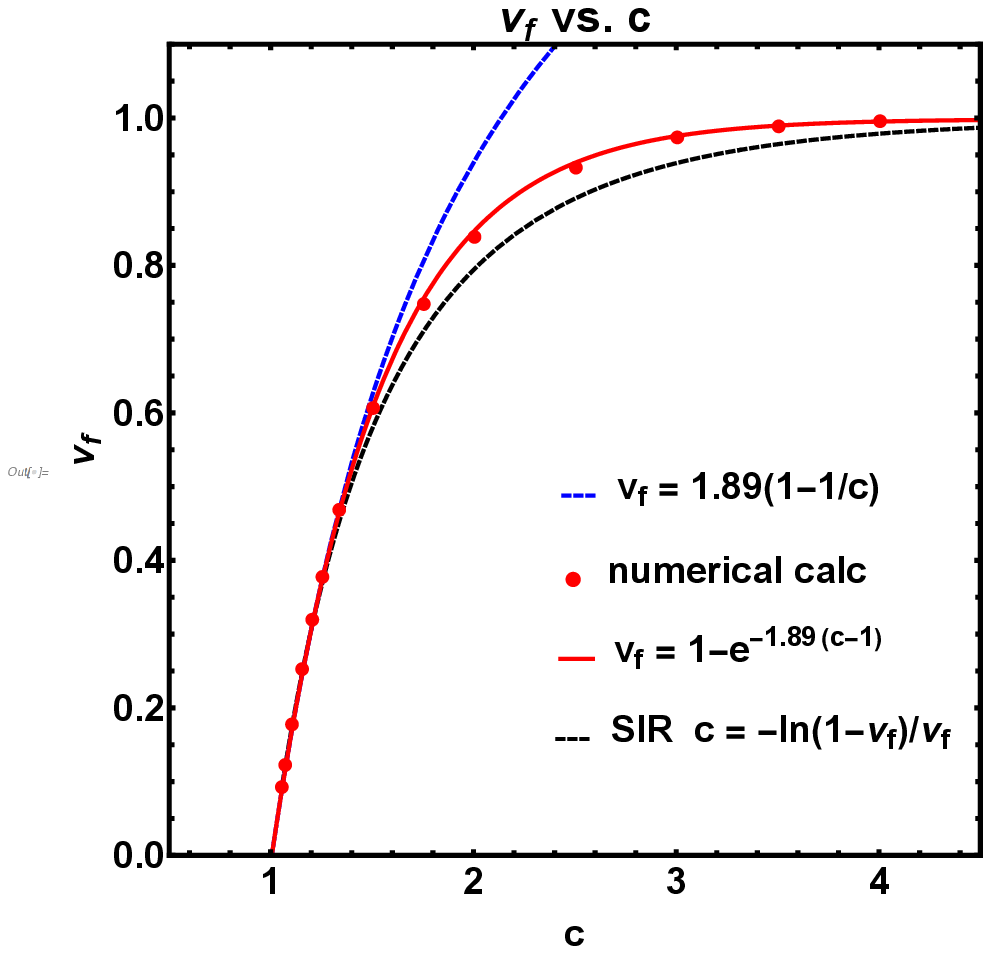}}}
 \caption{The final value, ${\nu}_{f}$, of the growth profile plotted versus the contagious
lifetime, $c$, comparing the results from the RGB-model (solid red and dashed blue curves)
and the SIR-model (black dashed curve) (cf. Figure~8).}
\end{figure}
\medskip

Next, we use Eq.(D7) to create a table of numerical values of $G(t)$, and then use Eqs.(D2) and
(D5) to create corresponding tables of $R(t)+G(t)$ and $R(t)$.  The results are plotted in
Figure~30 for values of $c=1.5, 2, 3$ and 10.  Referring to the corresponding RGB-plots in
Figure~15, we see again that the two models are similar for small values of $c$.  In each model,
for $c=1.5$ and $c=2$, the variation of $G$ is essentially the same as the variation of $R+G$,
but shifted $c$ units later in time, and the variation of $R$ is approximately $c$ times the
corresponding growth rate.  This starts to change when $c=3$, and by $c=10$, $R+G$
\begin{figure}[ht!]
 \centerline{%
 \fbox{\includegraphics[bb=88 235 552 716,clip,width=0.47\textwidth]{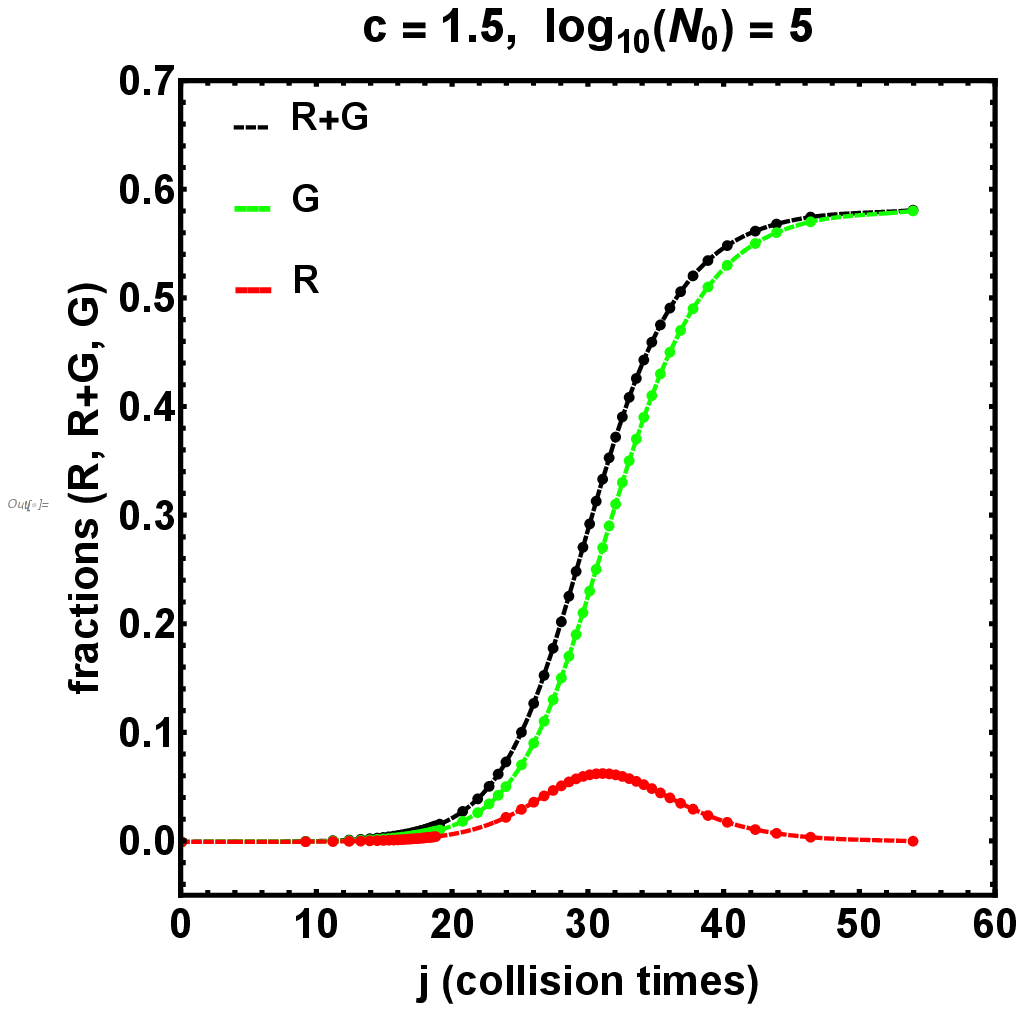}}
\vspace{0.01in}
 \fbox{\includegraphics[bb=88 240 552 716,clip,width=0.47\textwidth]{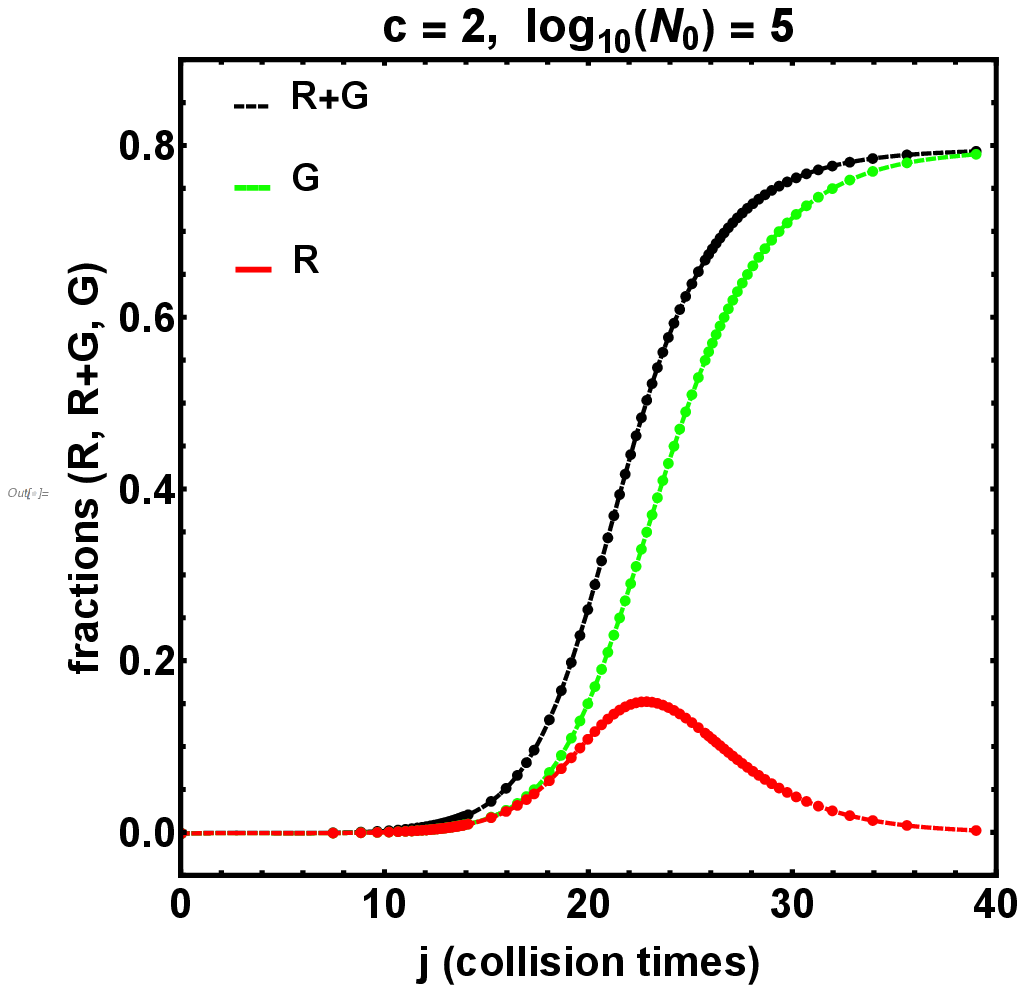}}}
 \hspace{0.01in}
\centerline{%
 \fbox{\includegraphics[bb=88 240 552 716,clip,width=0.47\textwidth]{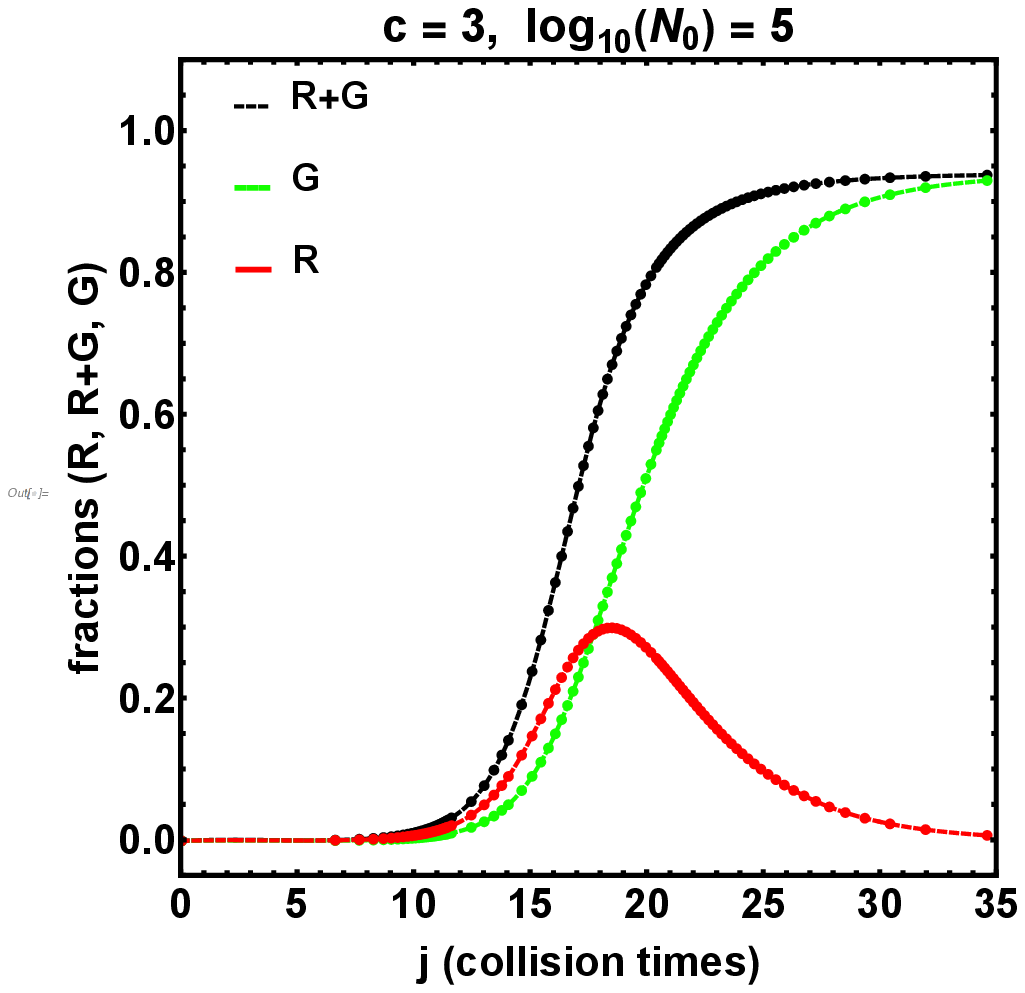}}
\vspace{0.01in}
\fbox{\includegraphics[bb=88 240 552 716,clip,width=0.47\textwidth]{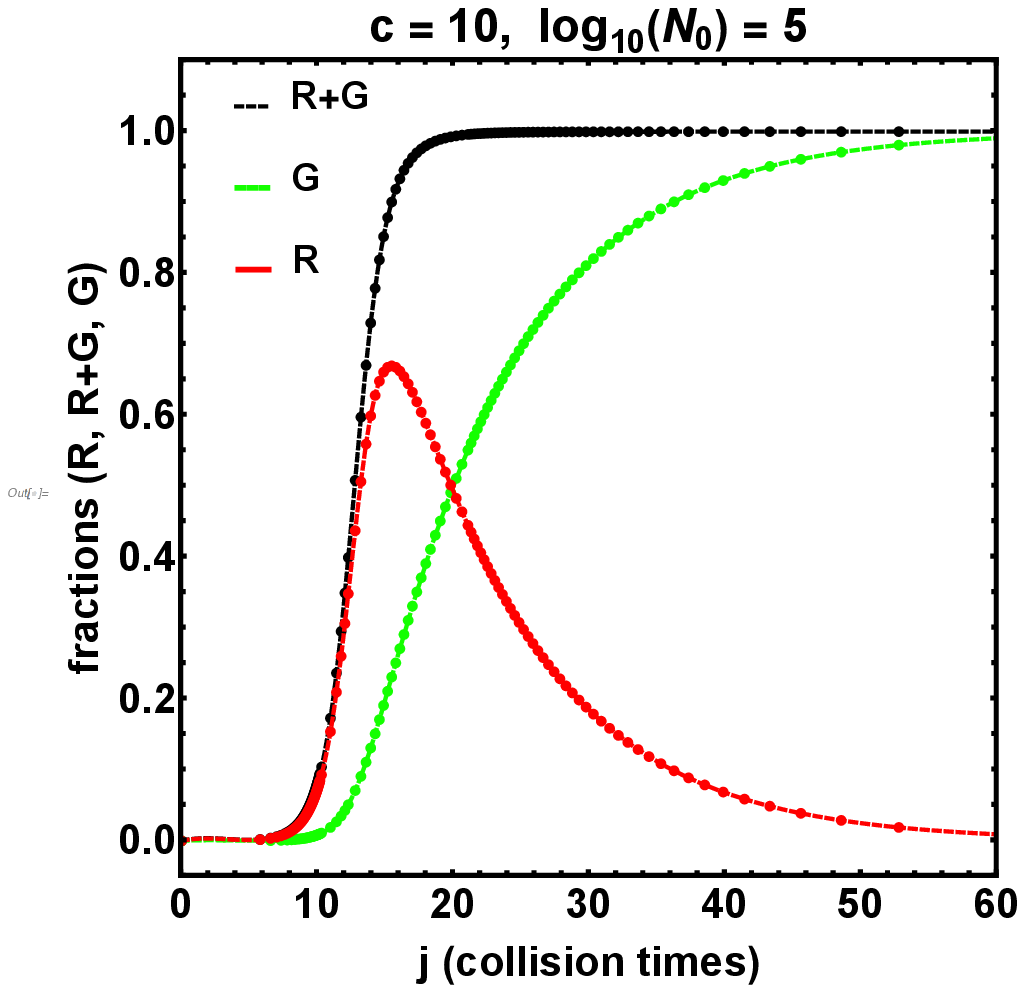}}
}
 \caption{The temporal evolution of R (red curve), G (green curve), and $R+G$ (black curve) populations
according to the SIR-model, for comparison with the corresponding evolutions in the RGB-model
(cf. Figure~15).  Note the scale changes between some of the images.}
\end{figure}
\noindent
increases so fast that it reaches about 0.67 before $G$ has begun its increase.  By this time,
$R$ no longer clings to the $R+G$ curve, and begins its descent like an exponential of the
form $e^{-t/{\tau}_{c}}$ while $G$ shows the corresponding exponential increase,
$1-e^{-t/{\tau}_{c}}$.  For the RGB-model in Figure~15, when $c=10$, $G$ continues to be a
delayed version of $R+G$.  Likewise, $R$ rises with $R+G$ until $G$ starts to increase, and
then falls to 0 as $G$ increases to 1.  Thus, when $c=10$ or more, the ultimate decline of the number
of red molecules reflects the steep rise of the number of green molecules in the $RGB$-model,
but it reflects the more gradual, exponential rise of the number of green molecules in the $SIR$-model.
\medskip

The exponential behavior of the SIR
model is shown even more clearly for the plots with $c=20$ and $c=50$ in Figure~31.  In the SIR model,
these plots show a sudden infection of the entire population followed by an exponential
decrease of the red population as it gradually turns green on the time scale, ${\tau}_{c}$.
\begin{figure}[ht!]
 \centerline{%
 \fbox{\includegraphics[bb=88 240 558 716,clip,width=0.46\textwidth]{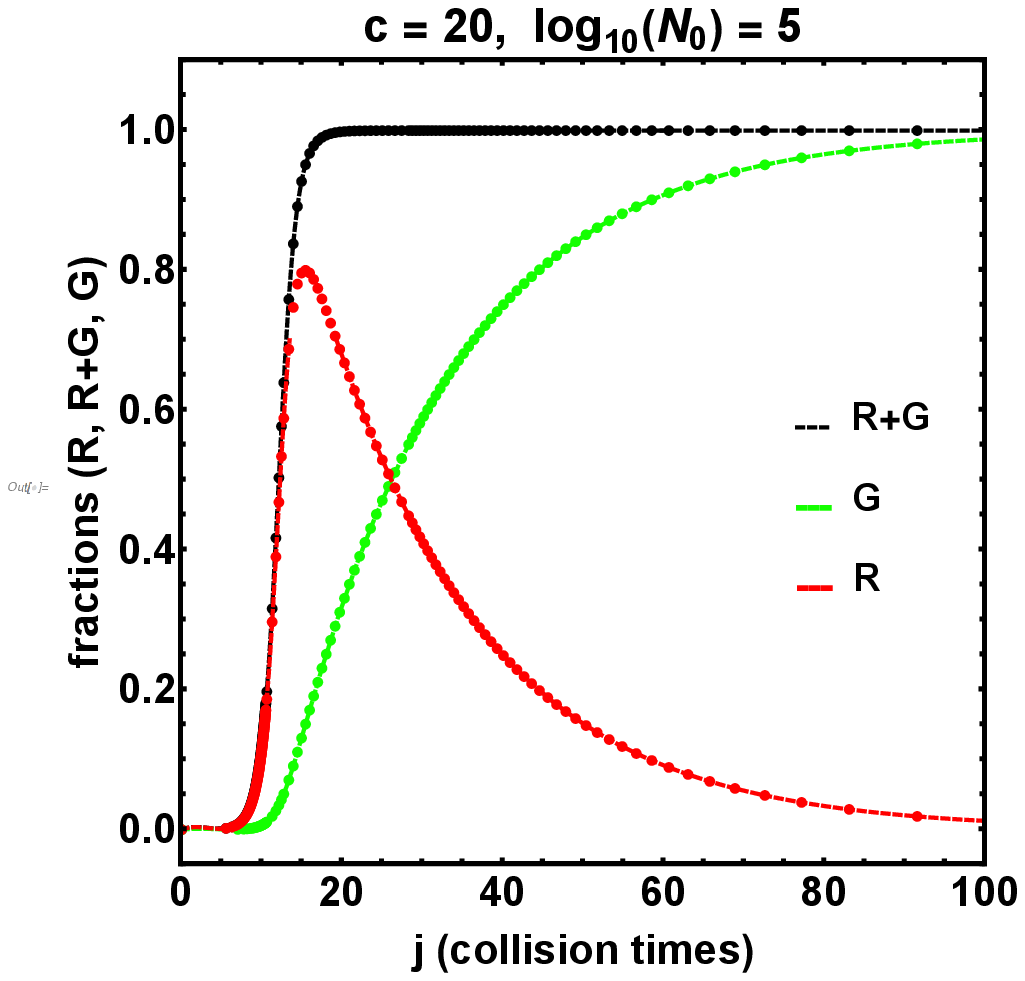}}
 \hspace{0.01in}
 \fbox{\includegraphics[bb=88 240 558 716,clip,width=0.46\textwidth]{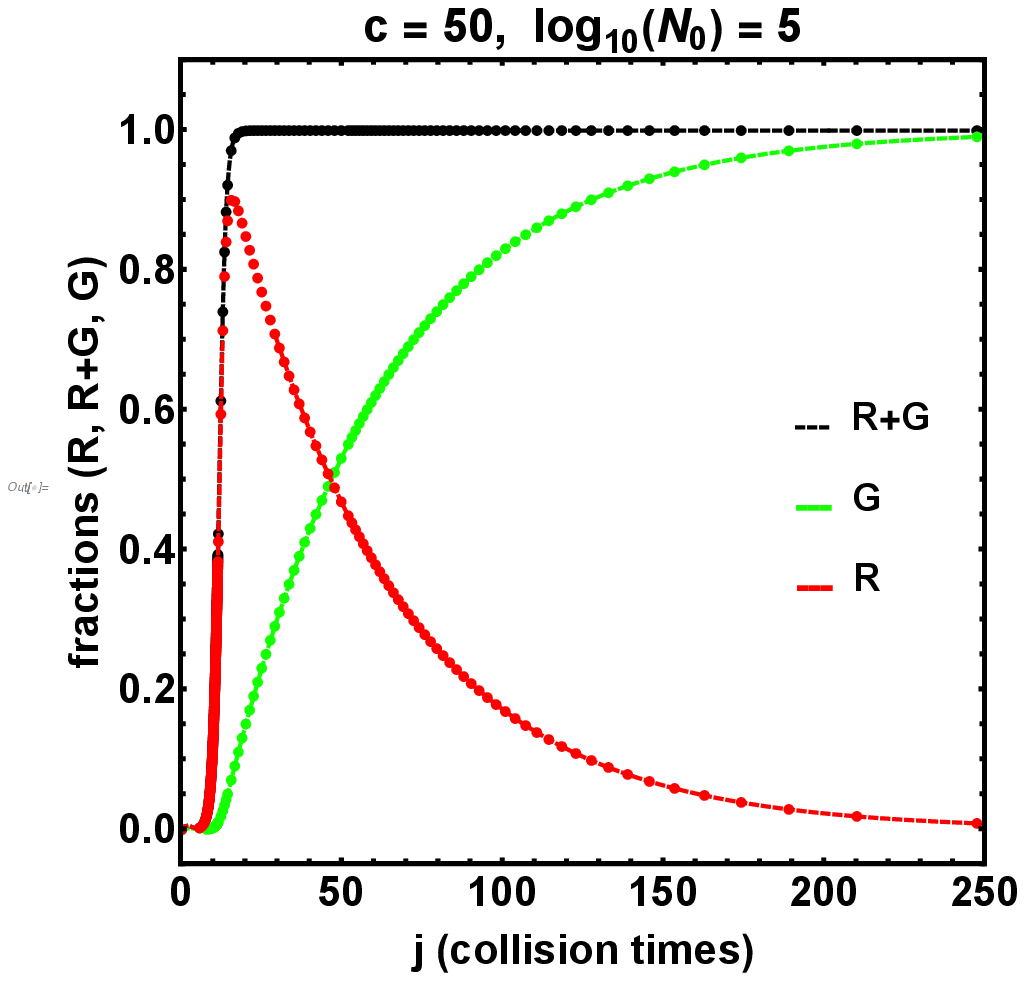}}}
 \caption{The extension of the SIR plots in Figure~30 to values of $c=20$ and $c=50$, showing
the steepening of $R+G$ and the exponential appearance of $G$ and $R$.}
\end{figure}
\medskip

Another approach is to display the SIR plots of $R+G$ and $d(R+G)/dt$ in the same way that we
displayed ${\nu}_{j}$ and ${\Delta}{\nu}_{j}$ in Figures~6 and 7.  Figures~32 and 33 show those
SIR plots.  At first glance, the plots look remarkably similar.  However, a closer look shows some
differences.  In Figure~32, the asymptotic values of $R+G$ are slightly less than the corresponding
values of ${\nu}_{f}$, as we have already seen in the plots of ${\nu}_{f}$ versus $c$
in Figure~29.  Also, in Figure~32, the `knees' of the plots of $R+G$ versus time are more rounded
than those for the RGB-model in Figure~6, especially for the larger values of $c$.  This causes the  
SIR plots of the growth rate in Figure~33 to return to 0 more gradually than the corresponding
RGB-plots in Figure~7.  Likewise the peak heights for the SIR growth rates are slightly less than
those for the RGB-model, except when $c<2$.  Again, this shows the general tendency for the models
to agree when $c$ is small and there is a lot of social distancing.  As a final part of our comparison
between the plots in Figures~32 and 33 and in Figure~6 and 7, we note that the SIR plots
of the growth rate reach their maximum values a few steps before the corresponding RGB growth
rates reach their peaks.
\clearpage
\begin{figure}[h!]
 \centerline{%
 \fbox{\includegraphics[bb=85 268 528 720,clip,width=0.90\textwidth]{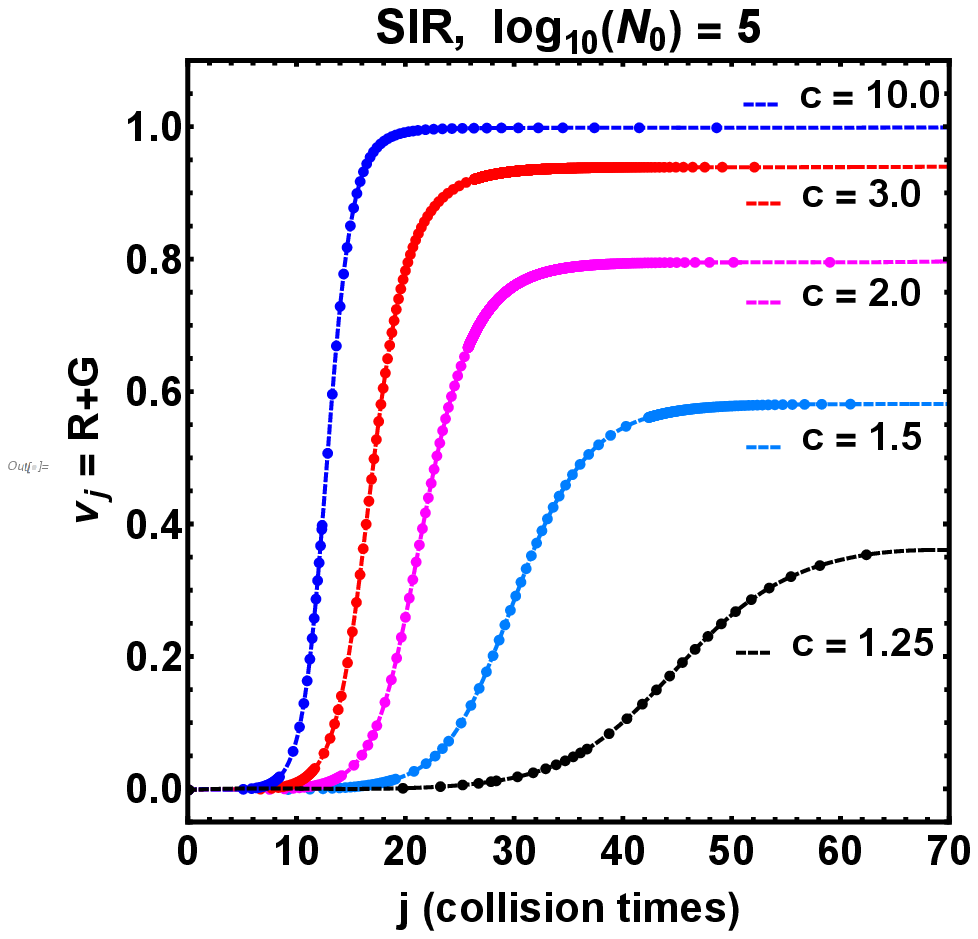}}}
 \caption{The fraction of infected population, $R+G$, calculated from the SIR-model using
Eqs.(D2) and (D7), for values of $c=1.25, 1.5, 2.0, 3.0$ and 10.  These curves are analogous
to the plots of ${\nu}_{j}$ in Figure~6 and show the delayed onset and
the reduced final fraction of the infected population as $c$ approaches 1.}
\end{figure}

\begin{figure}[h!]
 \centerline{%
 \fbox{\includegraphics[bb=85 268 540 720,clip,width=0.90\textwidth]{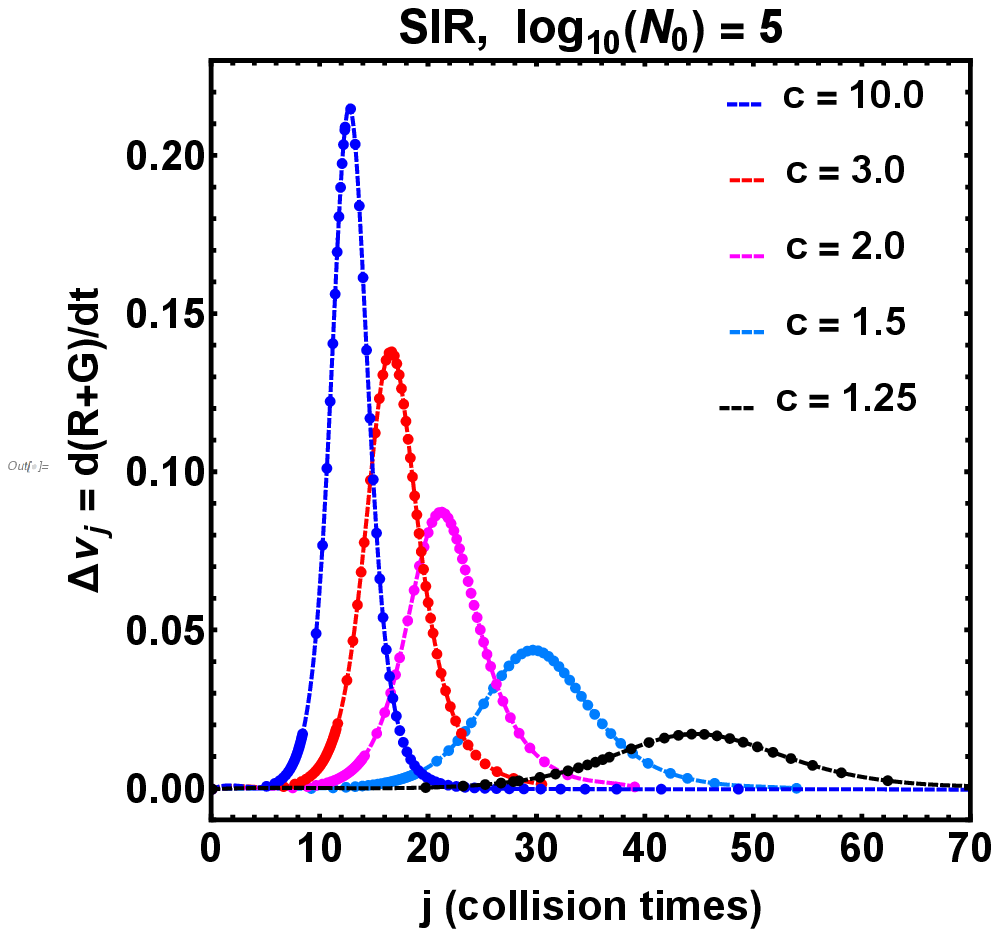}}}
 \caption{The growth rate, $d(R+G)/dt$, calculated from the SIR-model using Eq.(9.1a) with
Eqs.(D3), (D5), and (D7) for values of $c=1.25, 1.5, 2.0, 3.0$ and 10.  These curves are analogous
to the plots of ${\Delta}{\nu}_{j}$ in Figure~7, showing the weakening, widening, and shifting
of the growth-rate profiles as $c$ approaches 1.}
\end{figure}

In our discussion of `herd immunity' in section 6.2, we defined the threshold to be the value of
${\nu}_{j}$ where the number of red molecules, ${\nu}_{j}-{\nu}_{j-c}$, has its maximum value
and is starting to decline.  The resulting expressions are much simpler in the SIR-model than in
the $RGB$-model.  We begin with Eq.(9.1c) which tells us that $dG/dt = {\gamma}R$, and therefore
that $d^{2}G/dt^{2}={\gamma}dR/dt$.  Thus,  the herd-immunity condition, $dR/dt=0$, implies that
the slope, $dG/dt$, is a maximum when $R$ is a maximum.  (By way of contrast, in the $RGB$-model,
$R$ obtained its maximum value slightly before ${\nu}_{j-c}$ obtained its maximum slope.)
As Eq.(9.1b) indicates, $dR/dt = {\beta}RB-{\gamma}R$, so that $dR/dt=0$
also implies that $B=1/c$ and therefore that $R+G=1-1/c$.   Then, neglecting ${\epsilon}$ in Eq.(D3) 
for $B$, we find that $e^{-cG}=1/c$ which gives $G=ln(c)/c$ and $R=1-(1/c)-\{ln(c)/c\}$.  Summarizing
these results, in the $SIR$-model, $R$ has its maximum value when
\begin{subequations}
\begin{align}
B~=~\frac{1}{c},\\
R+G~=~1~-~\frac{1}{c},\\
G~=\frac{ln(c)}{c},\\
R~=~1~-~\frac{1}{c}~-~\frac{ln(c)}{c}.
\end{align}
\end{subequations}
These equations are identical to the classic herd equations for the $SIR$-model if we
regard $c$ to be the epidemiological parameter, $R_{0}$.
In particular, if $R_{0}=c=1.5$, then $B=0.67$,  $R+G=0.33$, $G=0.27$, and $R=0.06$ when
$dR/dt=0$.  If $R_{0}$ and $c$ are larger, say $R_{0}=c=3$, then $B=0.33$,  $R+G=0.67$,
$G=0.37$, and $R=0.30$ when $dR/dt=0$.  
\medskip

Figure~34 provides a graphic comparison of the herd-immunity threshold for the
\begin{figure}[h!]
 \centerline{%
 \fbox{\includegraphics[bb=88 246 552 720,clip,width=0.65\textwidth]{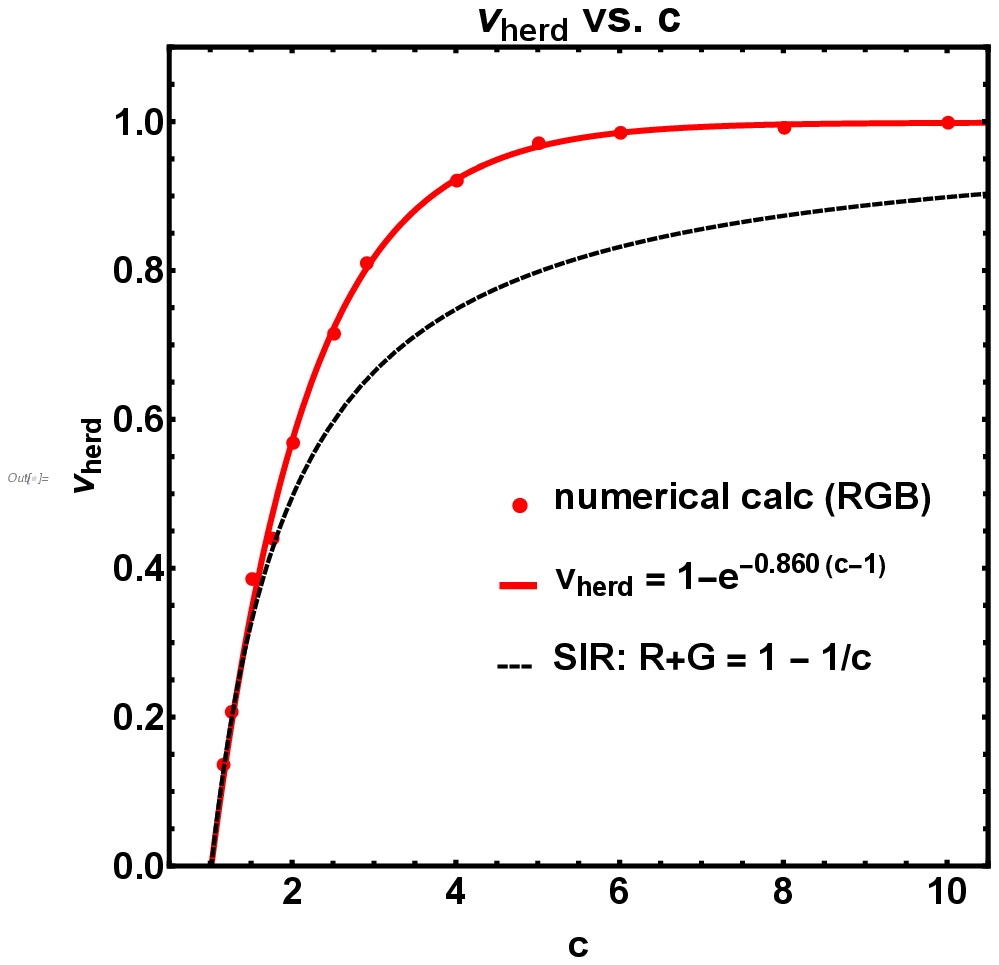}}}
 \caption{The value of ${\nu}_{herd}$, evaluated numerically for the RGB-model and from
$R+G=1-1/c$ for the SIR-model.  In each case, ${\nu}_{herd}$ is the total number of infections
($R+G$) when the number of contagious red molecules reaches its peak.}
\end{figure}
\noindent
SIR-model and
the RGB-model.  The red curve is the root-mean-square best fit to the computed
data points for
the RGB-model using the exponential given by
\begin{equation}
{\nu}_{herd}~=~1~-~e^{-0.860(c-1)}
\end{equation}
and the dashed black curve is the SIR-expression for $R+G=1-1/c$ given by Eq.(D9b).
Although the  two models show the same (normalized) number of infections for $c<1.5$,
the SIR-model shows a smaller number and a slower approach to 1 as $c$
becomes larger.
\medskip

In Figure~35, we show the difference, $j_{red}-j_{max}$, between the times that $R$ and
$R+G$ reach their respective peaks in the $SIR$-model.
\begin{figure}[h!]
 \centerline{%
 \fbox{\includegraphics[bb=88 246 552 720,clip,width=0.75\textwidth]{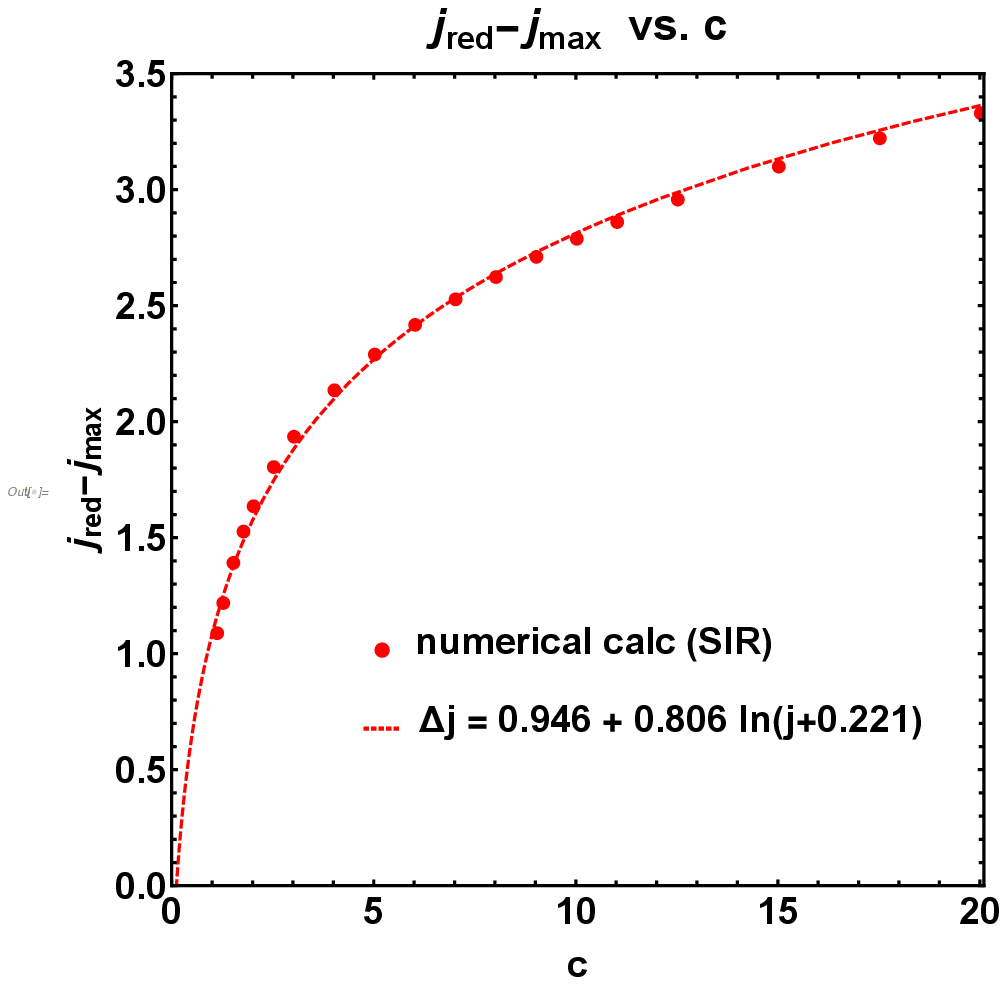}}}
 \caption{The time lag, $j_{red}-j_{max}$, between the location of the peak for infected
molecules, $R+G$, and the peak for contagious molecules, $R$, expressed in units of the
average time between collisions, ${\tau}_{s}$, and plotted as a function of
the contagious parameter, $c$, for the SIR-model.}
\end{figure}
\noindent
The $j$-values for each point were obtained by evaluating the
integral in Eq.(D7) for two different upper limits;  for $j_{red}$, we used the expression,
$ln(c)/c$, given in Eq.(D9c);  for $j_{max}$, we used the root of the equation
$G+2e^{-cG}=(c+1)/c$, which results from combining
$d^{2}(R+G)/dt^{2}~{\sim}~G''-c(G')^{2}=0$ with the expression for
$G'={\gamma}(1-G-e^{-cG})$ derived from Eqs.(9.1c) and (D5).  The dashed red curve is
the root-mean-square best fit to the data points, as given by the equation
\begin{equation}
j_{red}~-~j_{max}~=~0.946~=~0.806~ln(c+0.221).
\end{equation}
The data points used for this fit extend outward to $c=100$, which has
a $j$-value of $j_{red}-j_{max}=4.732$.  So although the lag does not increase rapidly with $c$,
it nevertheless has not reached a limit by $c=100$ (or even $c=1000$, when the lag is 6.934).
Like the corresponding lag determined for the RGB-model in Figure~19, this best-fit logarithmic
expression for the SIR-model has some curvature and increases very slowly with $c$.  By
comparison, `purely antisymmetric' models, like the hyperbolic-tangent and square-root
models described in the footnote to section (5.2) and in section (6.2), have lags that increase
linearly as $(c-1)/2$ before they saturate.
\medskip

Finally, we note that the epidemiological parameter, $R_{0}$, is often defined as the number of
secondary infections that are produced in a homogeneous population of susceptible blue
molecules by a single contagious red molecule before it loses its contagiousness and turns green.
In the $SIR$-model, this works out to $R_{0}={\beta}/{\gamma}$, which we have called $c$.
However, in the $RBG$-model, the number is $2^{c}-1$ because each infected molecule goes on
to infect other blue molecules before the original red molecule suddenly loses its contagiousness.
These secondary infections continue to infect blue molecules without the exponential
weakening that gives the average decay time, $1/{\gamma}$, in the $SIR$-model.  As mentioned
above, we can recover the classic $SIR$ expression
\begin{equation}
{\nu}_{herd}~=~1~-~\frac{1}{c}~=~1~-~\frac{1}{R_{0}}
\end{equation}
by substituting $c=R_{0}$ in Eq.(D9b).  However, the substitution of $2^{c}-1=R_{0}$ in $RGB$
Eq.(6.4) gives
\begin{equation}
{\nu}_{herd}~=~1~-~e^{-0.860(c-1)}~=~1~-~(\frac{2}{1+R_{0}})^{1.24}.
\end{equation}
These two equations give the value of ${\nu}_{herd}$ (or $R+G$) when the number of red
molecules reaches its peak.  In Figure~34 we saw that these thresholds occur for different
values of $c$.  Now, comparing Eqs. (D12) and (D13), we can also see that the thresholds
occur for different values of $R_{0}$ (except when $c=R_{0}=1$, so that ${\nu}_{herd}=0$
and there is no epidemic).
\pagebreak
\section{The Effect of Vaccinations}
In the previous sections, we learned how social distancing affects the spread of a virus.
Meanwhile, vaccines have been developed for covid-19 and a significant fraction of the
population has been vaccinated.  This motivates us to ask how vaccinations would affect
the spread of a virus according to the RGB-model, and, by implication, according to the
SIR-model.  For simplicity, we assume that the vaccinations are applied suddenly and
simultaneously to a fixed number of blue molecules which thereby turn green directly
without having to go through the intermediate state of being red.  This means that a fraction,
${\delta}$, of the total population of $N_{0}$ molecules suddenly changes from blue to
green at the vaccination time, $j_{v}$.  Depending on how many molecules are
vaccinated, the fraction, ${\delta}$, will range from 0 to a maximum value equal to $B_{jv}(0)$,
the fraction of blue molecules present at that time.
\medskip

\subsection{Equations for $\mathbf{{\nu}_{j}}$}

Mathematically, our objective is to solve Eqs (6.1) and (6.2) (equivalently Eq (4.3) in the limit of
${\nu}_{0}=1/N_{0}<<1$) for ${\nu}_{j}$, and then to deduce the fractions, $B_{j}$, $G_{j}$, and
$R_{j}$ from the array, ${\nu}_{j}$.  This is a relatively straightforward process outside the interval,
$[j_{v},j_{v}+c]$ where $1~{\leq}~j<j_{v}$ and $j>j_{v}+c$ because we can use the usual formulas
\begin{subequations}
\begin{align}
{\nu}_{j}~=~{\nu}_{j-1}~+~({\nu}_{j-1}-{\nu}_{j-1-c})(1-{\nu}_{j-1}),\\
B_{j}~=~1~-~{\nu}_{j},\\
G_{j}~=~{\nu}_{j-c},\\
R_{j}~=~1~-~B_{j}~-~G_{j}.
\end{align}
\end{subequations}
However, special care must be taken for the three cases that involve the interval $[j_{v},j_{v}+c]$
and its end points.  In particular, when $j=j_{v}$, we suddenly change ${\delta}$ molecules from blue
to green so that
\begin{subequations}
\begin{align}
{\nu}_{j}~=~{\nu}_{j-1}~+~({\nu}_{j-1}-{\nu}_{j-1-c})(1-{\nu}_{j-1})~+~{\delta},\\
B_{j}~=~1~-~{\nu}_{j},\\
G_{j}~=~{\nu}_{j-c}~+~{\delta},\\
R_{j}~=~1~-~B_{j}~-~G_{j}.
\end{align}
\end{subequations}
Also, to avoid other discontinuities, when $j_{v}<j<j_{v}+c$, we need to use
\begin{subequations}
\begin{align}
{\nu}_{j}~=~{\nu}_{j-1}~+~({\nu}_{j-1}-{\nu}_{j-1-c}-{\delta})(1-{\nu}_{j-1}),\\
B_{j}~=~1~-~{\nu}_{j},\\
G_{j}~=~{\nu}_{j-c}~+~{\delta},\\
R_{j}~=~1~-~B_{j}~-~G_{j},
\end{align}
\end{subequations}
and when $j=j_{v}+c$, we need to use
\begin{subequations}
\begin{align}
{\nu}_{j}~=~{\nu}_{j-1}~+~({\nu}_{j-1}-{\nu}_{j-1-c}-{\delta})(1-{\nu}_{j-1}),\\
B_{j}~=~1~-~{\nu}_{j},\\
G_{j}~=~{\nu}_{j-c},\\
R_{j}~=~1~-~B_{j}~-~G_{j}.
\end{align}
\end{subequations}
Now, with these formulas, we can generate the array, ${\nu}_{j}$, and deduce the corresponding
values of $B_{j}$, $G_{j}$, and $R_{j}$.  Figure~36 shows temporal plots of $B_{j}$, $G_{j}$, and
$R_{j}$, for $c=10$.
\begin{figure}[ht!]
 \centerline{%
 \fbox{\includegraphics[bb=88 250 542 716,clip,width=0.41\textwidth]{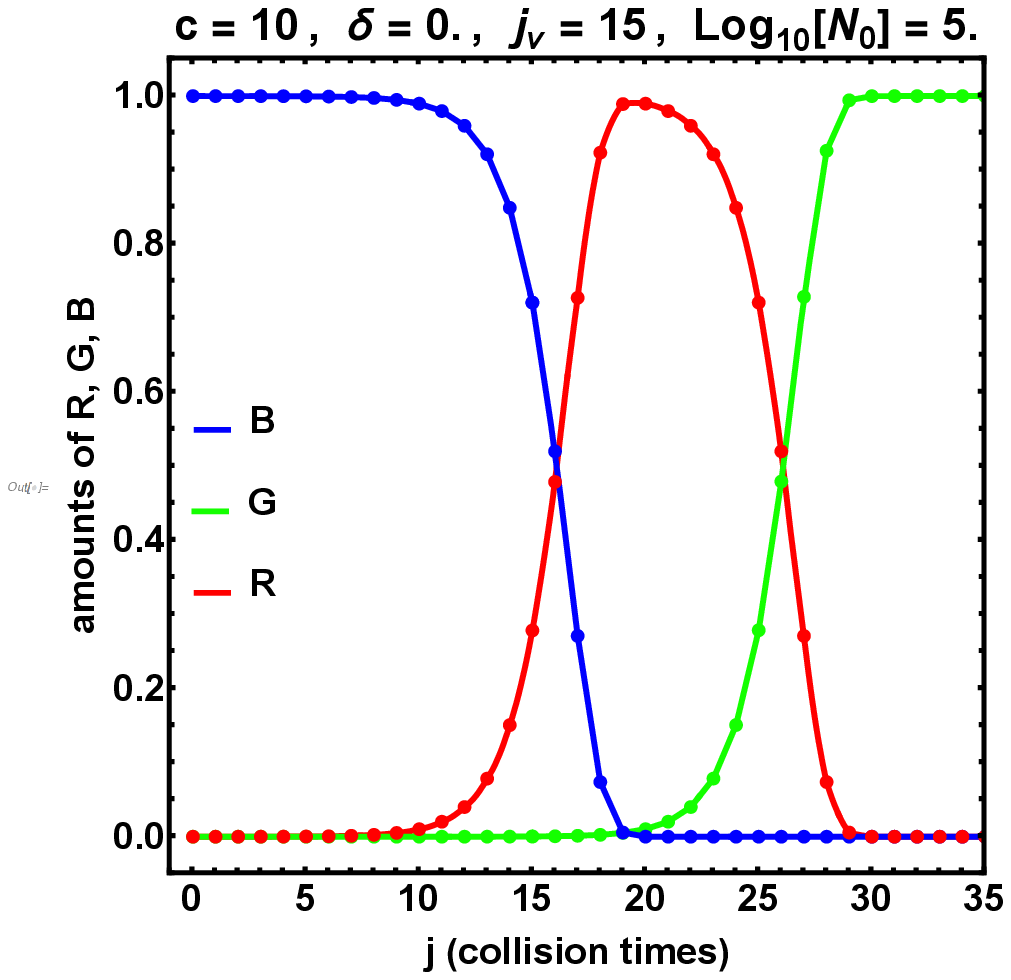}}
\vspace{0.01in}
 \fbox{\includegraphics[bb=88 250 542 716,clip,width=0.41\textwidth]{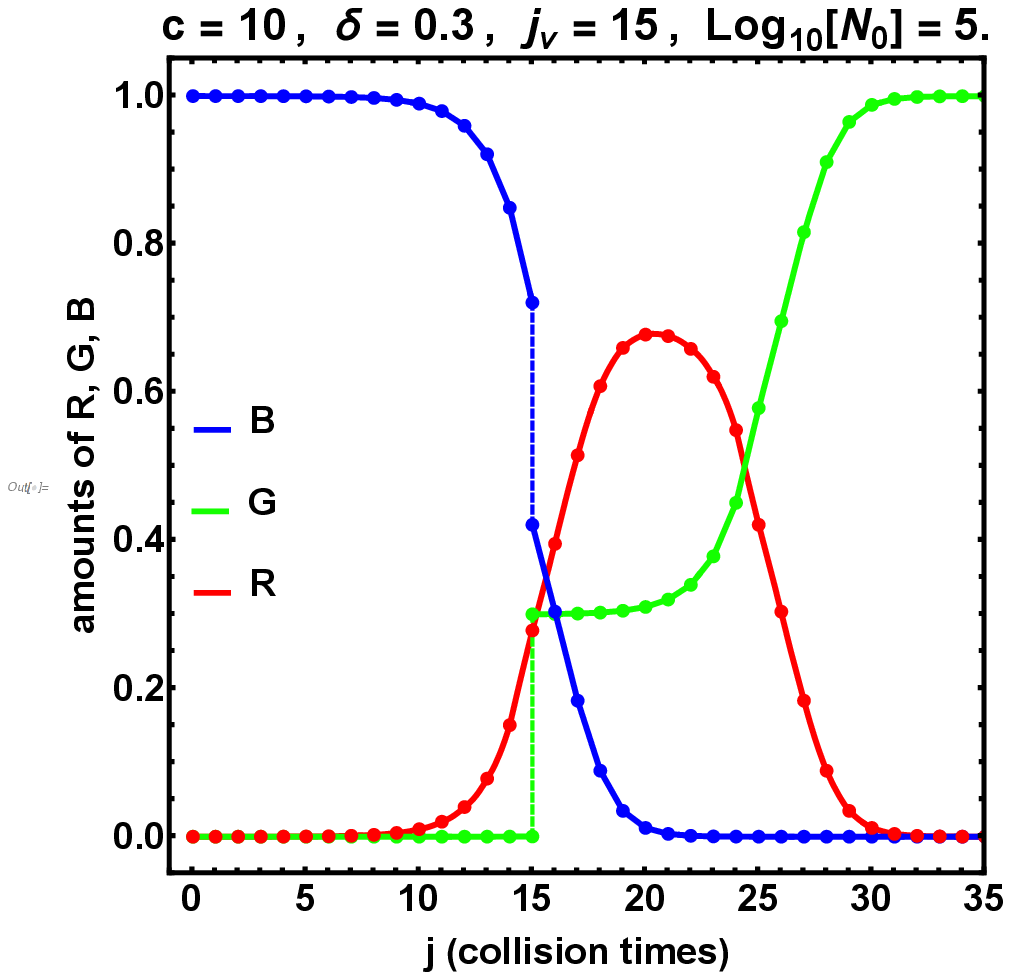}}}
 \hspace{0.01in}
\centerline{%
 \fbox{\includegraphics[bb=88 250 542 716,clip,width=0.41\textwidth]{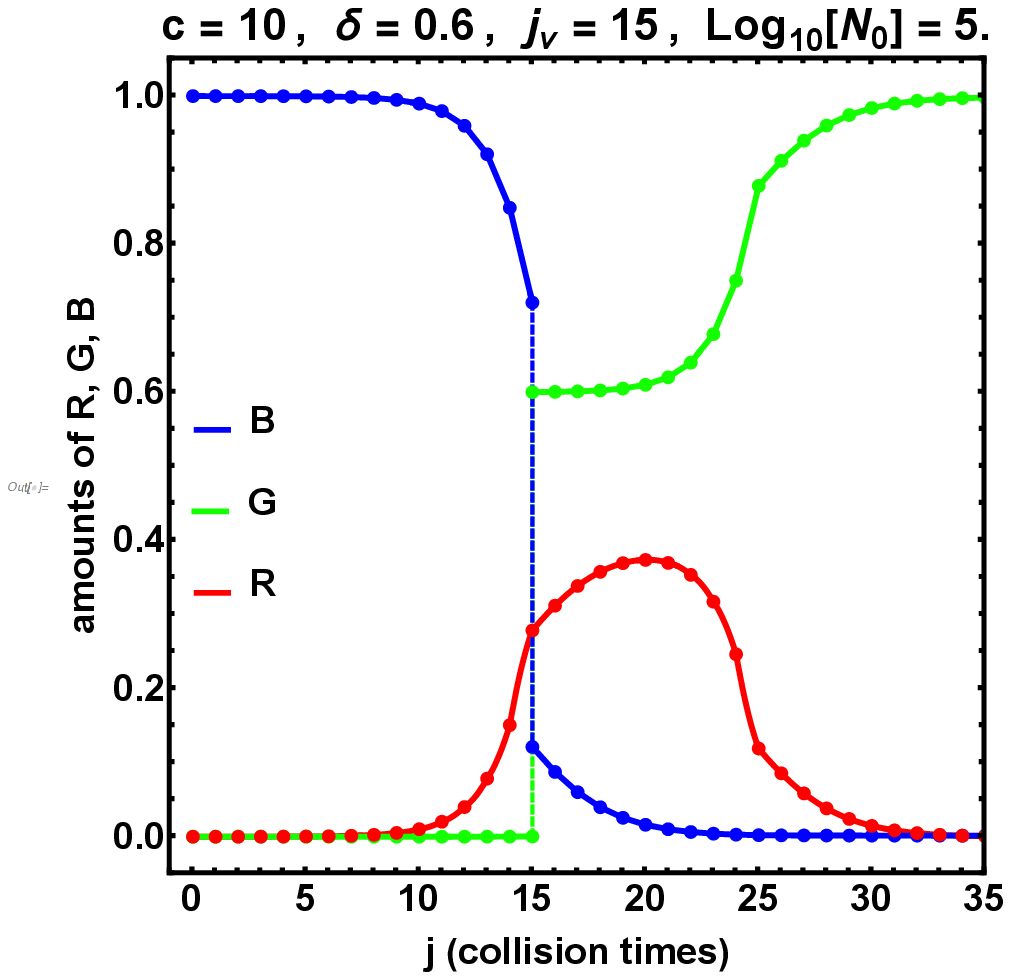}}
\vspace{0.01in}
\fbox{\includegraphics[bb=88 250 542 716,clip,width=0.41\textwidth]{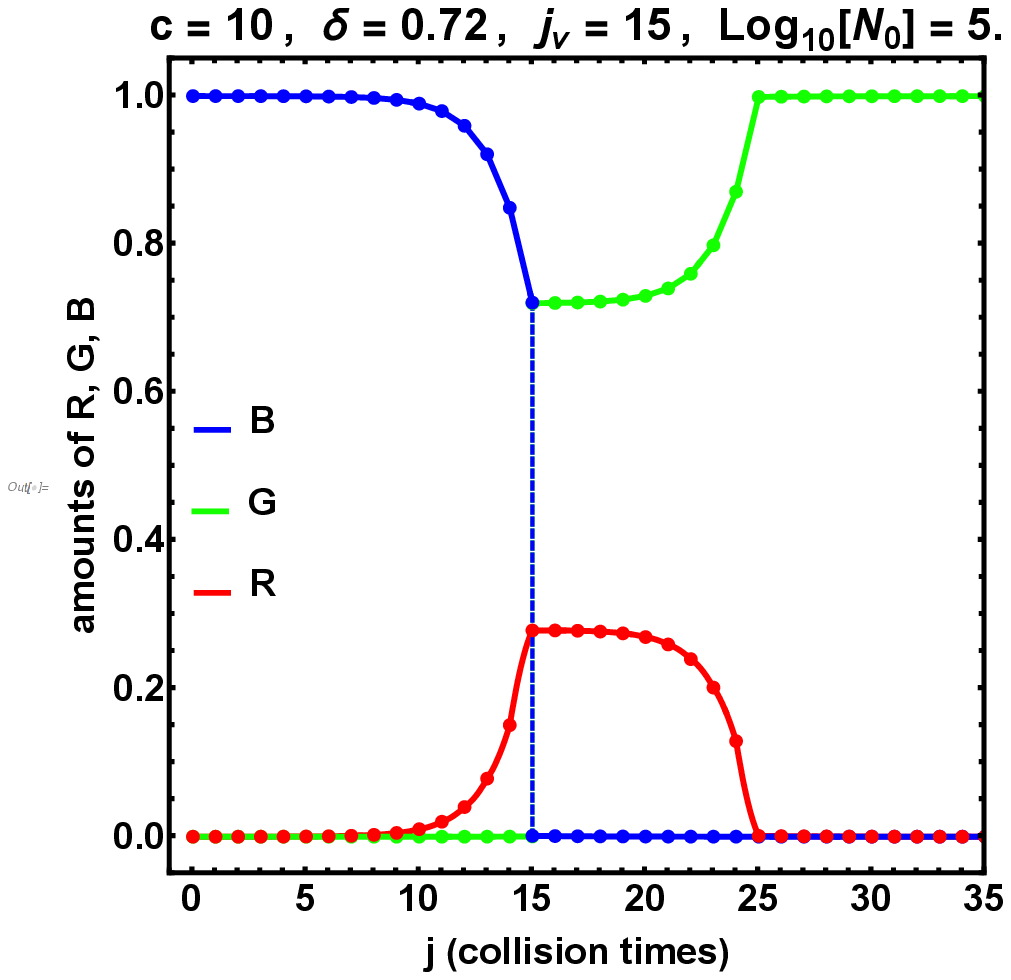}}
}
 \caption{The relative fractions of blue, green, and red molecules plotted versus time for $c=10$.
For this long lifetime of red molecules, there are no surviving blue molecules and the only molecules that escaped infection are the ${\delta}$ that were vaccinated at $j_{v}=15$.}
\end{figure}
\noindent
This relatively long contagious lifetime corresponds to virtually no social
distancing and no surviving blue molecules.  In the upper left panel, where no molecules were
vaccinated, all of the blue molecules became infected by the time $j=19$ and these red molecules
gradually
turned green during the subsequent 10 collisions.  Even when some molecules were vaccinated,
as indicated by the numbers ${\delta}=0.3$, 0.6, and 0.72 in the upper right and lower panels,
no blue molecules escaped infection. Only the vaccinated molecules escaped infection, and they
ended up as part of the final distribution of green molecules.
\medskip

Because the vaccinations were given at $j_{v}=15$ during the rising phase of the epidemic, a
relatively large number of molecules ($B_{jv}(0)=0.72$) were still blue and able to
be vaccinated before they could be infected by red ones.  When all of the remaining blue molecules
were vaccinated, as in the lower right panel, the infections stopped abruptly at $j=j_{v}=15$, leaving
a fraction of about 0.27 red molecules that gradually turned green
during the 10 steps between $j=15$ and $j=25$.  On the other hand, when some blue molecules
were not vaccinated, as in the upper right and lower left panels, the newly vaccinated green molecules
provided some additional shielding from the red molecules, and caused those unvaccinated blue
molecules to live slightly longer than they would have lived without any vaccinations (to
about $j=21$ and $j=23$ compared to $j=19$ in the upper left panel of Figure~36).

\subsection{Direct Equations for $R_{j}$, $G_{j}$, and $B_{j}$}

In an equivalent approach, we can solve directly for $R_{j}$, $G_{j}$, and $B_{j}$.
The equation for $B_{j}$ is simply
\begin{equation}
B_{j}~=~B_{j-1}~-~B_{j-1}R_{j-1},
\end{equation}
which says that the incremental decrease of $B_{j}$ is given by the product $B_{j-1}R_{j-1}$.  If we can
obtain an equation for $G_{j}$ we will be done because we could obtain $R_{j}$ from the relation
$R_{j}+G_{j}+B_{j}=1$.  The correct rule is that
\begin{equation}
G_{j}~=~G_{j-c}~+~R_{j-c}~+~vac
\end{equation}
where $vac$ refers to any vaccinations made in between $j-c$ and $j$, but $vac$ is 0 otherwise.
Thus, if $c=2$ and $j_{v} = 18$, then $G_{19}=G_{17}+R_{17}+{\delta}$ because ${\delta}$
vaccinations were given
during the intermediate step $j_{v}=18$.  Also, $G_{20}=G_{18}+R_{18}+0$ because $G_{18}$ was
defined to include the ${\delta}$.  Finally, as just mentioned, we can now obtain $R_{j}$ from
\begin{equation}
R_{j}~=~1~-~B_{j}~-~G_{j}.
\end{equation}
So, in principle, the three equations (E5)-(E7) can replace the 16 equations (E1a-d) - (E4a-d).  However,
in practice, I used Eqs.(E1)-(E4) for the calculations and plots shown here.
\medskip

Another approach is to eliminate $G_{j}$ as a dependent variable and express $R_{j}$, $G_{j}$, and
$B_{j}$ in terms of just $B_{j}$ and $R_{j}$.  After some algebra, we obtain
\begin{subequations}
\begin{align}
B_{j}~=~B_{j-1}~-~B_{j-1}R_{j-1}\\
R_{j}~=~(B_{j-c}~-~B_{j})~-~{vac}\\
G_{j}~=~(1~-~B_{j-c})~+~{vac},
\end{align}
\end{subequations}
where $vac$ refers to any vaccinations (${\delta}$) made between $j-c$ and $j$, but is
zero otherwise.  These equations are useful for examining the effect of the vaccination
when $c$ is relatively large and there is very little social distancing.
\medskip

Now, referring to the two panels of Figure~37 for which $c=5$ and
$c=10$, respectively, we
\begin{figure}[ht!]
\centerline{%
 \fbox{\includegraphics[bb=88 250 542 716,clip,width=0.47\textwidth]{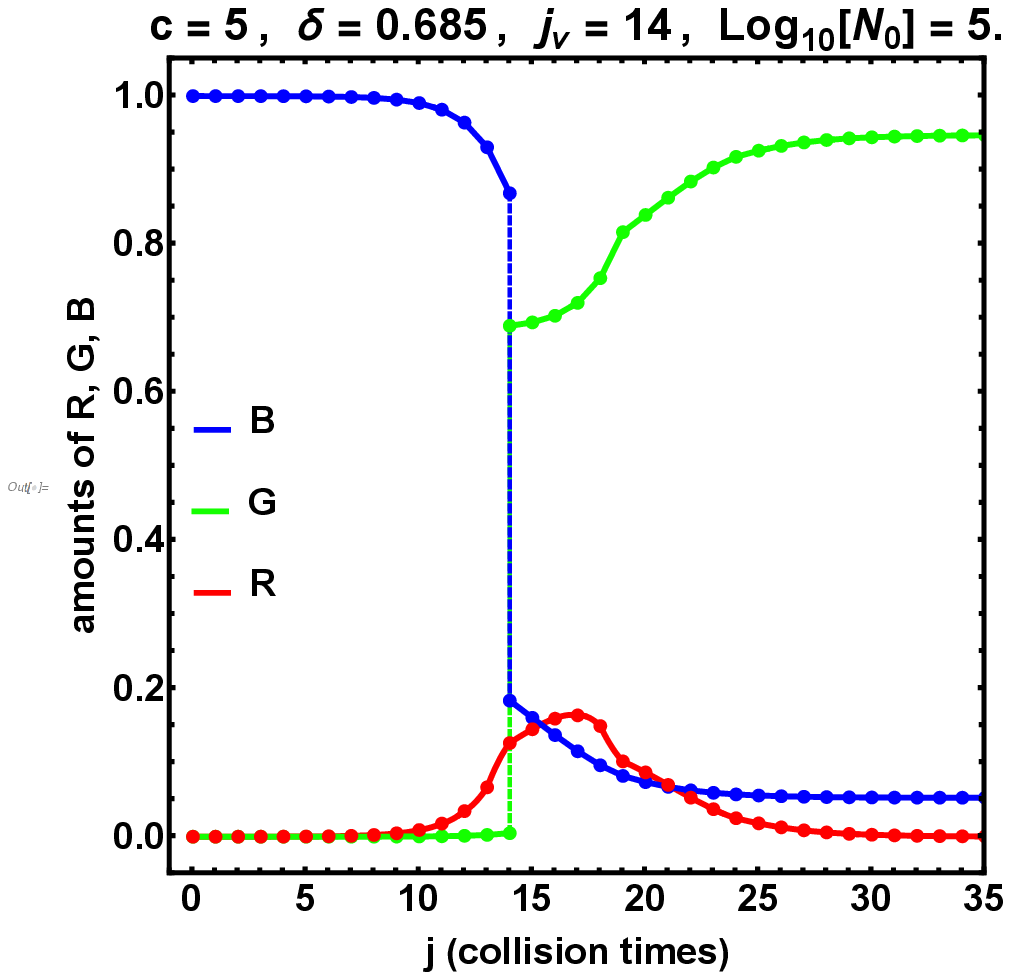}}
\vspace{0.01in}
\fbox{\includegraphics[bb=88 250 542 716,clip,width=0.47\textwidth]{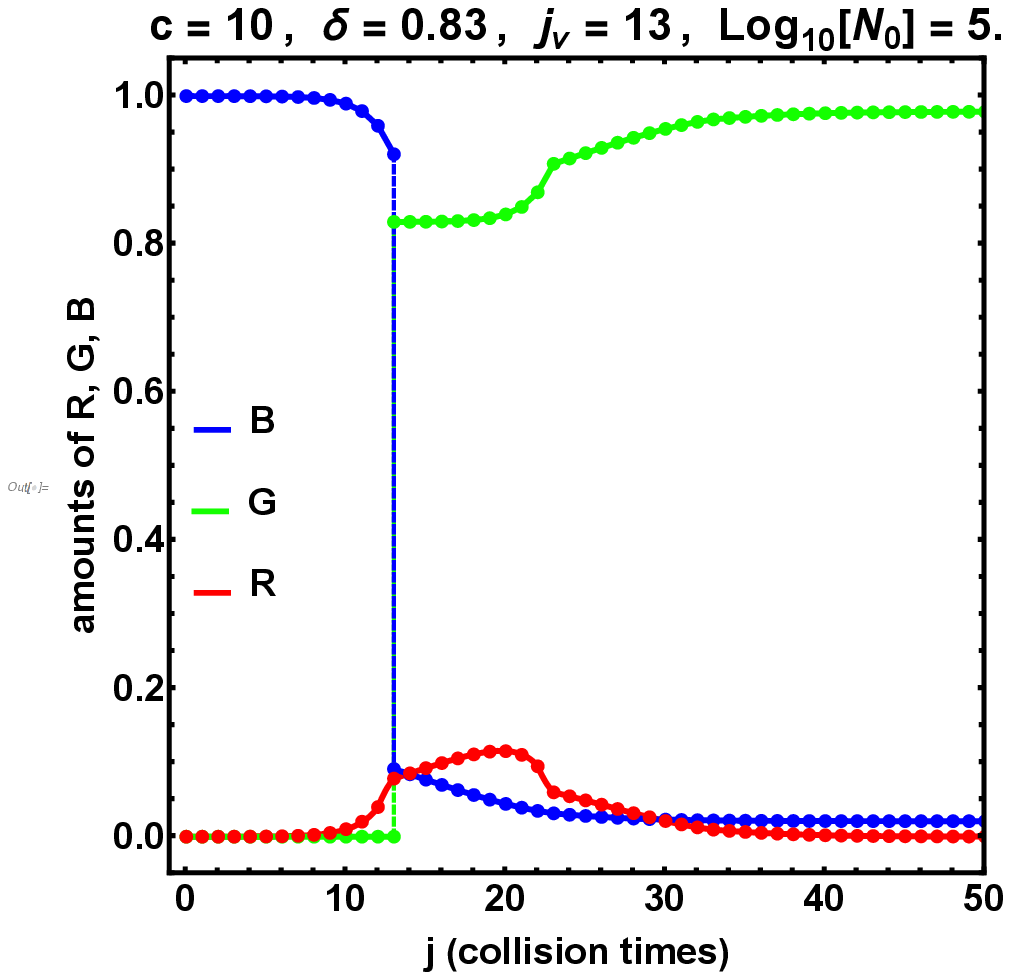}}}
\caption{The effect of vaccinations when the social distancing parameters are
$c=5$, and 10.  In each case, ${\delta}$ has been chosen to maximize $B_{f}$.  Without vaccinations,
the peaks of the red curves occurred  at
$j=18.9$, and 19.0, so these vaccinations with $j_{v}= 14$, and 13 were early in
the rising phase of the pandemic and produced appreciable changes in these curves.  Here,
complementary bulges of width, $c$, are visible after $j=j_{v}$ in the red and
green curves.}
\end{figure}
\noindent
can see that there are $c$ data points starting at the vaccination time, $j_{v}$, where the fractions
of red molecules become elevated and the fractions of green molecules become depressed
relative to their adjacent points.  These upward and downward `bulges' are places where the data
points are linked to points on the other side of the vaccination and therefore require $vac={\delta}$
for their evaluation in Eqs (E8b) and (E8c).  On the other hand, points outside these bulges do not
require this ${\delta}$-correction, and are calculated with $vac=0$.
\medskip

With this understanding, we can now follow the evolution of the pandemic in Figure~37.  Referring to
the right panel with $c=10$ as an example, we see that G and ${\Delta}G$ are both 0 prior to the
time of vaccination ($j_{v}=13$), so that the initial behavior is determined
entirely by collisions between the red and blue molecules, whose numbers start to change
rapidly between $j=10$ and $j=13$.  After the vaccinations, this collisionally dominated behavior
continues at a slower rate until the oldest red molecules `come of age' and turn green, causing
a sharp decrease in the fraction of red molecules and a corresponding sharp rise in the number
of green molecules.  The rise time is the same as the initial rise time of the red molecules (and the
fall time of the blue molecules) during $j=10$ to 13.  When $j=j_{v}+c=23$, these slopes suddenly
become smaller, reflecting the sudden deceleration of the red and blue curves at the time of the
vaccination.   Now, the relatively low number of blue molecules reduces the frequency of
red-blue collisions so much that these collisions can not replenish the loss of red molecules that are
turning green with age.  Consequently, the number of blue molecules reaches its final value, $B_{f}$,
and the pandemic gradually ends as the remaining red molecules turn green.
\medskip

\subsection{Vaccinations With More Social Distancing (Smaller $\mathbf{c}$)}
Next, we consider what will happen when the `social distancing' parameter $c$ is much less than 10
so that an appreciable fraction, $B_{f}$, of blue molecules survives when the pandemic is over and
all of the red molecules are gone.  This survival of some blue molecules occurs when $c$ is small
because it takes many of these small contagious times for red molecules to find blue ones toward
the end of the pandemic.  Consequently, those last remaining red molecules turn green, and the
hidden blue molecules survive.
In Section 5.1, Eq.(5.6) provides an empirical relation between $c$ and the final fraction of infected
molecules, ${\nu}_{f}$.  When ${\nu}_{f}$ is replaced by $B_{f}=1-{\nu}_{f}$, this relation becomes
\begin{equation}
B_{f}~=~e^{-1.890(c-1)}.
\end{equation}
Likewise, from Eq.(5.3), we find that
\begin{equation}
B_{j-1}~{\approx}~\frac{1}{c}
\end{equation}
when the `slope', ${\Delta}B_{j}$ is steepest.  Because this is close to the place that the number of red molecules, $R_{j}$, reaches its peak, we might expect that this relation would give an estimate of when the
pandemic reaches the herd-immunity threshold.  Although this is true for the SIR-model (as given by Eq (D9b)) and it is approximately true for our RGB-model when $c~{\lesssim}~2$ (see Figure~34), it
is not true for our RGB-model when vaccinations are involved.  In fact, we shall find that the $1/c$-rule
gives the approximate value of $B_{jv}$ when the fraction of surviving blue molecules has its maximum value, but that the herd immunity threshold (where $dR_{j}/dj=0$) occurs at a lower value of $B_{jv}$, corresponding to a larger amount of vaccination. 
\medskip

Here, our objective is to understand how the fraction of surviving blue molecules, $B_{f}$ depends on
vaccinations.  The outcome of the vaccination will depend on when those vaccinations are given and
how many molecules receive them.  For simplicity, we will give all of the vaccinations simultaneously.
If we do this early in the pandemic when the number of red molecules is in its rising phase, then we can prevent a large number of blue molecules from being infected.  However, if we wait until the number of
red molecules reaches its peak and the number of blue molecules is decreasing rapidly, then we cannot
save any more blue molecules than would escape by social distancing alone.  In that case, the vaccinated blue molecules would survive as uninfected green molecules.   Finally, if we perform the vaccinations during the declining phase of the pandemic, then we would decrease the number of surviving blue molecules below the level that would have escaped by social distancing alone.  However, this decrease
would be more than offset by the number of vaccinated green molecules, so that there will always be
a net saving when vaccinations are given.
\medskip

The upper left panel of Figure~38 shows the evolution of red, green, and blue molecules  
\begin{figure}[ht!]
 \centerline{%
 \fbox{\includegraphics[bb=88 250 542 716,clip,width=0.37\textwidth]{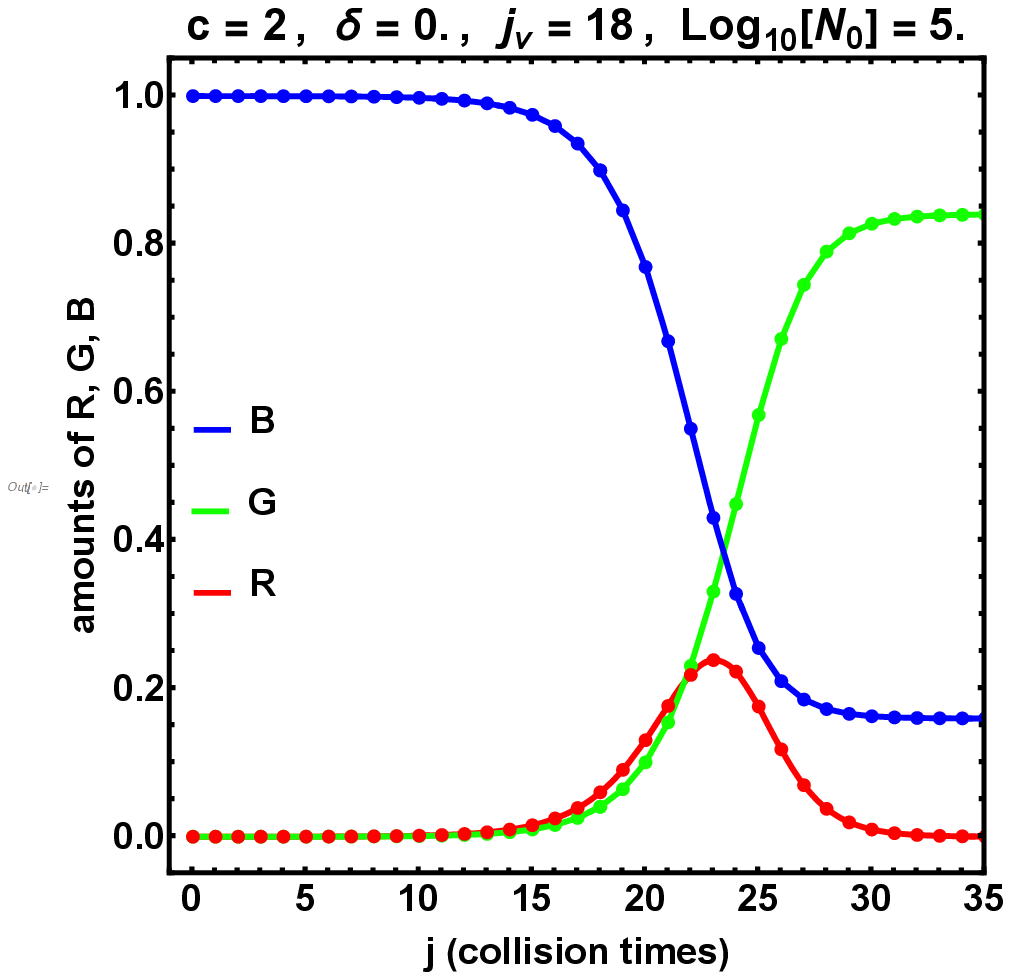}}
\vspace{0.01in}
 \fbox{\includegraphics[bb=88 250 542 716,clip,width=0.37\textwidth]{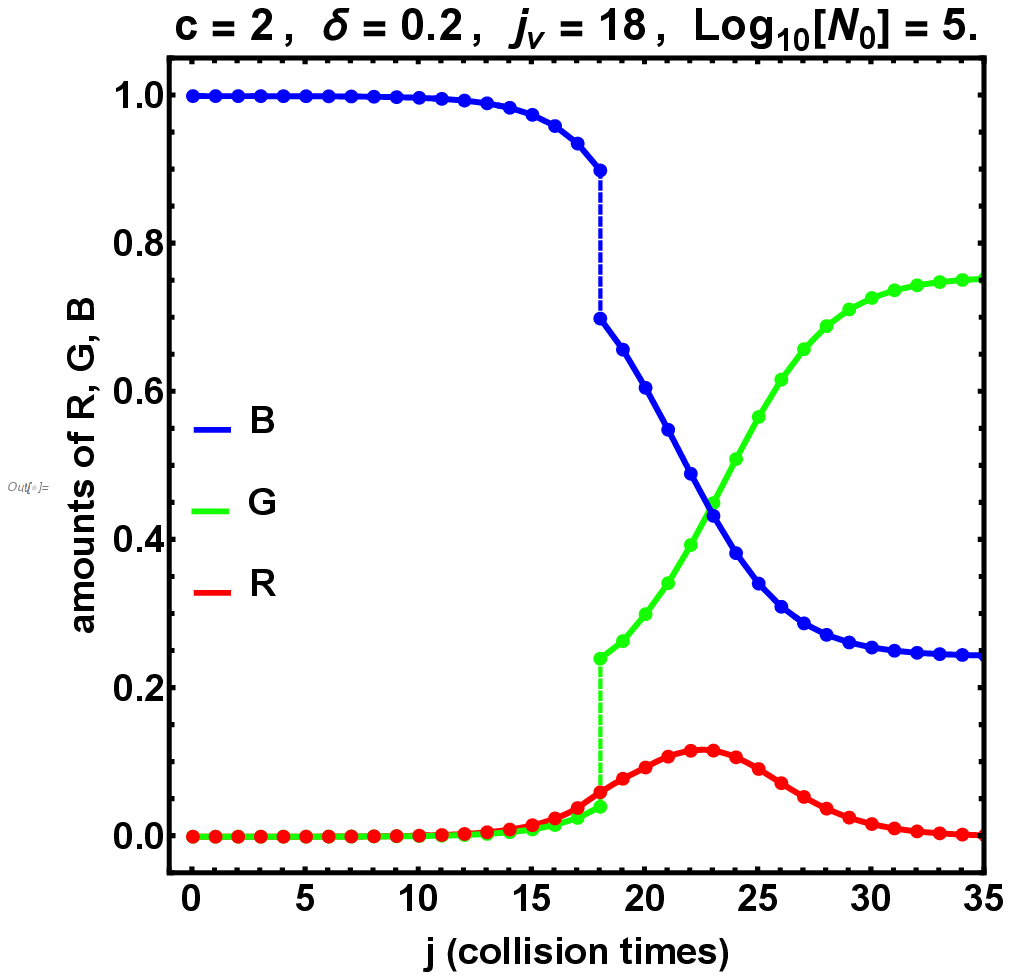}}}
 \hspace{0.01in}
\centerline{%
 \fbox{\includegraphics[bb=88 250 542 716,clip,width=0.37\textwidth]{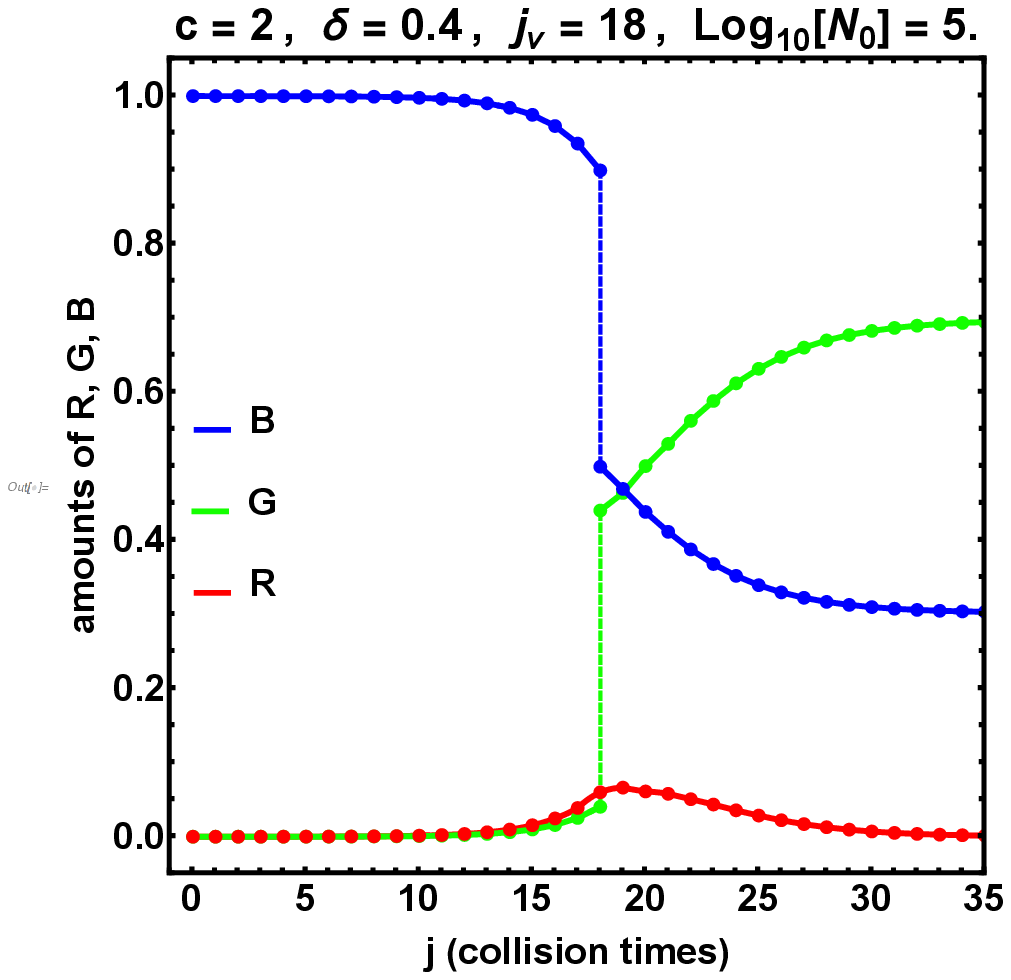}}
\vspace{0.01in}
\fbox{\includegraphics[bb=88 250 542 716,clip,width=0.37\textwidth]{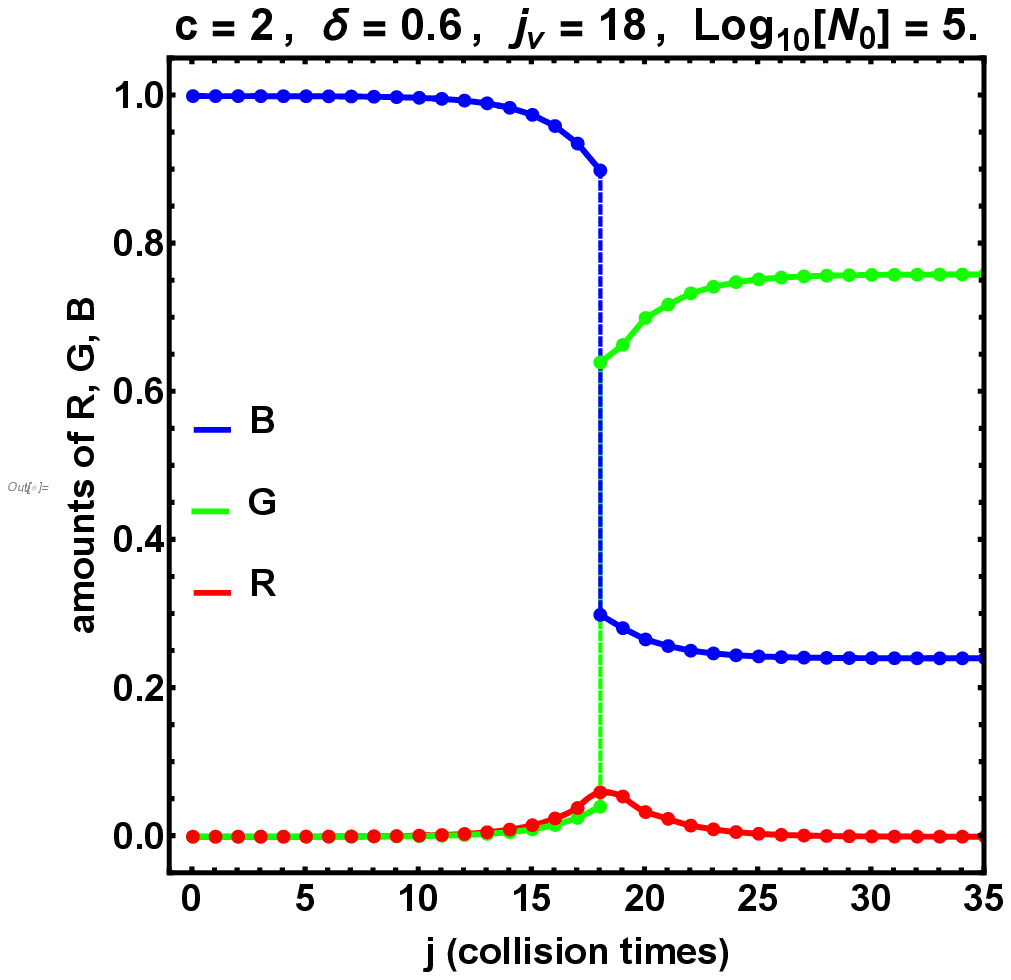}}
}
\caption{Fractions of blue, green, and red molecules versus time for $c=2$ and ${\log}_{10}N_{0}=5$, illustrating the effects of vaccinations on the evolving profiles.  Vertical shifts at $j_{v}=18$ correspond
to vaccinations of ${\delta}=0.2$, 0.4, and 0.6.}
\end{figure}
\noindent
 for a population
$N_{0}=10^5$ molecules and a value of $c=2$.  The final value of $B_{j}$ is $B_{f}=0.16$.  The remaining panels show the effect of applying vaccinations of ${\delta}=0.2$, 0.4, and 0.6 at the time $j_{v}=18$.
As ${\delta}$ increases from 0.2 to 0.4, the number of surviving blue molecules increases
to $B_{f}=0.30$ while the final number of green molecules decreases to 0.70.  As a result, the final
separation of the blue and green curves decreases.  For ${\delta}>0.4$, the effect reverses, as show
by the greater separation of the final green and blue curves when ${\delta}=0.6$.
\medskip

These results are independent of the value of $N_{0}$, provided that ${\nu}_0=1/N_{0}$
can be neglected in Eq.(4.3) for ${\nu}_{j}$ and in Eq.(4.4) for ${\Delta}{\nu}_{j}$.  As we have seen previously, the effect of $N_{0}$ is to shift the curves along the $j$-axis by the amount given by
Eq.(5.10).  This means that to achieve the same result, we would also have to shift $j_{v}$ by the amount
given by Eq (5.10).  Thus, if we used $N_{0}=10^6$ and $c=2$, we would have to use $j_{v}=18+5=23$ for the results to look the same as shown in Figure~38.  Likewise, if we used $N_{0}=10^6$ and $c=3$, we would have to use $j_{v}=18+4=22$.  Thus, for simplicity, we will set $N_{0}=10^5$ and study the effect of changing $j_{v}$ and ${\delta}$, the number of vaccinations expressed as a fraction of the total number of molecules, $N_{0}$.
\medskip

In the upper left panel of Figure~38, the peak of the red curve occurs at $j=23$.  Therefore, the
subsequent panels with $j_{v}=18$ show the effect of giving the vaccinations early in the
rising phase of the pandemic.  As one can see, by increasing ${\delta}$ from 0 to 0.2, the (normalized) number of red molecules decreases after $j_{v}=18$, and the asymptotic value of
the blue curve rises from 0.16 to 0.24.  Similarly, the final number of green molecules decreases
from 0.84 to 0.76, causing the ends of the blue and green curves to lie closer together.  However,
only 0.20 of those 0.76 green molecules are the uninfected product of
vaccinations.  The remaining 0.56 were contagious red molecules that turned green after 2 collision
times.  So by vaccinating 0.20 blue molecules, we caused the final number of uninfected molecules to
increase from the 0.16 blue molecules that would have occurred without vaccinations to the
combination of 0.24 blue molecules and 0.20 vaccinated green molecules, for a total of 0.44
uninfected molecules.  This is a net saving of 0.28 uninfected molecules over the original sample
of 0.16 surviving blue molecules.

Table~2 shows the results for other values of ${\delta}$ and $j_{v}$.  We have defined ${\Delta}B_{f}$ to
be the difference between $B_{f}$ and its value, $B_{f}^{(0)}=0.16$, when ${\delta}=0$.  Thus, ${\Delta}B_{f}$ is the net saving of blue
molecules due to vaccinations.   The quantity, ${\Delta}B_{f}+{\delta}$, is the net saving of all uninfected molecules (blue and green) over the number of surviving blue molecules when there are
no vaccinations.  As $j_{v}$ increases from 18 to 23,
$B_{f}$ and ${\Delta}B_{f}$ reach their maximum values at slightly lower values of ${\delta}$.  Finally, at
$j_{v}=23$, the maximum values of these two quantities occur at ${\delta}=0$.  At this point, the
numbers of surviving blue molecules are greatest in the absence of vaccinations.  Moreover, if
the vaccinations are given after $j_{v}=23$, the vaccinations actually reduce the number of surviving
blue molecules.  However, as mentioned above, these lost blue molecules are replaced by vaccinated
green molecules so that a net saving is always achieved by giving more vaccinations.

\begin{table}[h!]
\caption{Vaccination Trends For $c=2$}
\begin{tabular}{c c c c c}
\hline\hline
$j_{v}$ & ${\delta}$ & $B_{f}$ & ${\Delta}B_{f}=B_{f}-B_{f}^{(0)}$ & ${\Delta}B_{f}+{\delta}$  \\[1.5ex]
\hline\\
18   &    0.0    & 0.16   &    0.00  &    0.00 \\[0.3ex]
       &    0.1     & 0.20  &    0.04  &    0.14 \\[0.3ex]
       &    0.2    & 0.24  &    0.08  &    0.28 \\[0.3ex]
       &    0.3    & 0.28  &    0.12   &    0.32 \\[0.3ex]
       &    0.4    & 0.30  &    0.14  &    0.54 \\[0.3ex]
       &    0.5    & 0.29  &    0.13  &    0.42 \\[0.3ex]
       &    0.9    & 0.00  &  -0.16  &    0.74 \\[0.3ex]
\hline\\
20   &   0.0    & 0.16   &    0.00  &    0.00 \\[0.3ex]
       &    0.1     & 0.19   &    0.03  &    0.13  \\[0.3ex]
       &    0.2     & 0.21  &    0.05  &     0.25 \\[0.3ex]
       &    0.3     & 0.22  &    0.06  &    0.36 \\[0.3ex]
       &    0.4     & 0.21  &    0.05  &     0.45 \\[0.3ex]
       &    0.5     & 0.18  &    0.02  &     0.52 \\[0.3ex]
       &    0.6     & 0.12  &  -0.04  &     0.56 \\[0.3ex]
       &    0.7     & 0.05  &  -0.11   &     0.59 \\[0.3ex]
       &    0.77   & 0.00  &  -0.16  &     0.61 \\[0.3ex]
\hline\\
23	&	0.0    &	0.16  &    0.00  &    0.00	 \\[0.3ex]
	&	0.1 	   &	0.15  &  -0.01  &     0.09	 \\[0.3ex]
	&	0.2    &	0.12  &  -0.04  &    0.16	 \\[0.3ex]
	&	0.3    & 0.07  &  -0.09  &    0.21  \\[0.3ex]
       &	0.4    &	0.02  &  -0.14  &    0.26	 \\[0.3ex]
	&	0.43  & 0.00  &  -0.16  &    0.27  \\[0.3ex]
\hline\\
25	&	0.0    &	0.16  &    0.00  &    0.00	 \\[0.3ex]
	&	0.1 	   &	0.10  &  -0.06  &    0.04	 \\[0.3ex]
	&	0.2    &	0.04  &  -0.12  &    0.08	 \\[0.3ex]
	&	0.25  & 0.00  &  -0.16  &    0.09	 \\[0.3ex]
\hline 
\end{tabular}
\end{table}

This net saving is shown by
the positive values of ${\Delta}B_{f}+{\delta}$ in the last row of each group of $j_{v}$-values in Table~2.  However, these numbers, $(0.74, 0.61, 0.27, 0.09)$, trend toward 0 as the vaccination time is moved
toward the end of the pandemic, and there are fewer blue molecules left to vaccinate.  Conversely, if
$j_{v}$ were taken prior to the start of the pandemic, then the numbers would reach 0.84 at
the beginning of the sequence, corresponding to the difference between 1.00 vaccinated green
molecules and the 0.16 blue molecules that would  have survived without vaccinations. 
Thus, the largest increase of uninfected molecules is obtained by vaccinating all of the blue
molecules, and the smallest increase is obtained by social distancing alone.
\medskip

Vaccinations reduce the number of red molecules after the vaccination time, $j_{v}$.  The cropped
panels in Figure~39 show this effect clearly for $c=2$ and a range of ${\delta}$ from 0 to 
\begin{figure}[ht!]
 \centerline{%
 \fbox{\includegraphics[bb=88 435 542 716,clip,width=0.47\textwidth]{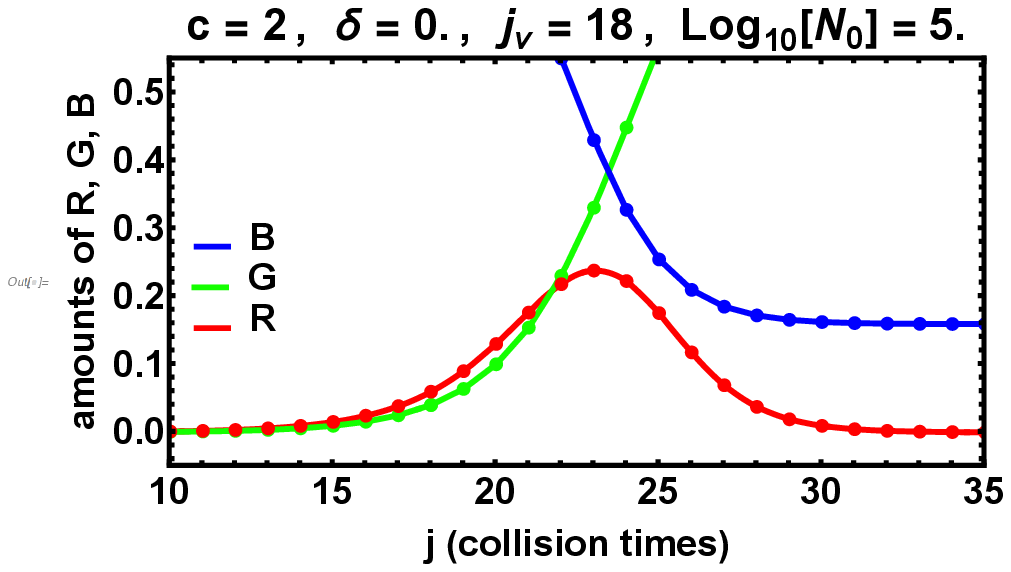}}
\vspace{0.01in}
 \fbox{\includegraphics[bb=88 435 542 716,clip,width=0.47\textwidth]{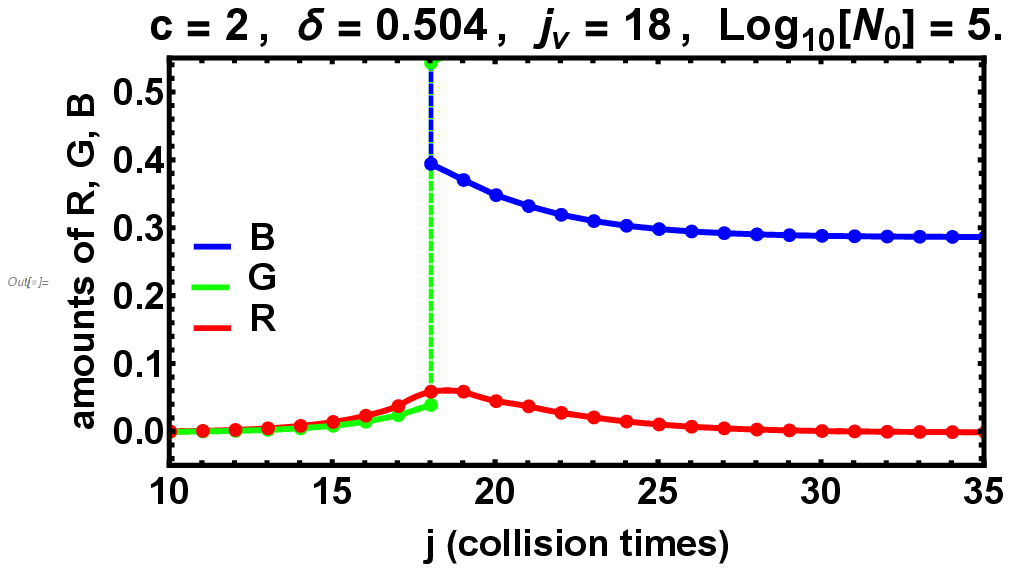}}}
 \hspace{0.01in}
\centerline{%
 \fbox{\includegraphics[bb=88 435 542 716,clip,width=0.47\textwidth]{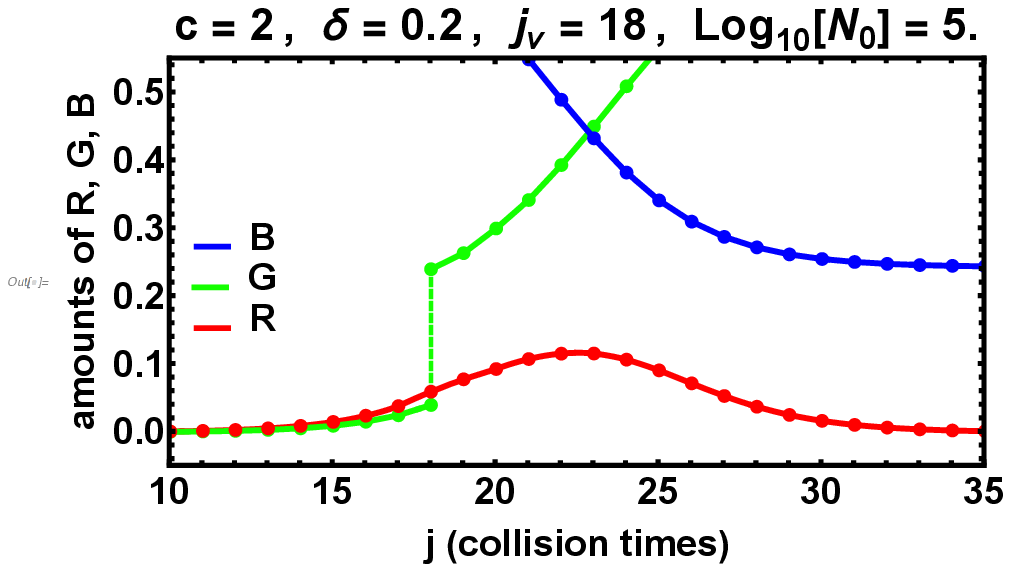}}
\vspace{0.01in}
\fbox{\includegraphics[bb=88 435 542 716,clip,width=0.47\textwidth]{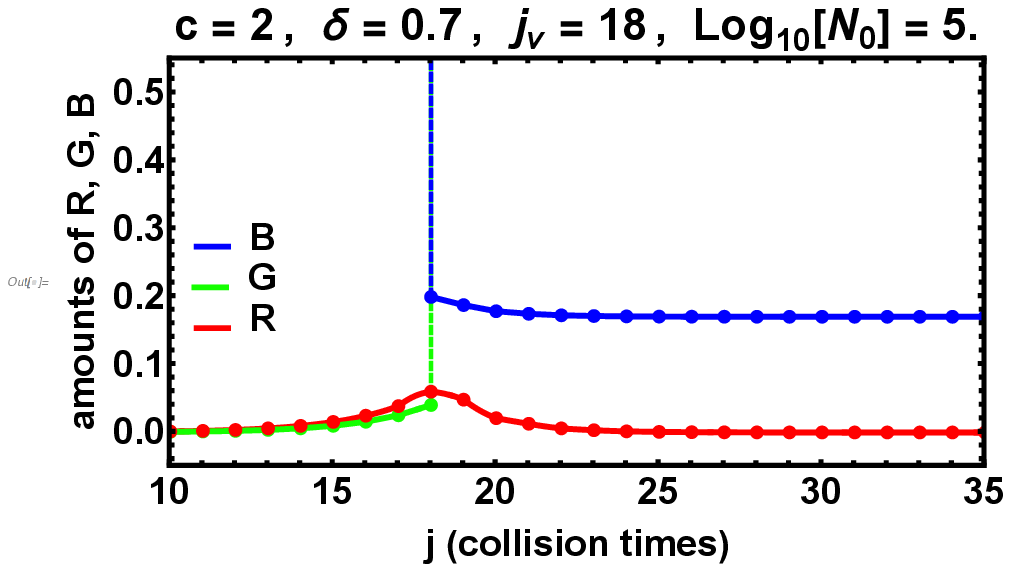}}}
 \hspace{0.01in}
\centerline{%
 \fbox{\includegraphics[bb=88 435 542 716,clip,width=0.47\textwidth]{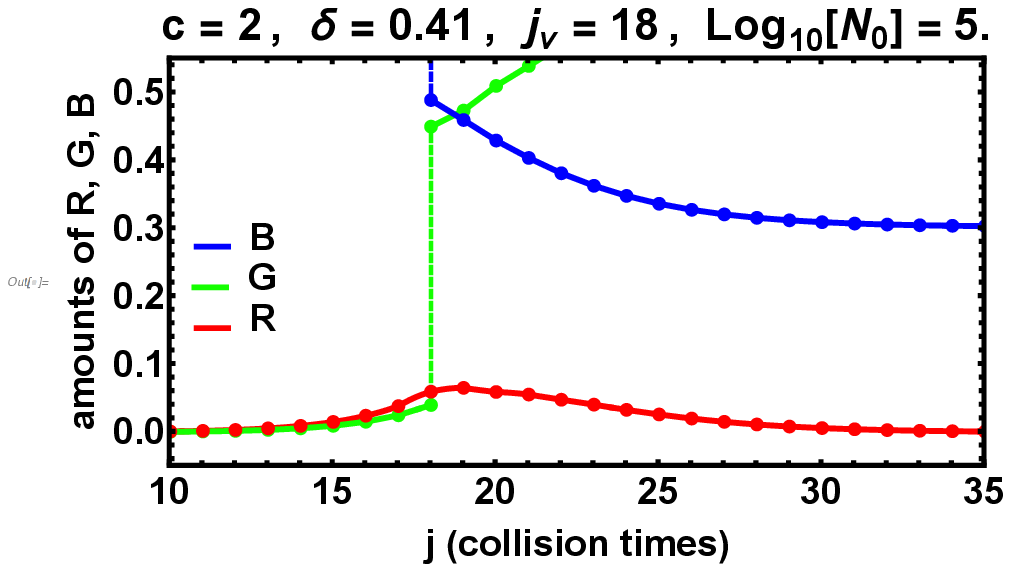}}
\vspace{0.01in}
\fbox{\includegraphics[bb=88 435 542 716,clip,width=0.47\textwidth]{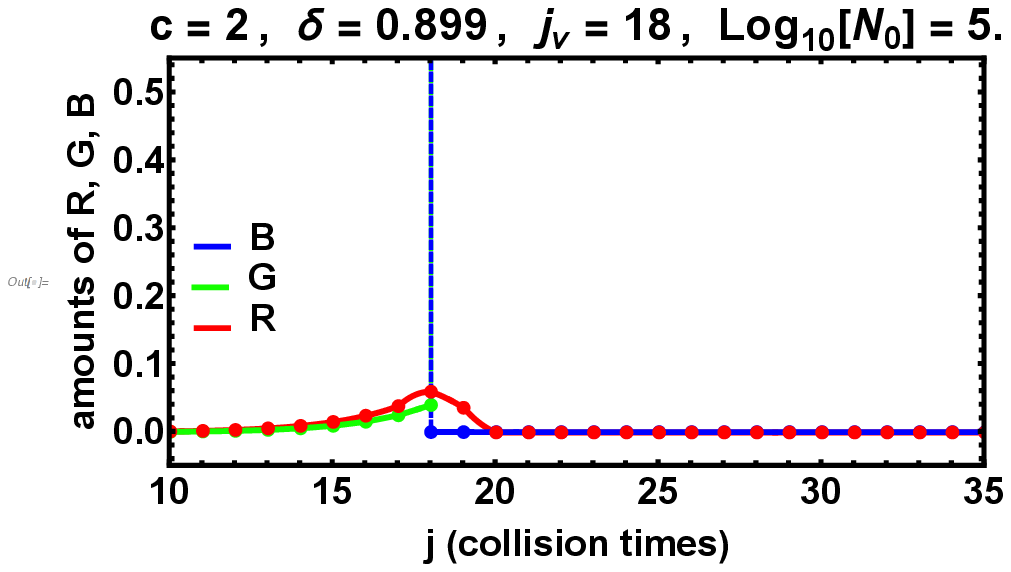}}}
\caption{Fractional numbers of blue, green, and red molecules for $c=2$, showing the
weakening of the red peak beyond $j_{v}=18$ and its shift toward $j_{v}$ for ${\delta}$ in the range $0-0.899$.}
\end{figure}
\noindent
0.899.  The panel with ${\delta}=0.41$ and $B_{jv}=0.5$ (which matches $1/c$) has the largest value
of $B_{f}=0.30$.  In this case, the value of $R_{jv}$ is slightly less than $R_{jv+1}$, which is the peak
of the red curve.  However, in the panel with ${\delta}=0.504$, the red curve flattens out with
$R_{jv}=R_{jv+1}=0.06$.  In this case, $B_{jv}=0.40$, which is significantly less than the value of 0.5 obtained when ${\delta}=0.41$.  (On the other hand,
$B_{f}$ was only about 0.01 smaller when ${\delta}=0.504$ than when ${\delta}=0.40$, so these
significantly different values of ${\delta}$ gave comparable values of $B_{f}$.)  In summary, as
${\delta}$ increased, the peak of the red curve weakened and shifted to lower values of $j$, reaching
$j_{v}$ when ${\delta}=0.504$, and then remaining fixed in both position and height for larger values of
${\delta}$.  At this point, the high-$j$ tail shrinks away as the remaining red molecules turn green and
the pandemic ends.
\medskip

The turning point where $dR{j}/dj=0$ is usually referred to as the herd immunity threshold.  This
condition is equivalent to $dG_{j}/dj=-dB{j}/dj$, which means that the number of blue molecules
decreases at the same rate that the number of green molecules increases.  Intuitively, this is what
we expect when $dR_{j}/dj=0$ because red molecules are produced in collisions with blue ones,
and red molecules are removed by turning green.
\medskip

The values of ${\delta}$ that produce this threshold can be calculated directly from data obtained prior
to administering the vaccinations.  To do this, we begin by rewriting Eq (E5) as
\begin{equation}
B_{jv}~=~\frac{B_{jv}-B_{jv+1}}{R_{jv}},
\end{equation}
and then express Eq (E6) as
\begin{equation}
R_{jv}~=~B_{jv-c}~-~B_{jv}(0).
\end{equation}
Combining these equations, we obtain
\begin{equation}
B_{jv}~=~\frac{B_{jv}-B_{jv+1}}{B_{jv-c}-B_{jv}(0)},
\end{equation}
where $B_{jv}(0)$ is evaluated before the vaccination and $B_{jv}$ is evaluated after the vaccination.
\medskip

The next step is to relate the post-vaccination variables in the numerator of Eq (E13) to
pre-vaccination variables using the herd immunity condition ${\Delta}R=R_{jv+1}-R_{jv}=0$.
Then, because ${\Delta}R+{\Delta}B+{\Delta}G=0$, it follows that
\begin{equation}
B_{jv}-B_{jv+1}~=~B_{jv-c}-B_{jv-c+1}.
\end{equation}
Combining Eqs (E13) and (E14), we obtain an expression for $B_{jv}$ in terms of the
pre-vaccination variables:
\begin{equation}
B_{jv}~=~\frac{B_{jv-c}-B_{jv-c+1}}{B_{jv-c}-B_{jv}(0)}.
\end{equation}
The amount of vaccination, ${\delta}$, is then obtained from the relation
\begin{equation}
B_{jv}~=~B_{jv}(0)-{\delta}.
\end{equation}

Together, Eqs (E15) and (E16) permit an advance determination of how much vaccination
is needed to reach the herd immunity threshold.  Thus, in the upper right panel of Figure~39,
I determined ${\delta}=0.504$ from pre-vaccination variables (obtained from
the calculation with ${\delta}=0$), and used it to make the plot with $R_{jv}=R_{jv+1}$.
Another example is contained in Figure~40, which shows a sequence of cropped plots
like those in Figure~39,
\begin{figure}[ht!]
 \centerline{%
 \fbox{\includegraphics[bb=88 435 542 716,clip,width=0.47\textwidth]{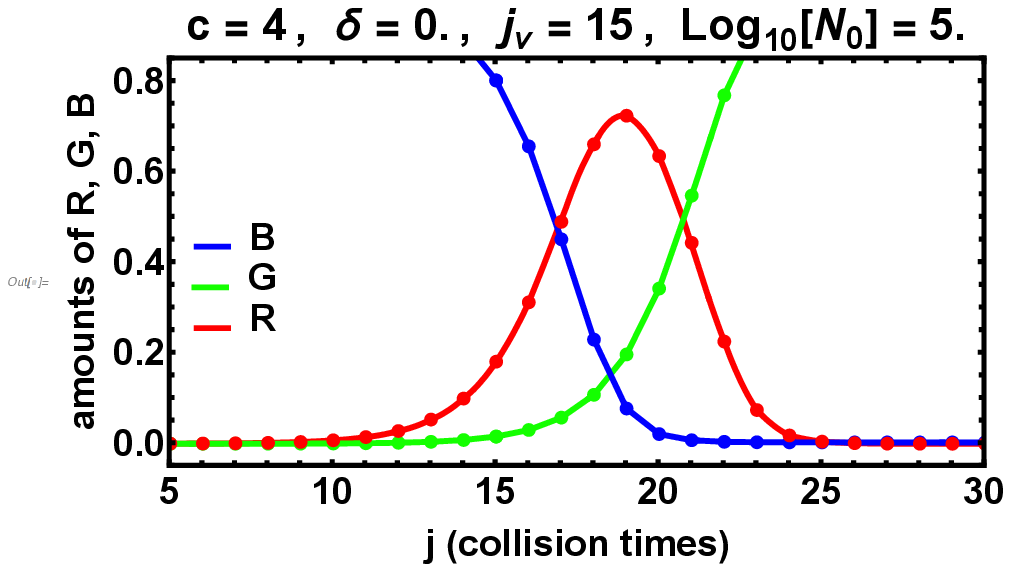}}
\vspace{0.01in}
 \fbox{\includegraphics[bb=88 435 542 716,clip,width=0.47\textwidth]{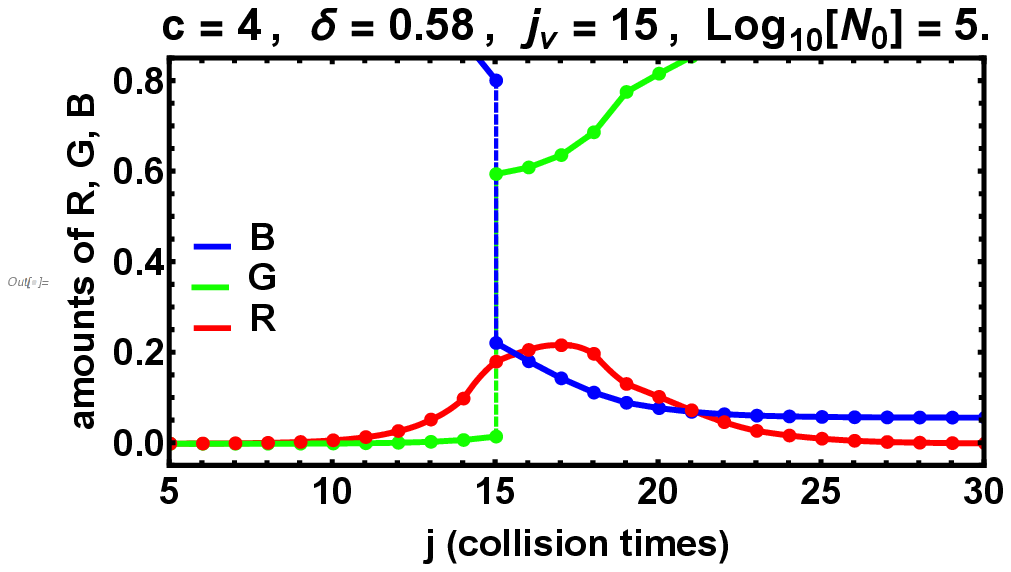}}}
 \hspace{0.01in}
\centerline{%
 \fbox{\includegraphics[bb=88 435 542 716,clip,width=0.47\textwidth]{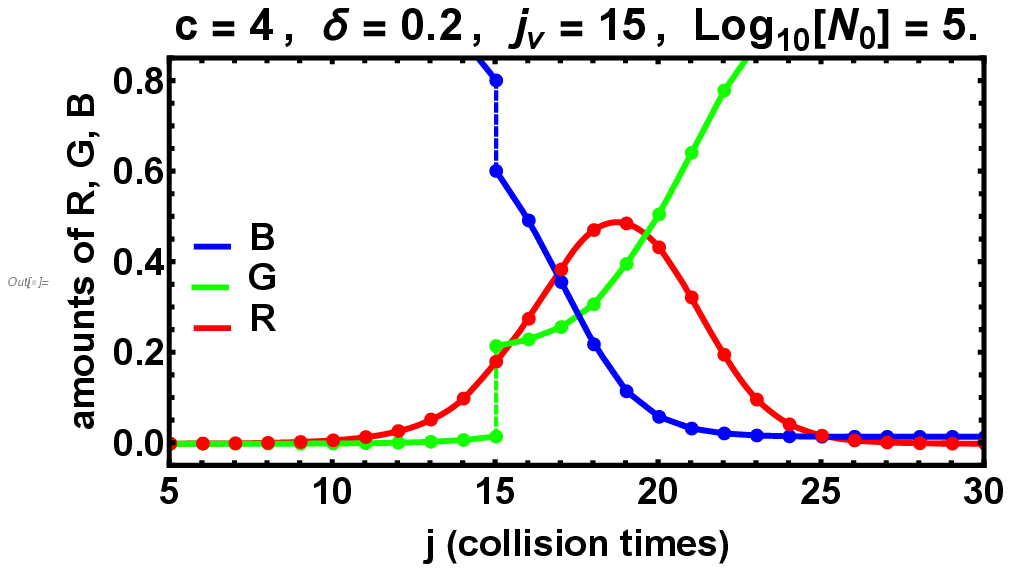}}
\vspace{0.01in}
\fbox{\includegraphics[bb=88 435 542 716,clip,width=0.47\textwidth]{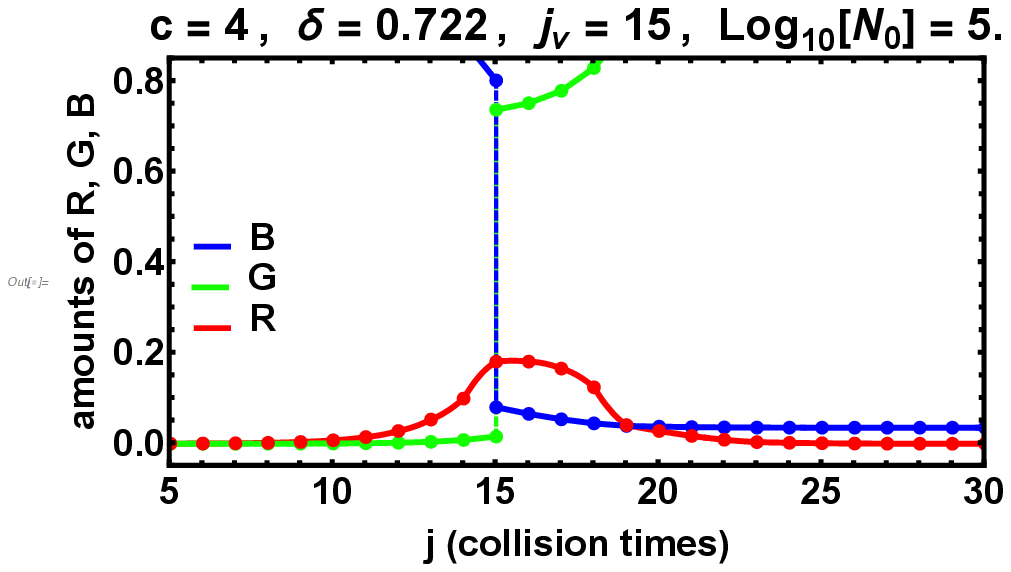}}}
 \hspace{0.01in}
\centerline{%
 \fbox{\includegraphics[bb=88 435 542 716,clip,width=0.47\textwidth]{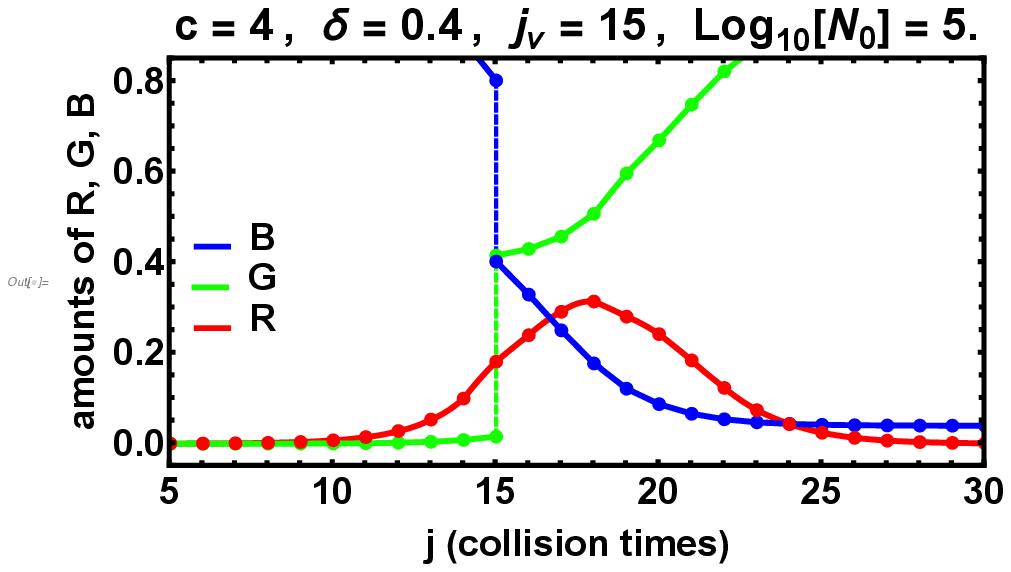}}
\vspace{0.01in}
\fbox{\includegraphics[bb=88 435 542 716,clip,width=0.47\textwidth]{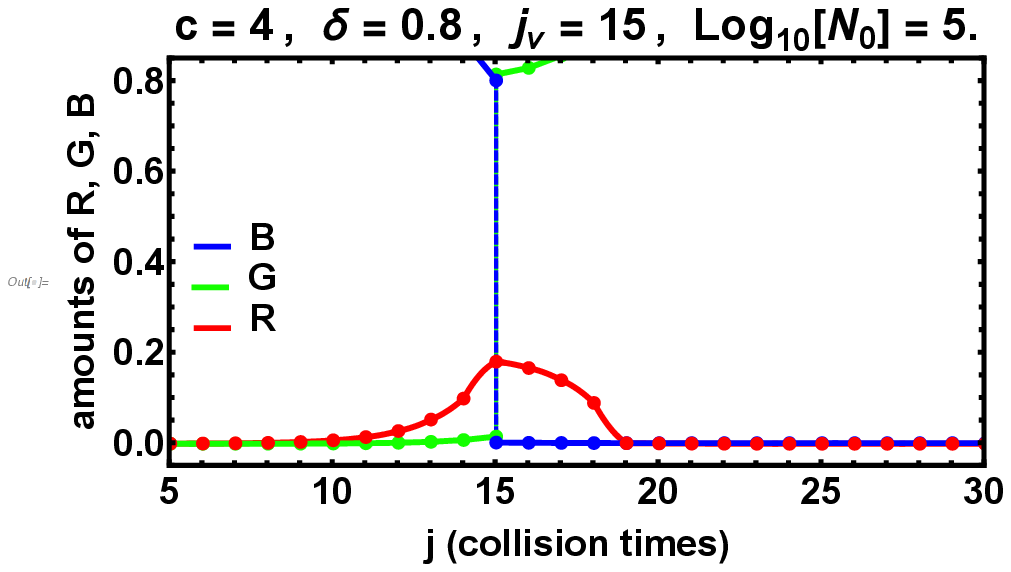}}}
\caption{Similar to Figure~39, except that $c=4$ and $j_{v}=15$, showing the
weakening of the red profile beyond $j_{v}=15$ for values of ${\delta}$ in the range 0-0.8.}
\end{figure}
\noindent
except that $c=4$ and $j_{v}=15$.  These plots with $c=4$ have
larger red peaks than the plots
with $c=2$, and therefore provide a more detailed view of the ${\delta}$-dependence of the red
curves.  In this case, the vaccination with ${\delta}=0.58$ (corresponding to $B_{jv}=0.22$) gives the maximum number of surviving blue molecules, and the vaccination with ${\delta}=0.722$
(corresponding to $B_{jv}=0.08$) gives
the flat-topped profile with $R_{jv}=R_{jv+1}$.  As before, the value of ${\delta}=0.722$
was calculated from pre-vaccination data obtained from the calculation with ${\delta}=0$.
\medskip
 
Although this technique is successful for determining the amount of vaccination needed to
reach the herd immunity threshold, it does not maximize $B_{f}$.  This can be seen in
Figure~41, which compares calculations obtained with incrementally different values of
${\delta}=0.58$ and 0.61 (left panel) and ${\delta}=0.722$ and 0.752 (right panel).  The
value of ${\delta}=0.58$ gives the maximum value of $B_{f}=0.06$ and the value of
${\delta}=0.722$ gives the flat-topped profile corresponding to the herd immunity threshold.
The coalescence of the blue curves in the left panel shows that $dB_{f}/d{\delta}=0$,
corresponding to a maximum number of surviving blue molecules.   The separation of the
blue curves in the right panel shows that $dB_{f}/d{\delta}<0$, corresponding to a smaller
number of surviving blue molecules on the downward leg of the $B_{f}({\delta})$ profile.  The
corresponding values of $B_{jv}$ are 0.22 (left panel) and 0.08 (right panel), illustrating that the
condition for maximum surviving blue molecules occurs at a larger value of $B_{jv}$
(and therefore smaller amount of vaccination, ${\delta}$) than the condition for the herd immunity
threshold.
\begin{figure}[ht!]
 \centerline{%
 \fbox{\includegraphics[bb=88 250 542 716,clip,width=0.47\textwidth]{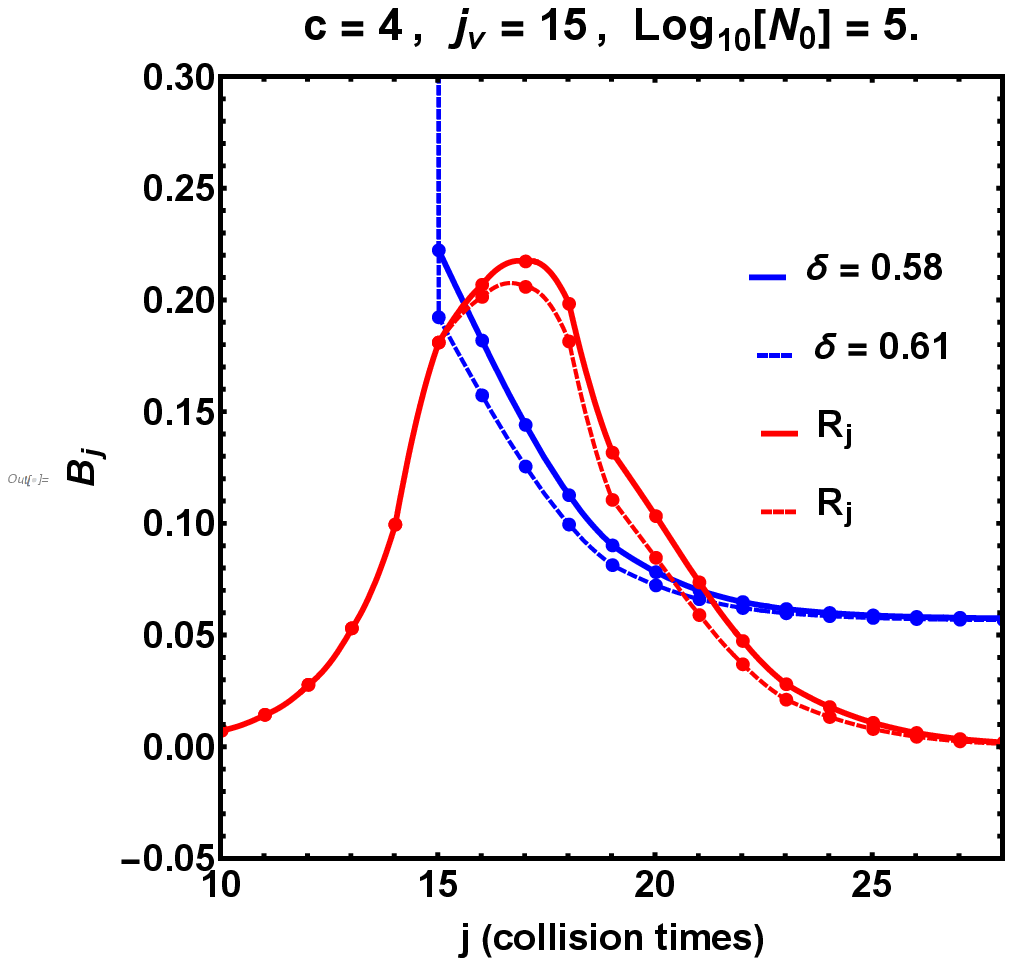}}
\vspace{0.01in}
 \fbox{\includegraphics[bb=88 250 542 716,clip,width=0.47\textwidth]{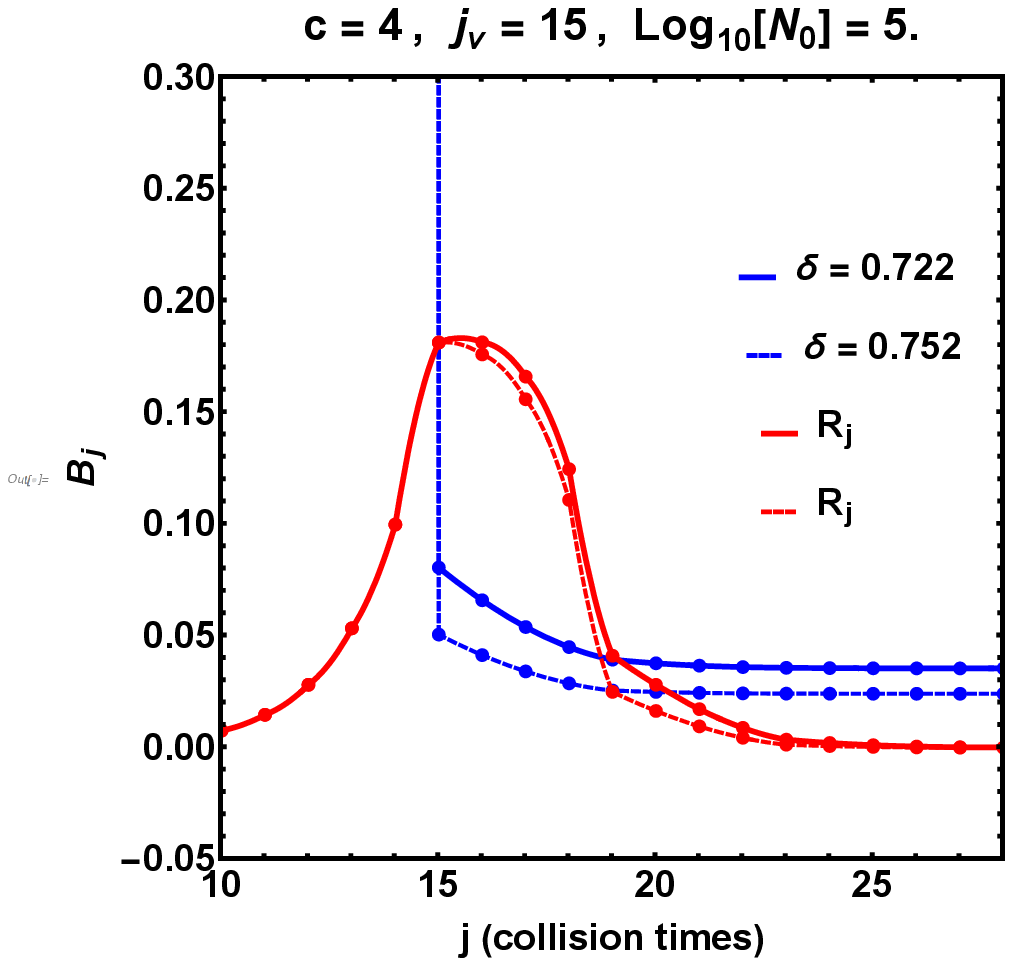}}}
\caption{Fractional numbers of red and blue
molecules when ${\delta}=0.58$ and 0.61 (left panel) and ${\delta}=0.722$ and 0.752 (right panel), showing different results when $B_{f}$ is a maximum and when $R_{jv+1}=R_{jv}$, respectively.   The
incrementally spaced curves show that $dB_{f}/d{\delta}=0$ when ${\delta}=0.58$ (left panel) and
$dB_{f}/d{\delta}<0$ when ${\delta}=0.722$ (right panel).}
\end{figure}

Even though the vaccination of ${\delta}=0.58$ produced more surviving blue
molecules than the vaccination of ${\delta}=0.722$ (0.06 compared to 0.035), this saving did
not nearly make up for the difference in the number of vaccinated molecules (0.58 compared
to 0.722).  Thus, there was a net saving of only 0.64 at the $B_{f}$ maximum compared to
0.76 at the herd immunity threshold.  (For $c=4$, the number of non-vaccinated blue survivors is
essentially 0 based on Eq (E6) and Figure~8, and therefore can be neglected in estimating the
net number of surviving molecules.) 
\medskip

Although I have not yet found an analytical expression for the value of $B_{jv}$ (and therefore
${\delta}$) that maximizes $B_{f}$, I have used Eqs (E2-4) and equivalently Eqs (E5-7) to
calculate $B_{f}$ as a function of $B_{jv}$ for a range of parameters, $j_{v}$.  The results are
plotted in Figure~42 for $c=2$, $\log_{10}(N_{0})=5$, and $j_{v}=09-23$.  In this figure, each
\begin{figure}[h!]
 \centerline{%
 \fbox{\includegraphics[bb=88 246 552 720,clip,width=0.80\textwidth]{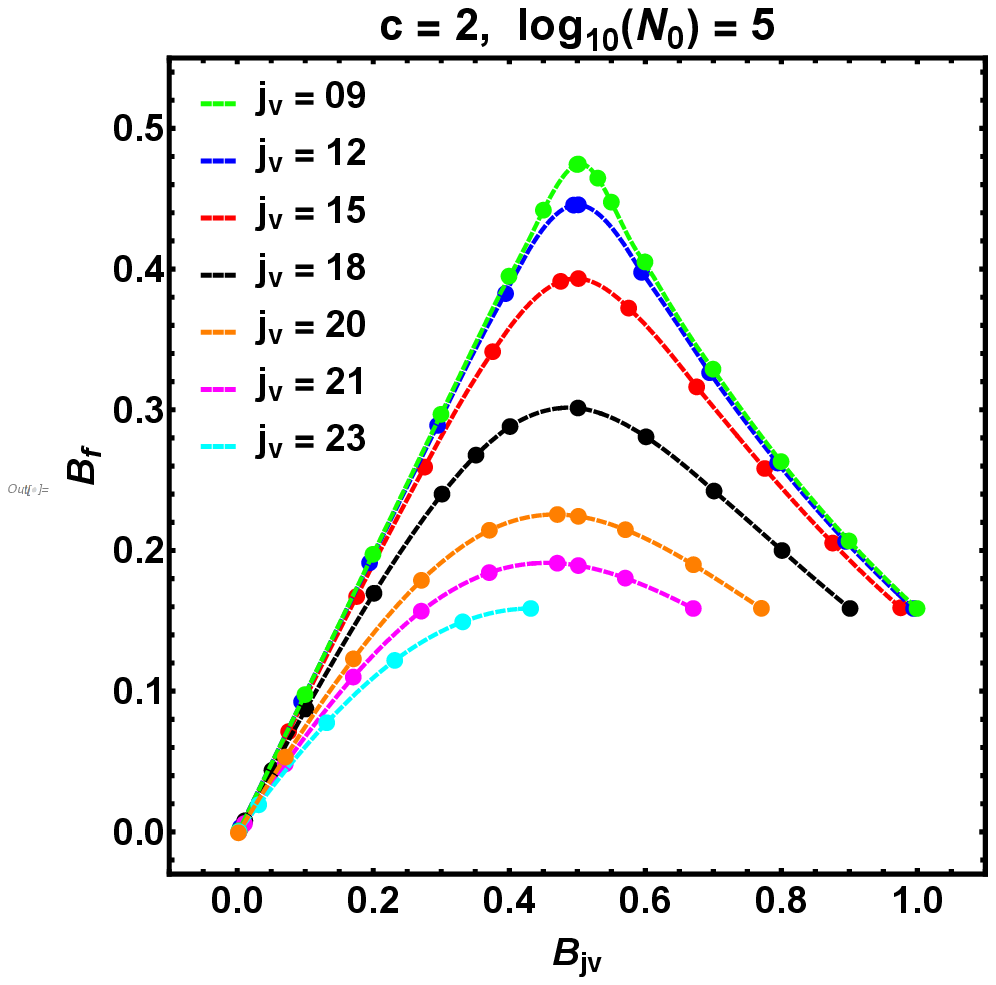}}}
 \caption{$B_{f}$ plotted versus $B_{jv}$ for values of $j_{v}$ in the range $09-23$, showing the
progression from sharp peaks at $B_{jv}=0.5$ when $j_{v}$ is small to broader and flatter peaks
at slightly smaller values of $B_{jv}$ as $j_{v}$ increases.}
\end{figure}
\noindent
curve refers to a set of points that were calculated for a given value of $j_{v}$ and fit by a dashed
line.  For each track, I usually calculated the points for evenly space values of ${\delta}=0$, 0.1, 0.2,
and so on.  However, I was not consistent in this choice of points.  Sometimes I added
an additional point
at $B_{jv}=0.5$ when the peak did not occur there.  Also, for the track with $j_{v}=09$, I added some
extra points near the sharp peak so that the fitting procedure would provide a smooth curve.  In most
cases, the reader can recognize such additions from their departure from even spacing along the curve.
So with this caveat, the plotted points indicate the amount of vaccination, ${\delta}$, as well as the
value of $B_{jv}$ along the curves.
\medskip

Another way to understand this plotting is from Eq (E16), whose form
$B_{jv}=B_{jv}(0)-{\delta}$ relates the value of $B_{jv}$ after vaccination to the amount
of vaccination ${\delta}$.  Thus, for each track in Figure~42, $j_{v}$ and $B_{jv}(0)$ are constants and
$B_{jv}$ decreases as ${\delta}$ increases.  Also, when ${\delta}$ reaches its maximum value of
$B_{jv}(0)$,  $B_{jv}$ will vanish and all of the available blue molecules will have been vaccinated.
For example, the black track with $j_{v}=18$ has $B_{jv}(0)=0.90$ so that $B_{jv}=0.90-{\delta}$.
As one moves upward from right to left along the black track, the points indicate values of
${\delta}=0$, 0.1, 0.2, 0.3, 0.4, ..., corresponding to values of
$B_{jv}=0.90$, 0.80, 0.70, 0.60, 0.50, ... on the horizontal axis.
\medskip

Thus, each curve starts at the same value of $B_{f}=0.16$ when ${\delta}=0$.  As ${\delta}$ increases in
steps of 0.1, the value of $B_{jv}$ decreases, and the value of $B_{f}$ increases as the curve
rises toward its maximum.  Also, as $j_{v}$ changes from 09 to 23, the peaks become lower and flatter
and their peak locations drift from $B_{jv}=0.50$ to $B_{jv}=0.43$.  The curve with $j_{v}=23$ starts at its maximum and bends downward as the vaccination is increased.  Although not shown here, curves with
$j_{v}>23$ would begin with a negative slope and look very similar to the converging ends of the tracks
plotted with $j_{v}<23$.
\medskip

In addition to the points at evenly spaced values of ${\delta}$, I have plotted points at $B_{jv}=0.5$
for the tracks with $j_{v}=20$ and 21.  This is to emphasize that the peaks of those curves lie
farther to the left at locations where $B_{jv}<0.50$. The curve with $j_{v}=23$ provides an extreme
case of this shift, in which the peak occurs at the start of the curve where $B_{jv}=B_{jv}(0)=0.43$
and ${\delta}=0$.
\medskip

Figure~43 shows the dependence of $B_{f}$ on $B_{jv}$ (and therefore ${\delta}$) when $c=3$.
In this case, the vaccinations were given for values of $j_{v}$ in the range $09-19$.  Compared to
Figure~42 where $c=2$, the peaks
are lower and occur at smaller values of $B_{jv}$, corresponding to larger values of ${\delta}$.
All of the tracks begin close to $B_{f}=0.025$, as Eq.(E8) predicts for $c=3$.  Also, as in Figure~42,
the points are evenly spaced with ${\delta}=0$, 0.1, 0.2, and so on, except for some points
that can usually be recognized by their deviation from constant spacing.  For $c=3$, the tracks
begin sharply peaked at $B_{jv}=0.33$ and become broader, lower, and shifted to smaller values of
$B_{jv}$ as $j_{v}$ increases from 09 to 19.  Unlike the final track in Figure~42, the final track in
Figure~43 does not begin at its peak, which occurs at $B_{jv}=0.25$ where
${\delta}~{\approx}~0.05$.  The reason for this difference is that for $c=3$ the red peak occurs at
a fractional value between $j_{v}=19$ and $j_{v}=20$.  Thus, the track that begins with zero slope
would be the one with $j_{v}~{\approx}~19.5$, and not the one with $j_{v}=19$, as shown.
\begin{figure}[h!]
 \centerline{%
 \fbox{\includegraphics[bb=88 240 552 720,clip,width=0.80\textwidth]{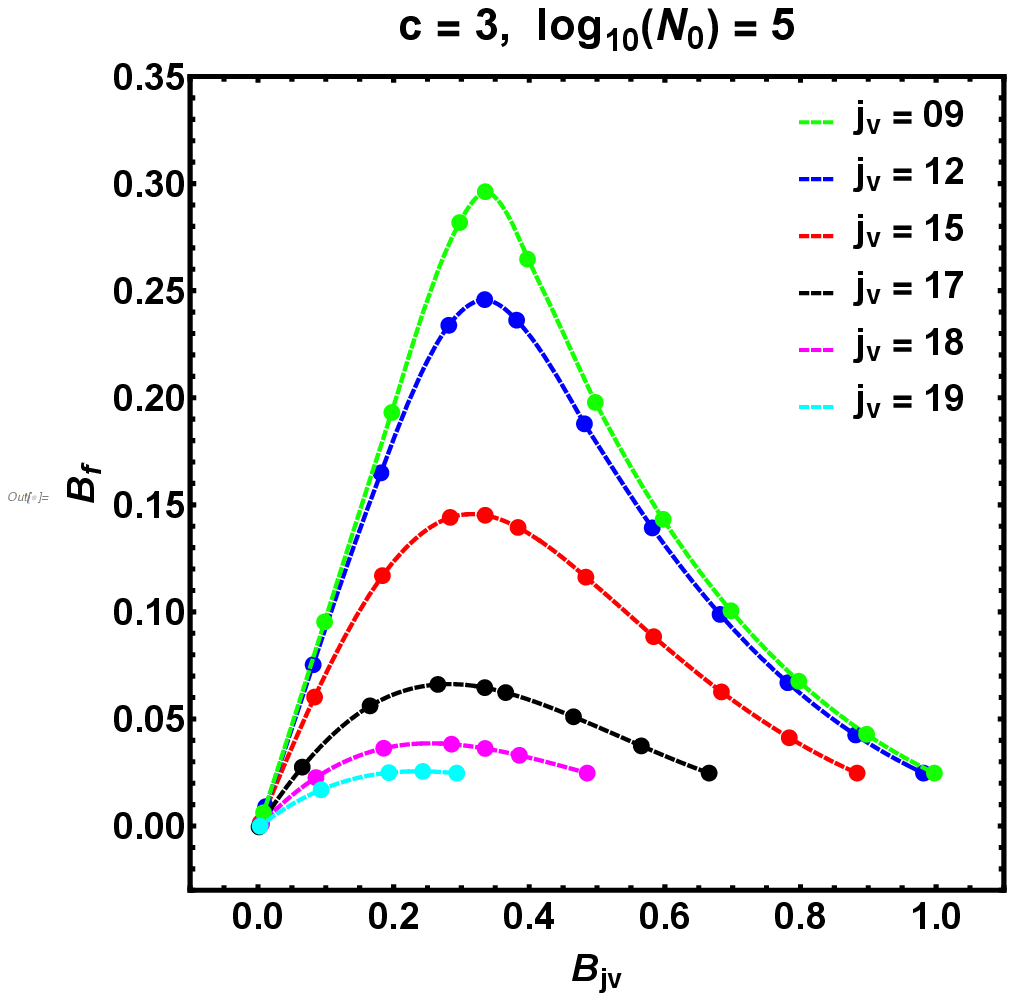}}}
 \caption{$B_{f}$ plotted versus $B_{jv}$ for values of $j_{v}$ in the range $09-19$, showing the
progression from sharp peaks at $B_{jv}=0.33$ when $j_{v}$ is small to broader and flatter peaks
at slightly smaller values of $B_{jv}$ as $j_{v}$ increases.}
\end{figure}

Figures~42 and 43 show that the peaks start at $B_{jv}=1/c$ and then gradually flatten and shift
to smaller values of $B_{jv}$ as $j_{v}$ becomes larger.  We can obtain some understanding of
this relation by returning to Eq (E13) and expressing the numerator and denominator in terms
of equivalent slopes of the blue curve.  In this case, Eq (E13) becomes
\begin{equation}
B_{jv}~=~\frac{1}{c}~\frac{B'_{jv+1/2}}{<B'_{jv-c/2}>},
\end{equation}
where the brackets indicate an average over the interval ($j_{v}-c$, $j_{v}$).
\medskip

Here, we think of the pre-vaccination slope as a constant, and ask how the post-vaccination slope
changes with the amount of vaccination, ${\delta}$.  For little to no vaccination, the post-vaccination
slope is larger than the pre-vaccination slope because the pandemic proceeds almost unabated and
the blue curve continues to steepen.  Consequently, $B_{jv}>1/c$, corresponding to the initial
rise of the tracks in Figures~42 and 43 (reading right to left).  However, as the amount of
vaccination increases, the post-vaccination slope decreases because the red molecules have
more difficulty finding the progressively smaller number of blue molecules.  Eventually, the
post-vaccination slope will equal the pre-vaccination slope and $B_{jv}$ will equal 1/c, corresponding
to the peak of the track, at least when $j_{v}$ is sufficiently small.  For more vaccination, the
post-vaccination slope will increase further and $B_{jv}$ will become less than 1/c.
\medskip

Referring again to Figures~42 and 43, we can see that the plot with the lowest value of $j_{v}=09$
seems to be an envelope surrounding the other plots.  After it reaches its maximum (again reading
right to left), $B_{f}$ seems to be given approximately by $B_{f}=B_{jv}$ whose ${\delta}$-dependence
is $B_{f}({\delta})=B_{jv}(0)-{\delta}$.  This is the dependence that each track seems to
approach in its declining phase as the remaining blue molecules become vaccinated.
\medskip

Also, because small values of $j_{v}$ correspond to vaccinations near the start of the pandemic
where $B_{jv}(0)~{\approx}~1$, the limiting values of $B_{f}$ and $B_{jv}$ at the top
of the curves in Figures~42 and 43 seem to approach the approximate relation
\begin{equation}
B_{jv}~=~B_{f}~=~1-{\delta}~=~\frac{1}{c}.
\end{equation}
In this case, the amount of vaccination ${\delta}=1-1/c$ gives the maximum possible value of $B_{f}$.
Thus, in our RGB-model, the vaccination of a social-distancing population will cause the final fraction of surviving blue molecules to range from $e^{-1.890(c-1)}$ when ${\delta}=0$, to $1/c$ when
${\delta}=1-1/c$ and the vaccination is applied near the start of the pandemic.
\medskip

Next, we consider what happens when the vaccination is applied when the pandemic has
progressed to its herd immunity threshold.  Figure~44 shows the evolution of the red and blue
molecules when $c=2$, ${\delta}=0$, and $j_{v}=23$.  In this case, $j_{v}=23$ is the time that
the fraction of red molecules reaches its peak and corresponds to the herd immunity threshold
of the unvaccinated population.  The solid red and blue curves mark the progress of the
pandemic without vaccinations, and the dashed curves refer to the progress when a very small
amount of vaccination, ${\delta}=0.05$, is applied at the time $j_{v}=23$.  By increasing ${\delta}$
from 0 to 0.05, the final value of $B_{j}$ is hardly changed, implying that $dB_{f}/d{\delta}=0$ at
this time.  Thus, when $j_{v}=23$, the unvaccinated value of $B_{f}=0.16$ is a maximum value,
and a further increase in the amount of vaccination at this time would only decrease the fraction
of surviving blue molecules.
\medskip

The same would be true for vaccinations with $j_{v}>23$, except that $dB_{f}/d{\delta}$ would be
negative instead of 0.  Consequently, the values of $B_{f}({\delta})$ would decrease rapidly from their
maximum value of 0.16, and approach the linear variation given by
$B_{f}({\delta})=B_{jv}({\delta})=B_{jv}(0)-{\delta}$ as $j_{v}$ becomes larger than 23, as discussed
above in relation to Figures~42 and 43.
\begin{figure}[h!]
 \centerline{%
 \fbox{\includegraphics[bb=85 235 550 725,clip,width=0.80\textwidth]{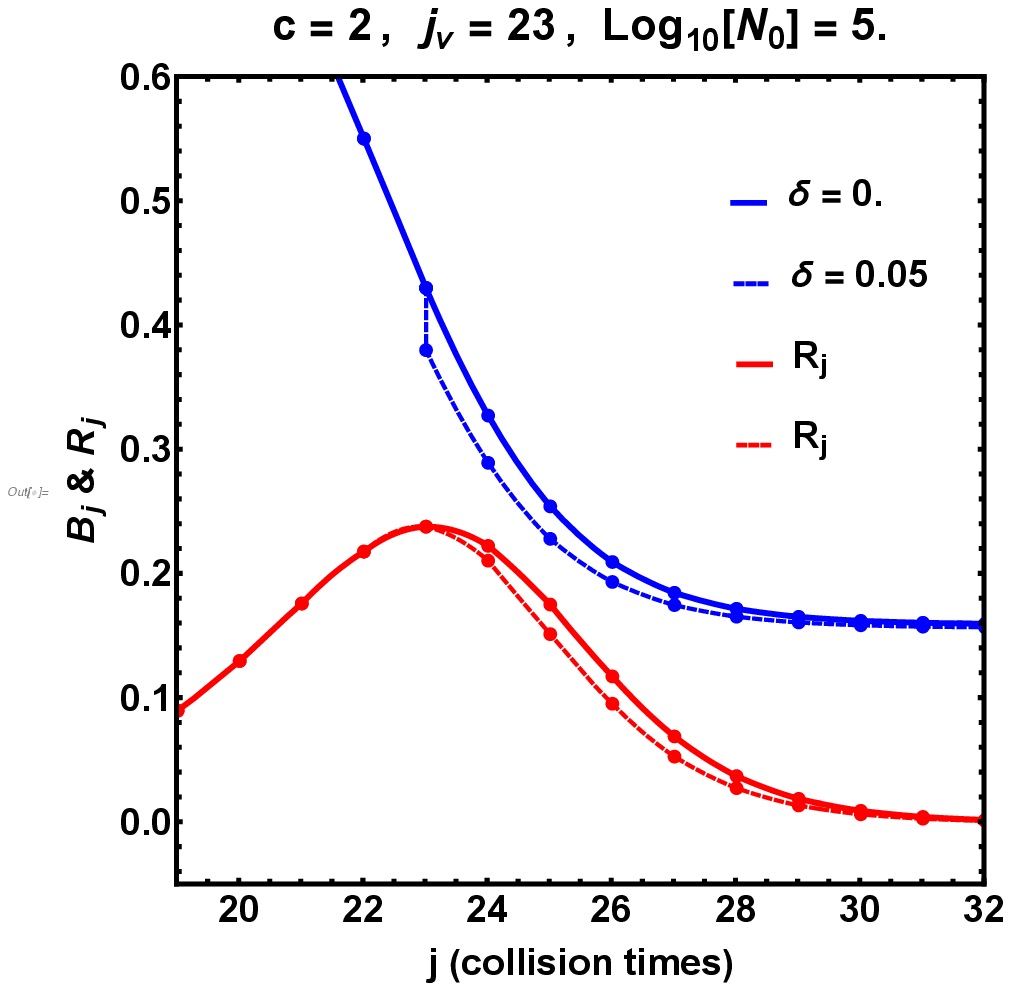}}}
 \caption{The evolution of red and blue molecules for ${\delta}=0$ and $c=2$ when $R_{j}$ has its maximum value at $j=23$ (solid curves), and similar plots for ${\delta}=0.05$ and $j_{v}=23$ (dashed
curves).  The coalescence of the blue curves for large $j$ implies that $dB_{f}/d{\delta}=0$.  Thus, $B_{f}$ is already a maximum, and further vaccinations would only decrease its value.}
\end{figure}

\subsection{Summary}
In the absence of vaccinations, the evolution of the pandemic depends on the amount of social distancing
as indicated by the value of the contagious parameter, $c$.  If $c={\infty}$, then the red molecules
remain contagious forever.  Consequently, they will eventually infect all of the blue molecules, so
that only red molecules are left.  If $c~{\gtrsim}~10$, then red molecules will remain contagious long
enough to infect all of the blue molecules before turning green, and after another 10 or more collision
times, even those newly infected molecules will turn green.
\medskip

If $c<10$, the red molecules will not remain contagious long enough to infect all of the blue molecules.
As time passes, the growing number of green molecules will reduce the probability
that red molecules will find blue ones before losing their contagiousness and turning green.  When
this happens, the pandemic will end, leaving only green molecules and some blue survivors at
the end of the pandemic.  In this case, the fraction of surviving blue molecules is related to the
social-distancing parameter, $c$, by the approximate formula $B_{f}=e^{-1.890(c-1)}$ given
by Eq (E9).  This means that $B_{f}$ is relatively small unless $c~{\lesssim}~2$, in which
case, $B_{f}~{\gtrsim}~0.15$.  Now we can see that the pandemic would be ended quickly, leaving
almost all of the blue molecules, if the social distancing were great enough.  In particular, if
$c$ were 1 or less, all of the red molecules would lose their contagiousness and turn green before
finding a blue molecule, and the pandemic would be over.  
\medskip

Vaccinations will accelerate this process.  Significantly more blue molecules will survive if
many other blue molecules are vaccinated, as they could be during the rising phase of the
pandemic.  This is because the vaccination turns blue molecules green directly without
sending them through the red phase of being infected.  By suddenly reducing the number of
blue molecules and increasing the number of green molecules, the vaccination reduces the
probability that red molecules can find (and infect) blue ones.  This would shorten the pandemic
and leave more surviving blue molecules at the end of the pandemic (unless very large vaccinations
started cutting into the blue survivors).  We found that the maximum fraction of surviving blue
molecules could be as large as $B_{f}=1/c$ if a vaccination of ${\delta}=1-1/c$ were applied near
the start of the pandemic where $B_{jv}(0)~{\approx}~1$.
\medskip

Of course, it may not be possible to begin vaccinations right away, so that we must
consider the effect of vaccinations during the rising phase of the pandemic.  In this case,
our simulations and analysis showed that the fraction of surviving blue molecules initially
increased with the amount of vaccinating.  At the same time, the peak in the number
of red molecules weakened and shifted from its initial location toward the time, $j_{v}$, of 
the vaccinations.
\medskip

As the amount of vaccinating increased, two points of interest occurred.
The first point was the place that the number of blue survivors reached its maximum value.  This
occurred where $B_{jv}~{\lesssim}~1/c$.  Consequently, $B_{jv}$ is large (and ${\delta}$ is small)
when $c$ is small, so that less vaccinating is required to reach this first point when there is a lot
of social distancing (low $c$) than when there is not.  As the amount of vaccinating increased,
$B_{f}$ started to decrease slowly.  But before it fell very
far, the profile of the number of red molecules flattened out so that $R_{jv}=R_{jv+1}$.  This is the
second point of interest and corresponds to the herd immunity threshold where red molecules
turn green with age at the same rate that they are being created by collisions with blue molecules.
\medskip

As the number of vaccinations increased further, the number of red molecules beyond $j_{v}$
continued to decrease, so that the point at $j_{v}$ became a fixed peak in the profile of the
number of red molecules.  Also, the number of surviving blue molecules continued to decrease
because those potential survivors were being vaccinated.  Of course, if every blue molecule were vaccinated, there would be no surviving blue molecules.  However, their loss would be more than offset
by the number of vaccinated green molecules, and the total number of uninfected survivors would be
largest when every molecule is vaccinated.
\medskip

Finally, if the vaccinations are given when $j_{v}$ is at the peak of the pandemic, there will be no
gain in the number of surviving blue molecules.  At this time, $B_{f}$ is maximum without vaccinations
(and will decrease quadratically with the amount of vaccination) and $R_{jv}$ is also maximum,
corresponding to the herd immunity threshold.  If the vaccinations are given during the declining
phase of the pandemic, then $B_{f}$ is also maximum, but $dB_{f}/d{\delta}<0$ so that $B_{f}$ will
decrease more steeply with the amount of vaccination, and $B_{jv}$ will remain the peak of
an increasingly distorted profile of red molecules.  In conclusion, a vaccination will provide the
greatest benefit to the total population if it is applied early in the pandemic and with a strength
that maximizes $B_{f}$ and preferably reaches the herd immunity threshold.

\end{appendix}

\section*{Acknowledgements}
I am grateful to Dr. Stephen Kennedy (Mathematical Association of America Press) and Dr. Pete Riley
(Predictive Science, Inc) for helpful comments related to the publication of an earlier version of this
manuscript.

\clearpage
\begin{center}
\section*{References}
\end{center}

\noindent
Anderson, R. M. and May, R. M. (1979), `Population Biology of Infectious Diseases: Part I', Nature 280, 361-367.

\noindent
Berra, Yogi, (1998), `The Yogi Book', Workman Publishing Co., New York

\noindent
Delamater, P. L., Street, E. J., Leslie, T. F., Yang, Y., \& Jacobsen, K. H., (2019). Complexity of the Basic Reproduction Number (R0). Emerging Infectious Diseases, 25(1), 1-4. https://dx.doi.org/10.3201/eid2501.171901.

\noindent
Fine, Paul; Eames, Ken; and Heymann, David L., (2011), ` `Herd Immunity': A Rough Guide', Invited article on vaccines, Stanley Plotkin (ed.) in Clinical Infectious Diseases 2011; 52 (7):911-916, Oxford Univ. Press.

\noindent
Huppert, A. and G. Katriel, 2013, `Mathematical modeling and prediction in infectious disease epidemiology',
Clinical Microbiology and Infection, Vol. 19, Issue 11, 999-1005.

\noindent
Jones, D. S. and Sleeman, B. D. (1983), Ch 14 in `Differential Equations and Mathematical Biology', Allen {\&} Unwin, London.

\noindent
Kermack, W. O. and McKendrick, A. G., 1927, `A Contribution to the Mathematical Theory of Epidemics.'
Proc. Roy. Soc. Lond. A 115, 700-721, 1927.

\noindent
Leighton, R. B., 1964, `Transport of Magnetic Fields on the Sun', Astrophys. J. 140, 1547.

\noindent
Ying Liu, Albert A Gayle, Annelies Wilder-Smith, Joacim Rocklov, 2020, The reproductive number of
COVID-19 is higher compared to SARS coronavirus, Journal of Travel Medicine, Volume 27, Issue 2,
March 2020, taaa021, https://doi.org/10.1093/jtm/taaa021

\noindent
Sanche, S., Lin, Y., Xu, C., Romero-Severson, E., Hengartner, N., \& Ke, R., 2020. High Contagiousness and Rapid Spread of Severe Acute Respiratory Syndrome Coronavirus 2. Emerging Infectious Diseases, 26(7), 1470-1477. https://dx.doi.org/10.3201/eid2607.200282.

\noindent
Smith, D. A. and Moore, L. C., (1996), in `Calculus: Modeling and Application',  D. C. Heath and Co., Lexington MA.

\noindent
Wang, Y. -M. and Sheeley, N. R., Jr., 1994, `The Rotation of Photospheric Magnetic Fields: A Random Walk Transport Model', Astrophys. J. 430, 399-412.

\noindent
Weisstein, Eric W., 2004, `Kermack-McKendrick Model.' From MathWorld--A Wolfram Web Resource.
https://mathworld.wolfram.com/Kermack-McKendrickModel.html




\end{document}